\documentclass[11pt,a4]{article}
\pdfoutput=1
\usepackage{graphicx,amssymb,amsmath,amsfonts}
\usepackage{hyperref}
\usepackage[margin=1.2in]{geometry}
\usepackage[noadjust]{cite}
\usepackage[utf8]{inputenc}

\def\be{\begin{equation}}
\def\ee{\end{equation}}

\numberwithin{equation}{section}
\def\bea{\begin{eqnarray}}
\def\eea{\end{eqnarray}}

\newcommand{\alp}{\ensuremath{\alpha^\prime}}

\newcommand{\ssb}{s\bar{s}}
\newcommand{\ccb}{c\bar{c}}

\newcommand{\bbb}{b\bar{b}}
\newcommand{\MEV}{\text{ MeV}}
\newcommand{\GEVm}{\text{ GeV}^{-2}}
\newcommand{\GEV}{GeV\(^{-2}\)}

\newcommand{\jp}[2]{\ensuremath{\frac{#1}{2}^{#2}}}
\newcommand{\jph}[2]{\ensuremath{#1/2^{#2}}}

\newcommand{\plm}{\ensuremath{\pm}}

\newcommand{\mud}{m_{u/d}}

\begin{document}
\begin{titlepage}
\title{\textbf{Excited mesons, baryons, glueballs and tetraquarks: Predictions of the Holography Inspired Stringy Hadron model}}

\author{\textbf{Jacob Sonnenschein} \\ \href{mailto:cobi@post.tau.ac.il}{cobi@post.tau.ac.il} \and \textbf{Dorin Weissman} \\ \href{mailto:dorinw@mail.tau.ac.il}{dorinw@mail.tau.ac.il}}

\date{\emph{The Raymond and Beverly Sackler School of Physics and Astronomy},\\
	\emph{Tel Aviv University, Ramat Aviv 69978, Israel} \\ \today}
	
%\emailAdd{}
%\emailAdd

\maketitle

\begin{abstract} 
In this note we collect and summarize the predictions of the Holography Inspired Stringy Hadron (HISH) model. Following a brief review of the model, we list the masses and widths of predicted excited states across the spectrum, based on placing the different hadrons on the non-linear Regge trajectories of a string with massive endpoints. Our predicted states include: (i) Light, heavy-light and heavy-heavy mesons. (ii) Baryons, including charmed, doubly charmed and bottom baryons. (iii) Glueballs, together with a method to disentangle them from flavorless mesons. (iv) Genuine tetraquarks, which are not ``molecules" of hadrons, and are characterized by their decay into a baryon and an anti-baryon.
\end{abstract}

\end{titlepage}
%\keywords{}
%\preprint{}

%&&&&&&&&&&&&&&&&&&&&&&&&&&&&&&&&&&&&&&&&&&&&&&
%\begin{document}
%\maketitle
\tableofcontents
\flushbottom

\section {Introduction}
The Holography Inspired Stringy Hadron (HISH) model \cite{Sonnenschein:2016pim} describes hadrons: mesons, baryons, glueballs and exotic hadrons in terms of bosonic strings in flat four dimensional space-time. The open strings have particles: ``quarks'', ``antiquarks'', or ``diquarks'' on their endpoints. The particles are massive and may also carry electric charges and/or spin.

The basic open string configuration with massive endpoints describes a meson. The corresponding trajectories, which we refer to as HISH modified Regge trajectories (HMRTs), were determined in \cite{Sonnenschein:2014jwa} and fitted to the observed meson spectrum. The mass of the quark as an endpoint particle $m_{sep}$ is neither the QCD current mass nor the constituent quark mass. A baryon corresponds to a string connecting a quark on one end and  a diquark on the other, the latter being composed of a baryonic vertex with two short strings attached to it. The spectra of the baryons and their fits were worked out in \cite{Sonnenschein:2014bia}. Folded closed strings describe glueballs \cite{Sonnenschein:2015zaa}. Exotic tetraquarks are formed of strings connecting baryonic vertices and anti-baryonic vertices \cite{Sonnenschein:2016ibx}. On top of determining the spectra of hadrons the HISH model enables us to compute the strong decay width of the various hadrons, both the total decay width as well as some branching ratios of the different decay channels \cite{Sonnenschein:2017ylo}.

An important feature of the HISH model is that it provides a unified description of both light and heavy hadrons. The endpoint masses allow us to describe the HMRTs even of hadrons containing \(c\) and \(b\) quarks, with the same slope as for the corresponding light hadrons' trajectories. Only for the bottomonium we find there are deviations from the universal slope. 
 
The HISH model may be viewed as  a ``renaissance'' of the old Regge model\footnote{The old stringy description of hadrons, which was one of the founding motivations of string theory and has been thoroughly investigated since the 1970s \cite{Collins:book}.} enhanced with several new features that follow from the string/gauge holographic duality. The model is built as a phenomenological model, and emphasizes comparison of its results with experimental data. Based on fits to data, we have extracted the best fit values of the basic physical parameters of the model which are the string tension (or the slope $\alp$) and the masses of the endpoint particles $m_{sep}$. In addition we fit all the intercepts $a$ associated with the different trajectories. The intercepts however are not free parameters and in principle should be calculable in the model. With these parameters we can predict the existence of additional yet undiscovered hadrons. 

In this note we summarize the predictions made in our previous papers on the HISH model, add additional predictions and further discussions, and point out certain issues which we believe are of special interest.

We list predictions for higher excited states for each of the leading meson trajectories, starting with the light mesons, the likes of the $\pi$ and \(\rho\), going through heavy-light mesons like the \(D\) and \(B\) and then all the way to heavy quarkonia including $\Upsilon$ and $\eta_b$. This is done for both angular momentum, for instance the states  $7^{--}$ and $8^{++}$ on the $\rho/a$ trajectory, as well as the radial excitations, for instance $n=4$ and $n=5$ on the $\Psi$ trajectory.  Unexpectedly, according to the HISH model there are certain excited states of the kaon that have not been confirmed, for instance $K$ with $3^+$.

Similarly we predict the masses and widths of higher states for the different baryon trajectories. The light flavor baryons admit a peculiar behavior that the states with even and odd angular momentum are on separate parallel trajectories. The odd and the even states have different intercepts. We still do not possess a theoretical explanation for this effect.\footnote{An attempt to explain this was made in \cite{Selem:2006nd}.} 
%It is still not clear if this phenomenon occurs also for the $\Omega$ baryons. If it does, we predict that the already discovered state with mass of $2250$ has to have $J=\frac{5}{2}$, and if it does not, there should be an additional state with mass of  $2050$.
As was discussed in \cite{Sonnenschein:2014bia} (section 6.2) there are a priori ambiguities about the structure of the diquark of the HISH baryon. First one must determine the content of the diquark. For instance in a doubly charmed baryon with quark content of $ucc$ it could be either $(uc)$ or $(cc)$. Second, the diquark masses as string endpoints have to be determined separately from the quark masses. For instance one may ask what is the mass of the \((cc)\) diquark. Naturally it is expected to be close to $2m_c$. However, in holography this is far from obvious, and it can have any value starting from $m_c$ and up to $2m_c$ or even higher. This depends on the location and tension of the baryonic vertex, which is a model dependent quantity \cite{Seki:2008mu,Dymarsky:2010ci}. We elaborate in this note about the various different predictions that follow from this situation. In particular we list the predictions of both options of the diquark in the \(\Omega_c\) who is either \((ss)c\) or \(s(sc)\), following its recently discovered first excited states. We also include predictions for excited bottom baryons: \(\Lambda_b\), \(\Sigma_b\), \(\Xi_b\), and \(\Omega_b\).

%%%%%%%%%%%%%%%%%%%%
Glueballs are described as closed strings, as described in works such as \cite{Bhanot:1980fx,BoschiFilho:2002ta,Lucini:2004my}. The lore about the glueballs is that there is no way to separate them from flavorless mesons. As was discussed in \cite{Sonnenschein:2015zaa} since mesons are open strings and glueballs are closed string, there should be a way to distinguish between them. The latter have a slope which is half that of the former. The slope of glueball trajectories can also be measured on the lattice as we did in \cite{Sonnenschein:2015zaa} or as in \cite{Meyer:2004jc,Athenodorou:2010cs}. In addition the low lying states of closed string should have only even angular momentum and even parity and charge conjugation. To find the possible trajectories of glueballs we use four candidates for the scalar glueball: \(f_0(980)\), \(f_0(1370)\), \(f_0(1500)\), and \(f_0(1710)\). They are chosen as the ground states of the glueball trajectories, while the rest of the flavorless states are placed on nearly linear trajectories for light mesons and  mass corrected trajectories of \(\ssb\). For each of these cases we predict the masses of the $n=2$ and $n=4$ excited scalar glueball states. In case that these states will be found for a particular ground state one will be able to declare an identification of a glueball trajectory and hence also of the individual glueball states. We similarly discuss tensor glueballs and the spectrum of \(f_2\) states. One supportive evidence for the existence of the glueball among these states is the fact that one cannot put all the flavorless states only on meson trajectories \cite{Sonnenschein:2015zaa}.

Another clue in identifying glueballs can come from their decays, as the closed string has to tear twice to produce two mesons unlike the open string which can tear only once. We work out the ratios of the widths for a decay of two vector mesons built from light quarks, for a decay into two $K^*$ and into two $\phi$ mesons. The presence of these three decay modes is a distinguishing feature of the closed string.

The fourth class of hadrons which we predict using the HISH model are tetraquarks. In \cite{Sonnenschein:2016ibx} we stated that in principle there should be a big zoo of exotic hadrons built from strings connecting baryonic vertices, anti-baryonic vertices, with various combinations of quarks and antiquarks. The focus there was on the simplest such states, the tetraquarks. We argued that if a hadron decays predominantly to a baryon and anti-baryon, it is a signal of a tetraquark. In fact this idea was raised already in the early days of hadron physics \cite{Rosner:1968si} and  has remained unresolved to this day. In recent years various experimental observations, specifically for hadrons including heavy quarks reignited this subject (see for instance \cite{Krokovny,Esposito:2014rxa,Maiani:2014aja}). We use  the  state denoted by $Y(4630)$ that has such a decay, namely to \(\Lambda_c\bar\Lambda_c\), as our prototype case. We have determined where its excited angular momentum and radial trajectory partners should be found. We predicted that a similar bottomonium-like state should have a decay process  to $\Lambda_b \bar \Lambda_b$ and its own trajectories of excited states. There may be also the hidden strange equivalent decaying to  $\Lambda  \bar  \Lambda $.
The $Y(4630)$ state and its bottom or strange analogues are all $1^{--}$ states, and hence should be found in $e^+e^-$ collisions. 
 We collected in this note the data about the predicted masses and widths for such tetraquarks. We also suggest the existence of a trajectory as a way to distinguish between a genuine tetraquark and a bound state/molecule of two mesons. To find the trajectory one should naturally look above the corresponding baryon-antibaryon threshold, although narrower below threshold states can also be predicted.
 
The paper is organized as follows: After this introduction we bring a brief summary of the HISH model in section \ref{sec:HISH}. We then describe the fitting model used to compare the outcome of the theoretical model with the experimental observations in section \ref{sec:Fitting}, where we write down all the formulas used to predicted excited states. We discuss the parameters of our model, namely the Regge slope $\alp$ , the various string endpoint masses $m_{sep}$ and the intercepts associated with the different trajectories.  We also discuss the uncertainties of our predictions.

The following sections are where we list all our predictions. In section \ref{sec:mesons} we discuss excited states on the meson trajectories. We have both orbital trajectories of states of high angular momentum and radial trajectories. For each trajectory we list the next two states. This is done for light mesons, light-heavy mesons and heavy-heavy charmonium and bottomonium. Similarly predictions about baryons are described in \ref{sec:baryons}. This include predicted masses for the excited states of $\Omega^-$ baryons, charmed baryons (\(\Lambda_c\), \(\Sigma_c\), \(\Xi_c\)), the doubly charmed baryon \(\Xi_{cc}^{++}\), and the various bottom baryons: $\Lambda_b$, \(\Sigma_b\), \(\Xi_b\), and \(\Omega_b\).

Next we present predictions and a method of identifying glueballs in section \ref{sec:glueballs}. Assuming that each of four candidates among the flavorless $f_0$ scalars, namely, \(f_0(980)\), \(f_0(1370)\), \(f_0(1500)\), and \(f_0(1710)\) are the ground state of the glueball trajectory, we predict the spectrum of excited glueballs that will follow. The glueball trajectories are characterized by a slope which is half that of the mesons. We also assign the rest of the flavorless states to normal meson trajectories. A similar procedure is applied also to tensor glueball states and the spectrum of the \(f_2\) states. We also discuss the total width of the glueballs and also the ratio between various different decay channels. Section \ref{sec:tetra} is devoted to predictions about tetraquarks. We analyze the $Y(4630)$ state that was observed to decay to $\Lambda_c\bar \Lambda_c$. We analyze the corresponding massive modified Regge trajectory and predict the states that reside on it. We perform a similar analysis for a bottomonium-like tetraquark, and tetraquarks in the light-quark sector. In the last part, section \ref{sec:summary}, we summarize and outline several additional types of predictions that could be extracted from the HISH model. 

%%%%%%%%%%%%%%%%%%%%%%%%%%%%%%
\section {A brief review of  the HISH model} \label{sec:HISH}
%%%%%%%%%%%%%%%%%%%%%%%%%%%%%%%%%%%%%%%%%%
The holographic duality is an equivalence between certain bulk string theories and boundary field theories. The original duality is between the ${\cal N}=4$ SYM theory and string theory in $AdS_5\times S^5$. The  ${\cal N}=4$ theory is obviously not the right framework to describe hadrons that resemble those found in nature. Instead we need a stringy dual of a four dimensional gauge dynamical system which is  non-supersymmetric, non-conformal, and confining. The  two main requirements on the desired string background is that it admits confining Wilson lines, and that it is dual to a boundary that  includes a matter sector, which is invariant under a chiral flavor symmetry that is spontaneously broken. There are by now  several ways to get a string background which is dual to a confining boundary field theory. For a review paper and a list of relevant references see for instance \cite{Aharony:2002up}.

Practically most of the applications of holography of both conformal and non-conformal  backgrounds are based on relating bulk fields (not strings) and operators on the dual  boundary field theory. This is based on the usual limit of $\alp\rightarrow 0$  with which we go, for instance, from a closed string theory to a theory of gravity.

However, to describe realistic hadrons we need strings rather than bulk fields since in nature the string tension, which is related  to $\alp$ via $T=(2\pi \alp)^{-1}$, is not very large. In gauge dynamical terms the IR region is characterized by $\lambda= g^2 N_c$ of order one rather than very large.

It is well known that in holography there is a wide sector of gauge invariant  physical observables  which cannot be faithfully described by bulk fields but rather require dual stringy phenomena. This is the case for  Wilson, 't Hooft, and  Polyakov lines.
 %low x DIS and also Entanglement entropy

In the  holography inspired stringy hadron (HISH) model \cite{Sonnenschein:2016pim}  we argue that in fact also  the  description of the spectra, decays and other properties of hadrons - mesons, baryons, glueballs and exotics - should be recast as a description in terms of holographic stringy configurations only, and not fields. The major argument against describing the hadron spectra in terms of fluctuations of fields, like bulk fields or modes on probe flavor branes, is that they generically do not reproduce the Regge trajectories (or more precisely the HMRT) in the spectra, namely, the (almost) linear relation between $M^2$ and the angular momentum $J$. Moreover, for top-down models with the assignment of mesons as fluctuations of flavor branes  one can get mesons only with $J=0$ or $J=1$. Higher $J$ states will have to be related to strings, but then there is a big gap of order $\lambda$ (or some fractional power of $\lambda$ depending on the model) between the low and  high $J$ mesons. Similarly  the attempts to get the observed linearity between $M^2$ and the excitation number $n$ are problematic whereas for strings it is an almost trivial property.

\begin{figure}[t!] \centering
\includegraphics[width= 0.9\textwidth]{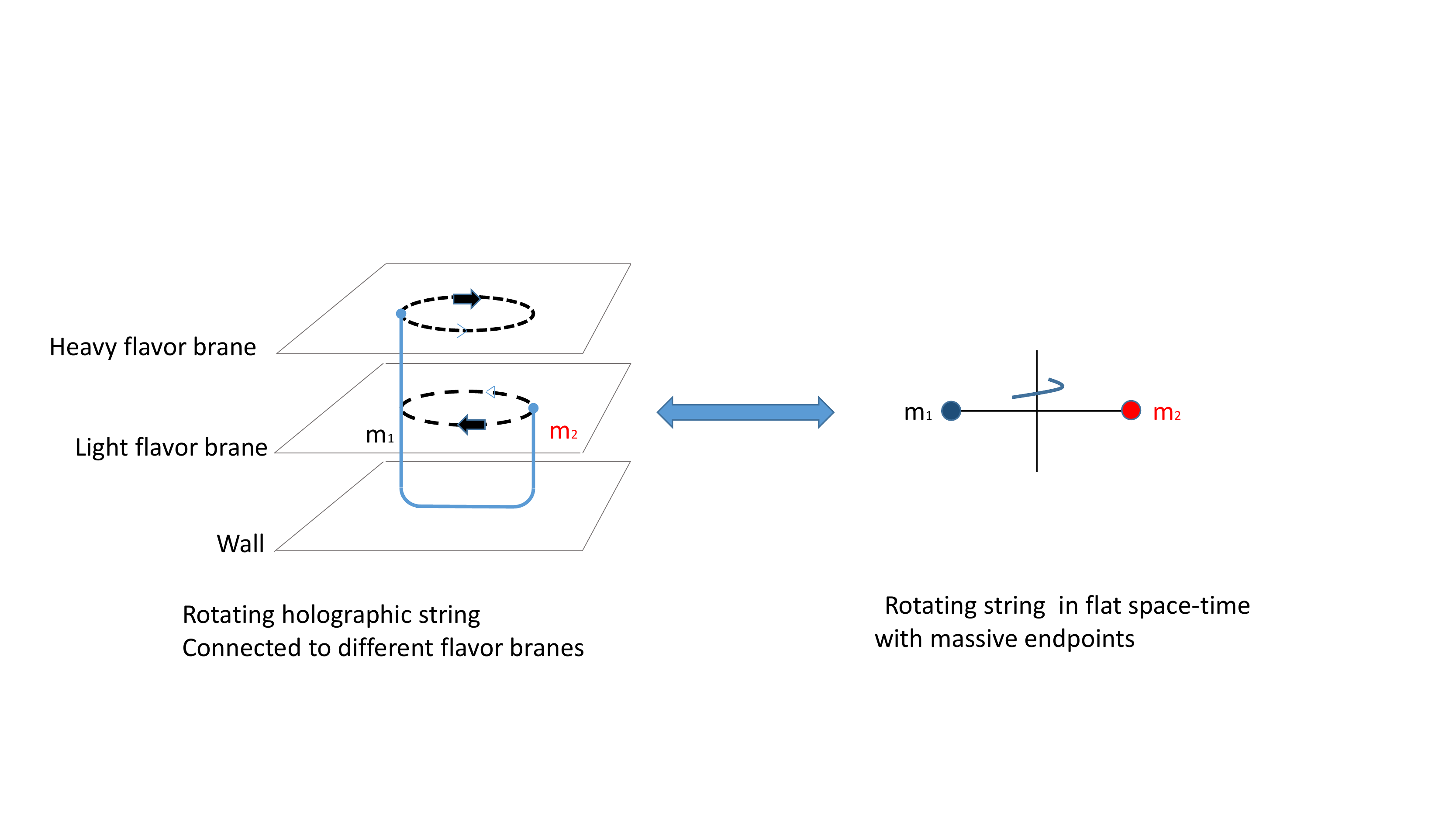}
					\caption{\label{fig:mapholflat} \textbf{Left:} Rotating holographic open string. \textbf{Right:} The corresponding open string with massive endpoints in flat spacetime. In this case we show a heavy-light meson.\label{hishmap}}
		\end{figure}

The construction of the HISH model is based on the following steps. (i) Analyzing string configurations in holographic string models that correspond to hadrons. (ii) Devising a  transition from the holographic regime of large $N_c$ and large $\lambda$ to the real world that bypasses formally taking the limits of $\frac{1}{N_c}$ and $\frac{1}{\lambda}$ expansions. (iii) Proposing a model of stringy hadrons in four flat dimensions that is inspired by the corresponding holographic strings. (iv) Confronting the outcome of the models with the experimental data (as was done in \cite{Sonnenschein:2014jwa,Sonnenschein:2014bia,Sonnenschein:2015zaa}) to measure the physical parameters and learn what additional features the stringy model should have.

Confining holographic models are characterized  by a ``wall'' that truncates in one way or another the range of the holographic radial direction. A common feature to all the holographic stringy hadrons is that there is a segment of the string that stretches along a constant radial coordinate in the vicinity of the ``wall'', as in figure \ref{hishmap} \cite{PandoZayas:2003yb,Kruczenski:2004me}. For a stringy glueball it is the whole folded closed string that rotates there and for an open string it is part of the string, the horizontal segment, that connects with vertical segments either to the boundary for a Wilson line or to flavor branes for a meson or baryon. This fact that the classical solutions of the flatly  rotating strings reside at fixed radial direction is a main rationale behind the map between rotating strings in curved spacetime and rotating strings in flat spacetime described in figure \ref{fig:mapholflat}. 

A key player of the map is the ``string endpoint mass'', $m_{sep}$, that provides in the flat spacetime description the dual of the vertical string segments. It is important to note that this mass is neither the QCD mass parameter (the current quark mass) nor the constituent quark mass, and that the massive endpoint as a map of an exactly vertical segment is an approximation that is more accurate the longer the horizontal string is.

The stringy picture of mesons has been thoroughly investigated in the past (see \cite{Collins:book} and references therein). In recent years there have also been attempts to describe hadrons in terms of strings. Papers on the subject that have certain overlap with our approach are \cite{Baker:2002km,Schreiber:2004ie,Bigazzi:2006jt,Iengo:2006gm,Bigazzi:2007qa,Armoni:2009zq,Hellerman:2013kba,Zahn:2013yma,Hellerman:2014cba,Dubovsky:2015zey,Dubovsky:2016cog,Rossi:2016szw}. A somewhat different approach to the stringy nature of QCD, at least conceptually, is the approach of low-energy effective theory on long strings reviewed in \cite{Aharony:2013ipa}. 

%%%%%%%%%%%%%%%%%%%
%%%%%%%%%%%%%%%%%%%

\begin{figure}[t!] \centering
					\includegraphics[width=.90\textwidth]{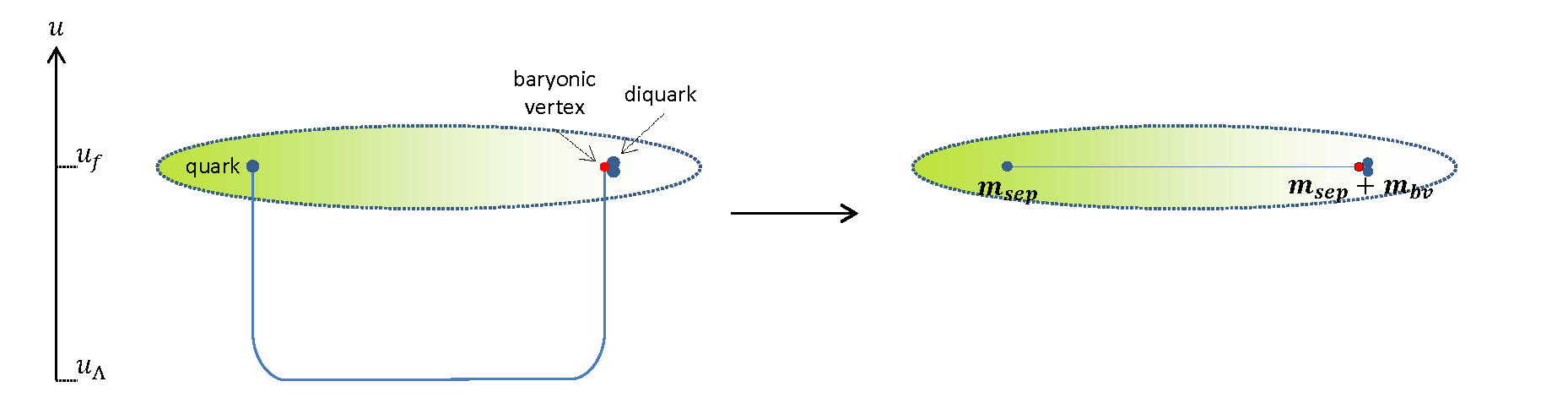}
					\caption{\label{holtoflat2}  The holographic setup (for \(N_c=3\)) of a quark and a diquark is mapped to a similar configuration in flat spacetime. The vertical segments of the holographic string are mapped into masses of the endpoints. The mass of the baryonic vertex also contributes to the endpoint mass.}
\end{figure}

The HISH model describes hadrons in terms of the following basic ingredients:
\begin{itemize}
\item
Open strings which are characterized by a tension $T$, or equivalently a slope $\alp=(2\pi T)^{-1}$). The open string generically has a given energy/mass and angular momentum associated with its  rotation. The latter gets contribution from the classical configuration and in addition there is also a quantum contribution  characterized the intercept $a$. 
%An essential property of the HISH intercept is that it must always be negative, $a<0$. 
When considering trajectories of the orbital angular momentum (not including the spin) as a function of  $M^2$ it is a universal  experimental fact that the corresponding intercepts are always negative $a<0$. This property is responsible to the fact that the ground state of the stringy hadrons is not tachyonic.  The negative intercept also provides a repulsive Casimir force that guarantees that even a non-rotating stringy hadron has a finite length. We do not have yet a theoretical explanation to this essential property of  the hadronic spectra. The quantization of the HISH string including the determination of the intercept and the spectrum of excited states  was performed in \cite{Sonnenschein:2018aqf}. 
  
\item
Massive particles - or ``quarks'' - attached to the ends of the open strings   can have four\footnote{There is no a priori reasoning behind assuming \(m_u = m_d\) in the HISH model, but the difference between the two masses is too small to be relevant (or measurable) in the current work.} different values $m_{u/d}$, $m_s$, $m_c$, $m_b$. The latter are determined by fitting the experimental spectra of hadrons. One of the successes of the HISH model is the fact that all the HMRT can be described by this set of universal values of the masses. This applies both for mesons and for baryons.  The endpoint particles obviously  contribute to the energy and angular momentum of the hadron of which they are part. Moreover, the endpoint particles of the string can carry electric charge, flavor charges and spin.
These properties affect the value of the intercept as is reflected by the differences of the values of the intercept obtained for trajectories of hadrons with different quark content and spin.

\item
``Baryonic vertices'' (BV) which are connected to a net number of \(N_c=3\) strings. In holography the BV is built from a $D_p$ brane wrapping a \(p\)-cycle connected to flavor branes by $N_c$ strings. A priori there could be different metastable configurations
of the $N_c$ strings. In the HISH we take that the preferable string configuration is that of a single long string and two very short ones. Since the endpoints of the two short strings are one next to the other we can consider them as forming a diquark. There are two arguments in favor of this string setup. It was shown that the Y-shape configuration, which is the most symmetric form when $N_c=3$, is in fact unstable and an introduction of a small perturbation to a rotating Y-shaped string solution will cause it to evolve into the string between quark and diquark. Secondly, had the Y-shape string been stable, the baryon trajectory slope $\alp_B$ should have been $\alp_B= (2/3) \alp_m$, where $\alp_m$ is that of a meson. However as was shown in \cite{Sonnenschein:2014bia} the slope of the baryonic trajectory is within $5\%$ the same as that of a meson trajectory. 

The setup of a holographic baryon and its map to HISH model are depicted in figure \ref{holtoflat2}. We emphasize that unlike other models which have quarks and diquarks as elementary particles (such as \cite{Friedmann:2009mx}), in the HISH model the diquark is always attached to a baryonic vertex. A BV can be also connected to any of more complex combinations of quarks and anti-baryonic vertices as discussed in \cite{Sonnenschein:2016ibx}.

\item Closed strings which have an effective tension that is twice the tension of the open string ($\alp_{closed}= \frac12 \alp_{open}$). They can have non-trivial angular momentum by taking a configuration of a folded rotating string. The excitation number of a closed string is necessarily even, and so is the angular momentum (on the leading Regge trajectory). The close string intercept, the quantum contribution to the angular momentum, is also twice that of an open string ($a_{closed}=  2 a_{open}$). These different properties of closed strings, the slope, the fact that the angular momentum has to be even, and the intercept, can serve as a way to disentangle between flavorless mesons and glueballs\cite{Sonnenschein:2015zaa}.
	\end{itemize}
	
Hadrons, namely mesons, baryons and glueballs are being constructed in the simplest manner from the HISH building blocks.
 \begin{itemize}
\item
A single open string attached to two massive endpoint particles corresponds to a meson.
\item
A single string that  connects on one end to a quark and on the other hand a baryonic vertex with a diquark attached to it is the HISH description of a baryon.
\item
A single closed string is a glueball.
\item
A  single string connecting  a baryonic vertex with a diquark on one end and an anti-baryonic vertex and anti-diquark, is a stringy description of a tetraquark which may be realized in nature.
\end{itemize}

\section{Fitting model and parameters \label{sec:Fitting}}
All the different hadrons described by the HISH model will be grouped into the HISH modified Regge trajectories of strings with massive endpoints (HMRTs). From the relation between angular momentum and energy of such a string we can derive the equations \cite{Sonnenschein:2016pim}
\begin{align} M =& \sum_{i=1,2}\left(\frac{m_i}{\sqrt{1-\beta_i^2}} + T\ell_i\frac{\arcsin{\beta_i}}{\beta_i}\right)\,, \\
J + n -a  =& \sum_{i=1,2}\left(\frac{m_i \beta_i \ell_i}{\sqrt{1-\beta_i^2}} + \frac12T\ell_i^2\frac{\arcsin\beta_i-\beta_i\sqrt{1-\beta_i^2}}{\beta_i^2}\right)\,. \label{eq:MJ}\end{align}
Here \(T\) is the string tension, \(\beta_i\) is the velocity of the endpoint with the mass \(m_i\), and \(\ell_i\) the distance of the mass from the center of mass around which the endpoint particles rotate. The endpoint velocities \(\beta_i\) are related to each other from the condition that the angular velocity is the same for both arms of the string, implying
\be \omega = \frac{\beta_1}{\ell_1} = \frac{\beta_2}{\ell_2}\,,\ee
while the boundary conditions of the string imply
\be \frac{T\ell_i}{m_i} = \frac{\beta_i^2}{1-\beta_i^2}\,. \ee

The last two equations provide enough conditions that together with the equations for \(M\) and \(J\) they define the function \(J(M)\) which is the classical HMRT of the string with massive endpoints. To fit the mesons and baryons we add the quantum intercept to the classical expressions by taking \(J\to J-a\). To discuss also radial excitations we take \(J\to J+n-a\), where \(n = 0,1,\ldots\) is the radial excitation number. As we have shown in our more recent work \cite{Sonnenschein:2018aqf}, this is not the exact form of the quantum corrections, but it works well as a first approximation, especially considering we leave the intercept as a free parameter. Note also that we define the trajectory in terms of the orbital angular momentum and not the total one, and if there is additional spin in the hadron the intercept is shifted accordingly. In other words, all trajectories start from \(N = J_{\mathrm{orb}}+n = 0\).

In the high energy or low mass limit where the endpoint masses are ultra-relativistic with \(\beta \rightarrow 1\), we can write an expansion in \(\frac mE\) (we write here the symmetric case \(m_1=m_2\) only for simplicity):
		\be J + n - a  = \alp E^2\left(1-\frac{8\sqrt{\pi}}{3}\left(\frac{m}{E}\right)^{3/2} + \frac{2\sqrt{\pi^3}}{5}\left(\frac{m}{E}\right)^{5/2} + \cdots\right) \label{eq:lowMass}\ee
		from which one can see that the linear Regge behavior is restored in the limit \(m\rightarrow 0\), and that the first correction is proportional to \(\alp m^{3/2}E^{1/2}\). The Regge slope \(\alp\) is related to the string tension as usual by \(\alp = (2\pi T)^{-1}\). The opposing low energy limit, \(\beta \rightarrow 0\), holds when \((E-2m)/2m \ll 1\). Then the expansion is
		\be J + n - a  = \frac{4\pi}{3\sqrt{3}}\alp m^{1/2} (E-2m)^{3/2} + \frac{7\pi}{54\sqrt{3}} \alp m^{-1/2} (E-2m)^{5/2} + \cdots \label{eq:highMass} \ee
		The HMRT describes then the interpolation from this highly non-linear behavior of \(J \propto (E-2m)^{3/2}\) for high endpoint masses to the familiar linearity of \(J \propto E^2\) for light quark masses.

The width of a string state is given by the formula
\be \Gamma/\Phi(M,M_1,M_2) = \frac{\pi}{2} A T L(J,m_1,m_2,T,a)\,, \label{eq:Gamma}\ee
Here \(L\) is the string length, which is a function of the mass (or angular momentum) of the state and depends on string tension \(T\) and endpoint masses. It also receives quantum corrections which are manifested as a dependence on \(a\). In addition to it we have a phase space suppression factor \(\Phi\), which we introduce to fit those states that are just above the threshold for decays.\footnote{The example where this factor is most relevant is the \(\phi\), which has to decay into \(K\bar K\) and is just above the threshold to do so, making it narrower than the formula without the extra suppression factor would suggest.} The factor is given by
\be \Phi(M,M_1,M_2) \equiv 2\frac{|p_f|}{M} = \sqrt{\left(1-(\frac{M_1+M_2}{M})^2\right)\left(1-(\frac{M_1-M_2}{M})^2\right)}\,.\ee
where \(M_1\) and \(M_2\) are the masses of the outgoing states for a given channel. 

Note that for the predictions for higher states listed here we do not use this phase space suppression, taking it to be one,  but it is used in some of the fits to the lower states from which the free parameter \(A\) is determined. This \(A\) is in fact the dimensionless asymptotic ratio of \(\Gamma/M\) at high energies. For mesons it was found to be \(A\approx 0.1\) for most trajectories examined.

The statement that the decay width is proportional to the string length is true in a frame where the string is static. For the rotating string we have to adjust for the position dependent time dilation along the string, which affects the probability to tear and decay at each point:
\be \Gamma_{rot} = \int_{-\ell_1}^{\ell_2} d\sigma \frac{1}{\gamma(\sigma)}(\frac{\Gamma}{L})_{stat} \ee
Therefore the length \(L\) of the string that enters in formula \ref{eq:Gamma} is in fact adjusted for this time dilation and given by
\be L = \int_{-\ell_1}^{\ell_2} d\sigma\sqrt{1-\omega^2\sigma^2} = \sum_{i=1,2}\frac{\ell_i}{2}\left(\sqrt{1-\beta_i^2}+\frac{\arcsin\beta_i}{\beta_i}\right)\,. \ee
The effective \(L\) differs from the length of the string \(\ell_1+\ell_2\) by a factor which is a function of \(\beta_i\), and this factor goes from \(1\) for \(\beta_i \ll 1\) to \(\pi/4\approx0.79\) as \(\beta_i\to1\).

For the decays the intercept is also introduced by taking \(J\to J-a\). This will introduce a quantum ``zero point length'' of the string, and so will give a finite length even to those states with zero orbital angular momentum whose length vanishes classically. The rationale behind choosing this fitting formula for the decays was discussed at length in section 8.1 of \cite{Sonnenschein:2017ylo}. Note that the decay widths in \cite{Sonnenschein:2017ylo} were calculated for the leading (\(n=0\)) orbital trajectories. In this note we assume the replacement \(J\to J+n-a\) is sufficient to give estimates for the widths of states on radial trajectories, even though the dependence of the widths on \(n\) was not fully examined. In any case, we distinguish in the following the widths determined by our model as in \cite{Sonnenschein:2017ylo} and more naive estimates where given.

For glueballs (closed strings), the situation is simpler since we do not have the endpoint masses. We use 
\be J + n - a = \frac12\alp M^2 \ee
The slope of the closed strings is half of the slope of open strings \(\alp\). Here \(J+n\) takes only even values since left and right moving modes must have equal total excitation number for a closed string. The length of the folded closed string \(\ell\) is related to the mass via
\be M = \frac{\pi}{2} T\ell \ee
The effective length adjusted for time dilation that affects the decay width is simply
\be L = \frac\pi4 \ell = \frac{M}{2T} \ee
so the decay width of a glueball is expected to be proportional to its mass. Note that by taking \(\Gamma\propto M\) we implicitly assume how quantum corrections affect the string length. For example, the ground state with \(J = n = 0\) which has a classically vanishing length will have \(M^2 = -2a/\alp\) and thus a ``quantum length'' proportional to \(\sqrt{\alp |a|}\). Whether this is the correct way of defining the quantum length of the closed string is an open question.

\subsection{Parameters}
For the endpoint masses and quarks we take the values obtained from the fits in \cite{Sonnenschein:2014jwa,Sonnenschein:2014bia} and most recently in \cite{Sonnenschein:2017ylo}. The quark masses we use are
\be m_u = m_d = 60 \MEV \qquad m_s = 400 \MEV \qquad m_c = 1490 \MEV \qquad m_b = 4700 \MEV \ee
The heavy quark masses, \(m_c\) and \(m_b\) are rather strongly determined as their constituent quark masses. The mass of the \(c\) quark is half the mass of the lightest charmonium, and likewise for the \(b\). This is what gives the optimal fits in those cases. In \cite{Sonnenschein:2014jwa} the mass of the \(s\) quark was seen to be anywhere between 200 and 400 MeV, with some trajectories being fitted better by lower values of \(m_s\) and other with higher masses. The additional measurements from the decay width fits of \cite{Sonnenschein:2017ylo} are more consistent with the higher values in that range, so we use 400 MeV here. The light quark masses are not uniquely determined. They could be anywhere between zero and up to 100 MeV. The string tension is approximately \(T = (440 \MEV)^2\) so the ratio \(m_u^2/T\) is small in any case. The value 60 MeV used here is an average value that fits well both the spectrum and decay widths of the various mesons.

The Regge slope is largely universal, with some variations. The most common value
\be \alp_{\text{meson}} = 0.88 \GEVm \ee
is common to all the meson trajectories in the \((J,M^2)\) plane, with the bottomonium trajectories being a sole exception. In the \((n,M^2)\) plane the slope is typically somewhat lower, approaching \(0.80 \GEVm\) in some cases \cite{Sonnenschein:2014jwa}. The slope shared by the baryon trajectories was seen \cite{Sonnenschein:2014bia} to be a little higher
\be \alp_{\text{baryon}} = 0.95 \GEVm \ee

For each HMRT there is also the \emph{intercept}, which we leave at this point as a free parameter, to be determined for each trajectory separately by fitting to formula \ref{eq:MJ}.

\subsubsection{Diquark masses}
Our basic assumption is that the baryon can be described as a single string between a quark and a diquark. For the meson we only had to measure the quark masses and string tension, but for the baryons we also need to determine which of the quarks in the baryon form the diquark, and then also the mass of any possible diquark configuration. From the cases examined in \cite{Sonnenschein:2014bia} we only have measurements for diquarks containing at least one \(u\) or \(d\) quark. Our fits indicate that the mass of such a diquark is roughly the same as the mass of the heavier quark in the diquark, that is
\be m_{(qu)} \approx m_{(qd)} \approx m_{q} \ee
In other words, the baryonic vertex and the short string stretched from it to the \(u/d\) flavor brane do not appear to have a contribution to the endpoint mass that is sufficiently large for us to measure. To give an example of how that works, take the charmed baryon with the quark content \(udc\). There are three a priori configurations for the quark-diquark in this baryon, \(u(dc)\), \(c(ud)\), or \(d(cu)\), but for all of them we will only see a string with the mass \(m_c\) at one of the endpoints and \(\mud\) at the other. This could also be an indication that the light quarks themselves should be nearly massless, though we do not use that conclusion where lone light quarks are concerned.

In this work we address also baryons which may have an \((ss)\), \((cs)\), or \((cc)\) diquark. Unfortunately there are too few states from which we can measure the mass of these diquarks. However, our predictions are sensitive to the diquark masses, so any future measurements of the states we predict should help determining them. For these diquarks we assume, for the purpose of providing predictions and as a zeroth order approximation, that the quark masses sum: \(m_{(ss)} = 2m_s\), \(m_{(cs)} = m_c + m_s\), etc. We note again that this seemingly trivial property is not fully apparent in the holographic models.

\subsubsection{Uncertainties}
The predictions presented below are not expected to be fully accurate. The great degree of precision to which some hadrons' masses are known is clearly not achievable in a model as simple as that of a string with massive endpoints. There is then a theoretical systematic error that is dominant when describing the spectrum through modified Regge trajectories alone. A certain deviation of the observed hadron masses from our predicted formulas was seen in our previous works, and is usually at most two to three percent in the mass, with a few exceptional points deviating more. This can also be seen in the plots of the fitted HMRTs in figures \ref{fig:mesons_light}, \ref{fig:mesons_heavy}, and \ref{fig:baryons_J}. A similar deviation of a few percent can also be expected for the masses of the states that we predict below.

\begin{figure}[p!] \centering
	(a)\includegraphics[width=0.44\textwidth]{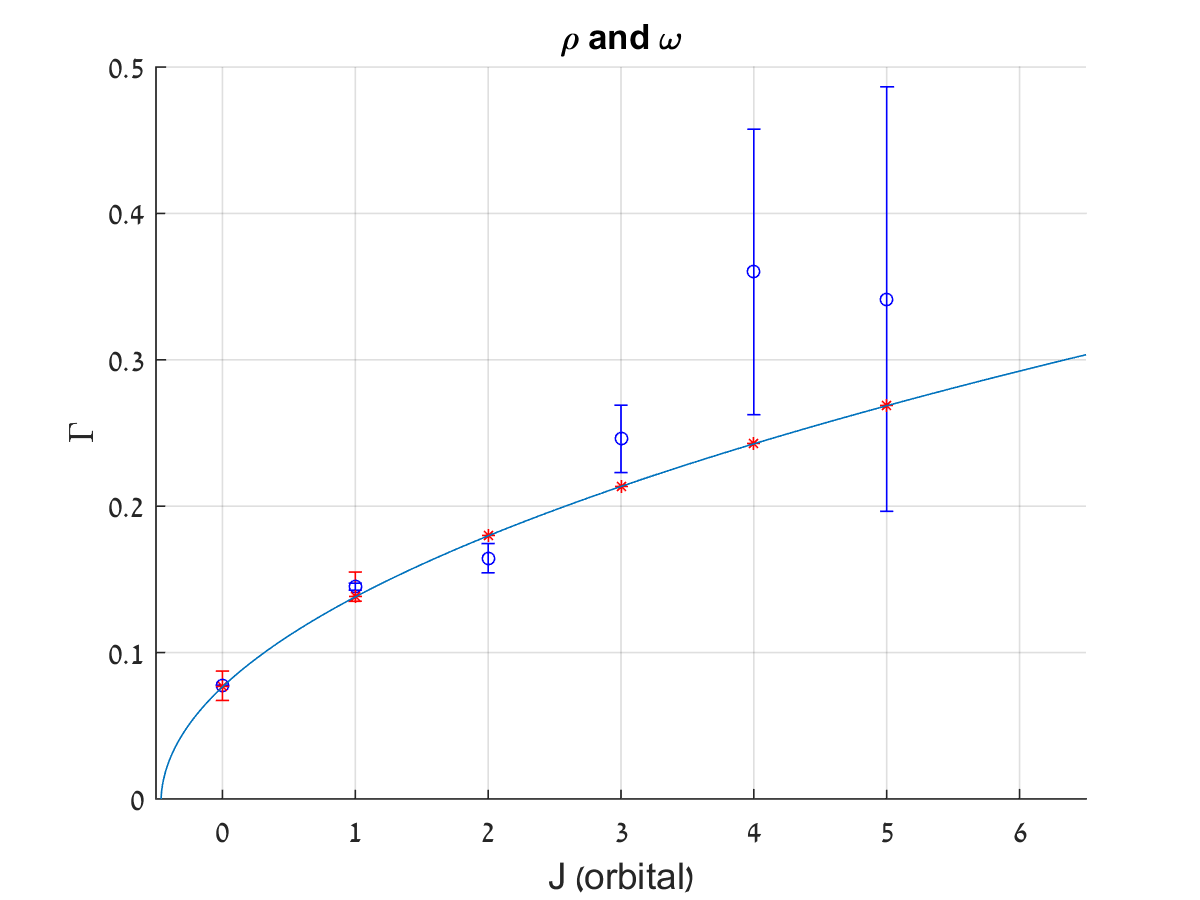}
	(b)\includegraphics[width=0.44\textwidth]{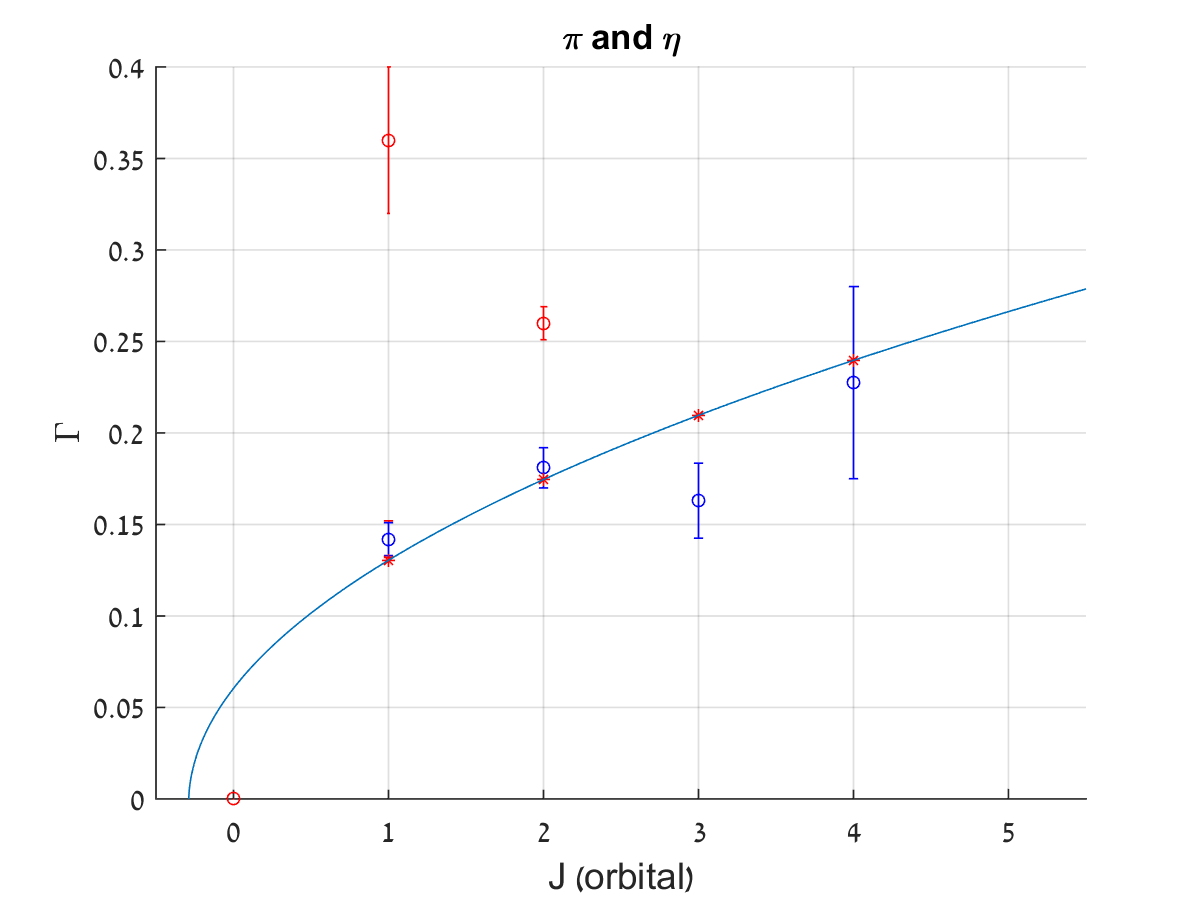} \\
	(c)\includegraphics[width=0.44\textwidth]{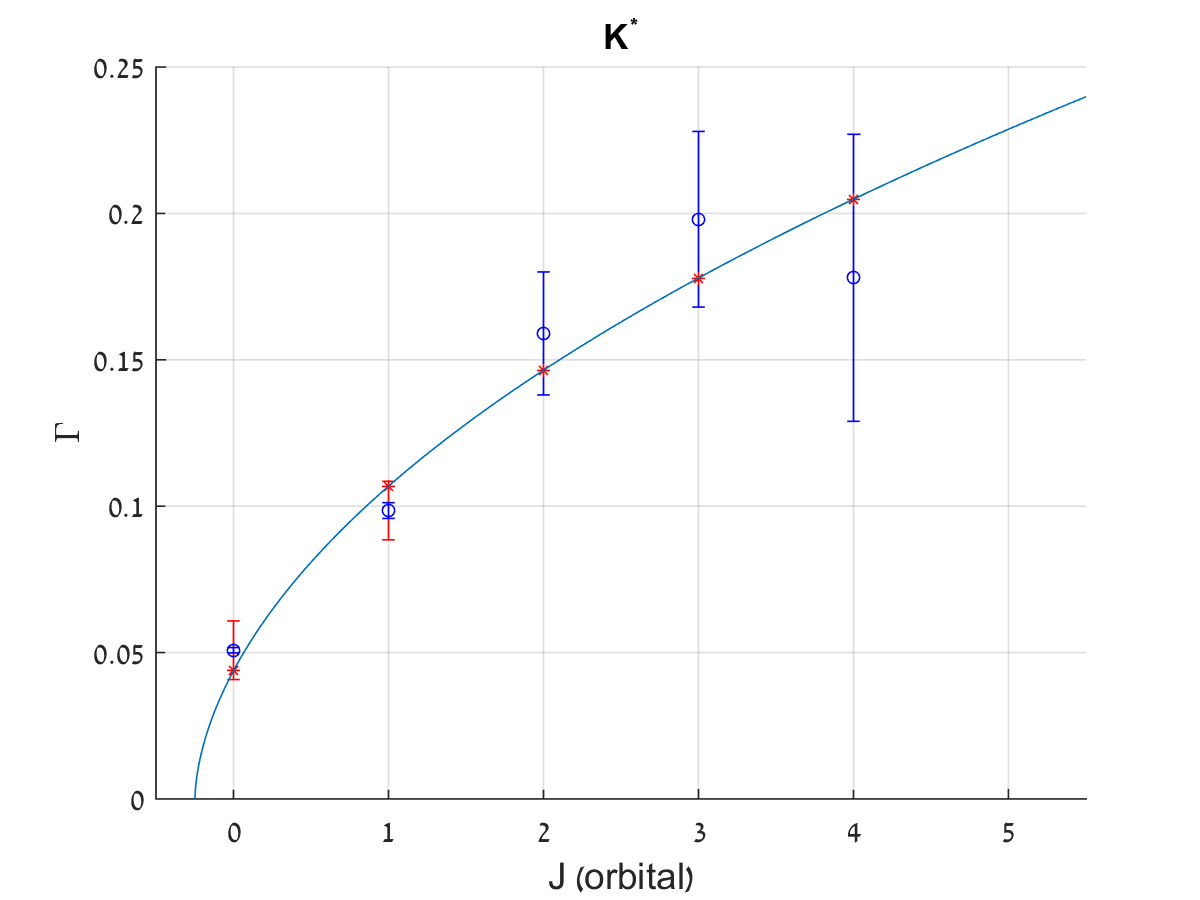}
	(d)\includegraphics[width=0.44\textwidth]{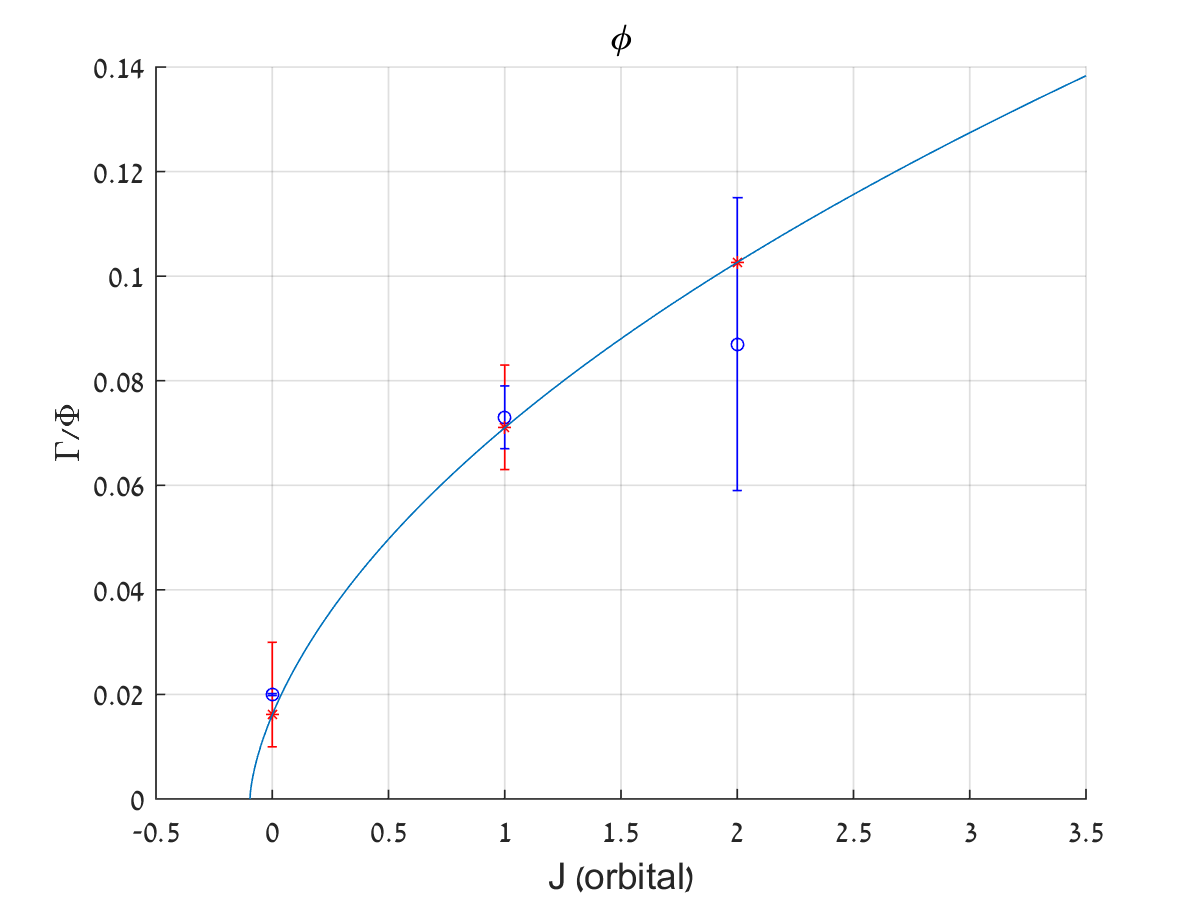} \\
	(e)\includegraphics[width=0.44\textwidth]{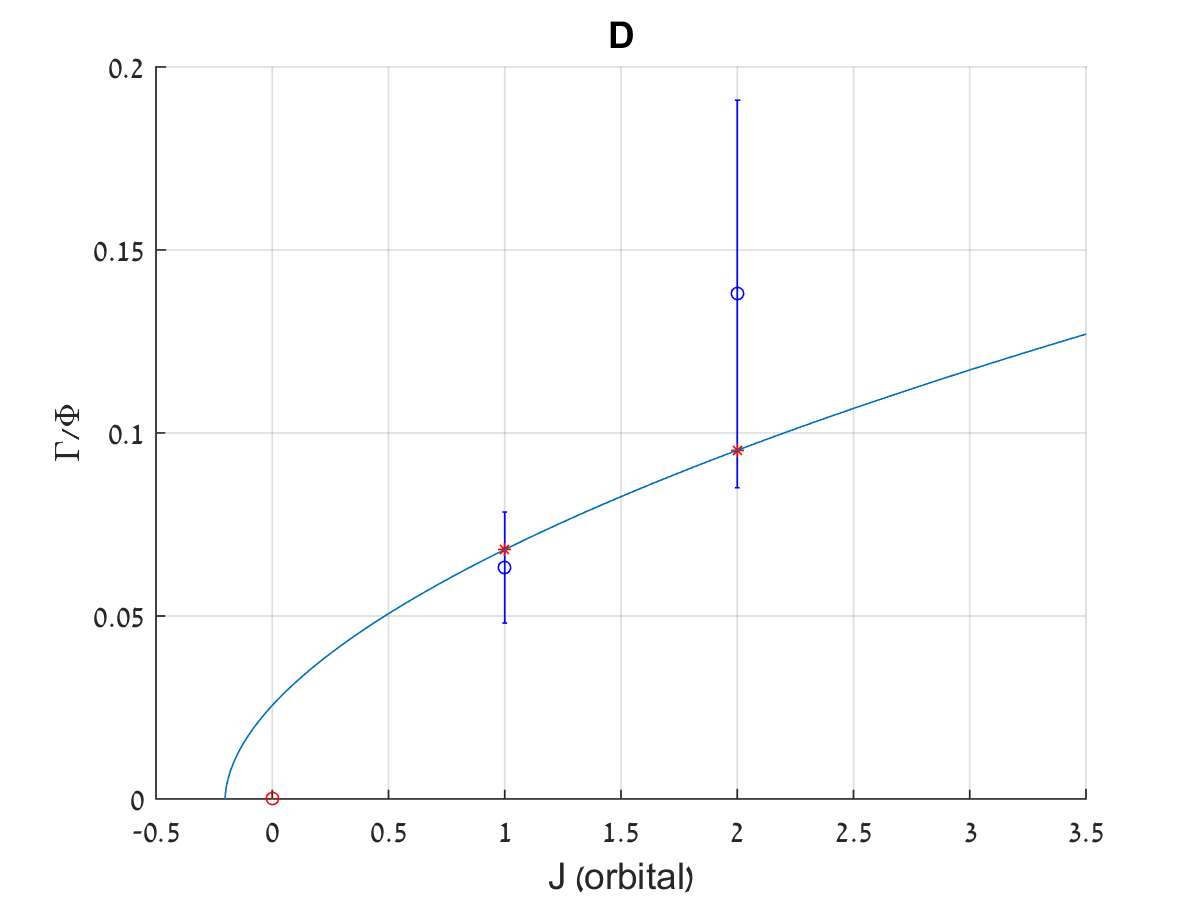}
	(f)\includegraphics[width=0.44\textwidth]{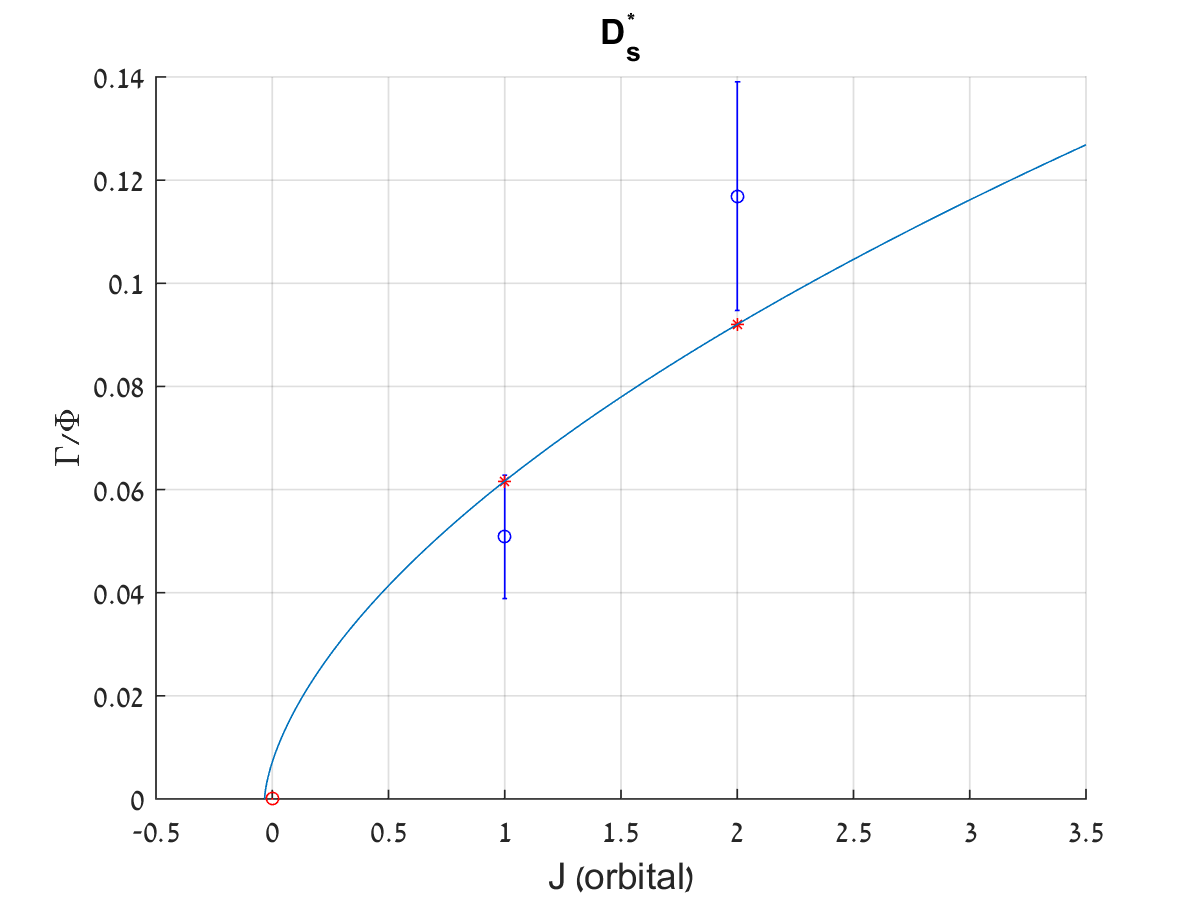}
	\caption{\label{fig:mesons_decay} The meson decay width fits of \cite{Sonnenschein:2017ylo}. We plot the decay width \(\Gamma\) (in some cases normalized by phase space factor as \(\Gamma/\Phi\)) as a function of the orbital angular momentum. In blue are the data points, and the red asterisks along the trajectory are the theoretical values. Some exceptional states were excluded from the fits and are marked in red.}
	\end{figure}

For estimating the decay widths we use less data and the deviations can be larger, in particular for the baryons where our model was less successful in describing the widths of the known states. For states on the orbital trajectories of mesons our calculated widths usually fall within the experimental range of error, but deviations are more common (see figure \ref{fig:mesons_decay} for the fits of the meson widths). The predicted widths listed in table \ref{tab:mesons_J} for higher excited mesons on the orbital trajectories are expected to be the most accurate. For states predicted outside that table, including the baryons and the radial trajectories of the mesons, the widths are provided as an estimate.

%%%%%%%%%%%%%%%%%%%%%%%%%%%%%%
\clearpage
\section{Predictions for meson states} \label{sec:mesons}
\begin{table}[ht!] \centering
		\begin{tabular}{|c|c|ccc|ccc|} \hline
						
		Trajectory & Quarks & \(J^{P[C]}\) & Mass & Width & \(J^{P[C]}\) & Mass & Width \\ \hline
						
		\(\pi/b\) & \(I = 1\) & \(5^{+-}\) & 2480 & 240 & \(6^{-+}\) & 2700 & 270 \\
		
		\(\eta/h\) & \(I = 0\) & \(5^{+-}\) & 2470 & 260 & \(6^{-+}\) & 2690 & 290 \\
						
		\(\rho/a\) & \(I = 1\) & \(7^{--}\) & 2720 & 260 & \(8^{++}\) & 2920 & 280 \\
		
		\(\omega/f\) & \(I = 0\) & \(7^{--}\) & 2710 & 320 & \(8^{++}\) & 2910 & 350 \\
		
		\(K\) & \(s\bar q\) & \(3^+\) & 2050 & 220 & \(4^-\) & 2330 & 250 \\
						
		\(K^*\) & \(s\bar q\) &\(6^+\) & 2620 & 230 & \(7^-\) & 2840 & 250 \\
						
		\(\phi\) & \(s\bar s\) & \(4^{++}\) & 2260 & 130 & \(5^{--}\) & 2520 & 150 \\
						
		\(D\) & \(c\bar q\) & \(3^+\) & 3030 & 70 & \(4^-\) & 3270 & 90 \\
		
		\(D^*\) & \(c\bar q\) & \(4^+\) &  3070 & 100 & \(5^-\) & 3310 & 120 \\
		
		\(D_s\) & \(c\bar s\) & \(2^-\) & 2890 & - & \(3^+\) & 3160 & - \\
		
		\(D^*_s\) & \(c\bar s\) & \(4^+\) & 3160 & 120 & \(5^-\) & 3400 & 140 \\
								
		\(\Psi\) & \(c\bar c\) & \(4^{++}\) & 4020 & 90 & \(5^{--}\) & 4230 & 130 \\
		
		\(\eta_c\) & \(c\bar c\) & \(2^{-+}\) & 3790 & - & \(3^{+-}\) & 4030 & - \\
		
		\(B\) & \(b\bar q\) & \(2^-\)  & 5980 & - & \(3^+\)  & 6210 & - \\
		
		\(B^*\) & \(b\bar q\) & \(3^-\)  & 6000 & - & \(4^+\) & 6230 & - \\
		
		\(B_s\) & \(b\bar s\) & \(2^-\) & 6080 & - & \(3^+\) & 6320 & - \\
		
		\(B^*_s\) & \(b\bar s\) & \(3^-\) & 6100 & - & \(4^+\) & 6330 & - \\
						
		\(\Upsilon\) & \(b\bar b\) & \(4^{++}\) & 10420 & Narrow & \(5^{--}\) & 10630 & - \\
		
		\(\eta_b\) & \(b\bar b\) & \(2^{-+}\) & 10180 & Narrow & \(3^{+-}\) & 10410 & Narrow \\
						
		\hline \end{tabular}
		\caption{\label{tab:mesons_J} Predictions for the next states on the leading HMRTs  in the \((J,M^2)\) plane of various types of mesons. \(q\) signifies a light quark (\(u/d\)). Widths are not written where there is not enough data to determine them.}
		\end{table}
					
					\begin{table}[ht!] \centering
	\begin{tabular}{|c|c|c|ccc|ccc|} \hline
						
	Traj. & Quarks & \(J^{PC}\) & \(n\) & Mass & Width & \(n\) & Mass & Width \\ \hline
						
	\(\pi\) &  \(I = 1\) & \(0^{-+}\) & 5 & 2610 & 300 & 6 & 2830 & 330 \\
						
	\(\pi_2\) & \(I = 1\) &  \(2^{-+}\) & 3 & 2520 & 300 & 4 & 2740 & 350\\
					
	\(a_1\) & \(I = 1\) &  \(1^{++}\) & 2 & 1990 & 350 &  4 & 2520 & 390 \\
						
	\(h_1\) & \(I = 0\) &  \(1^{--}\) & 4 & 2470 & 400 &  5 & 2700 & 450 \\
					
	\(\omega\) & \(I = 0\) &  \(1^{--}\) & 5 & 2560 & 360 & 6 & 2780 & 390 \\
						
	\(\omega_3\) & \(I = 0\) &  \(3^{--}\) & 3 & 2510 & 230 & 4 & 2740 & 250 \\
						
	\(\phi\) & \(s\bar s\) &  \(1^{--}\) & 2 & 2000 & 100 & 4 & 2570 & 120 \\
	
	\(\eta_{c}\) & \(c\bar c\) &  \(0^{-+}\) & 2 & 4020 & - &  3 & 4330 & - \\ 
						
	\(\Psi\) & \(c\bar c\) &  \(1^{--}\) & 4 & 4620 & 110 &  5 & 4860 & 120 \\
	
	\(\chi_{c1}\) & \(c\bar c\) &  \(1^{++}\) & 1 & 3920 & - &  2 & 4240 & - \\ 
						
	\(\Upsilon\) & \(b\bar b\) &\(1^{--}\) & 6 & 11310 & 90  & 7 & 11510 & 100 \\
						
	\(\chi_{b1}\) & \(b\bar b\) &\(1^{++}\) & 3 & 10800 &  - & 4 & 11040 & - \\						
	\hline \end{tabular}
	\caption{\label{tab:mesons_n} Predictions for the next states in the \((n,M^2)\) plane for flavorless mesons. Note that in our convention trajectories start from \(n=0\). Widths are determined using the fitting model of \cite{Sonnenschein:2017ylo}, although there the behavior of the width as a function of \(n\) was not determined. In some cases widths are not written since there is not enough data to determine them.}
	\end{table}
	
In \cite{Sonnenschein:2014jwa} we studied the spectrum of mesons, and showed that both light and heavy mesons can be placed on the modified Regge trajectories predicted by the HISH model. There are two types of trajectories, the orbital trajectories of increasing angular momentum, or trajectories in the \((J,M^2)\) plane, and secondly the radial trajectories in the \((n,M^2)\) plane, where \(n\) is the quantum radial excitation number. We list predictions for both types of trajectories. In \cite{Sonnenschein:2017ylo} we calculated the decay widths of mesons on leading orbital trajectories, based on the linear relation between the decay width of a string state and its length. The fits done there are used to calculate the width of the mesons on these trajectories. In this note we add predictions for many states not found in \cite{Sonnenschein:2014jwa}, specifically including more mesons with heavy quarks.

Table \ref{tab:mesons_J} collects all the predictions for the \((J,M^2)\) trajectories, for higher \(J\) states. In table \ref{tab:mesons_n} we list the predictions for the \((n,M^2)\) trajectories, for highly excited states with fixed \(J^{PC}\). The orbital trajectories are plotted in figures \ref{fig:mesons_light} and \ref{fig:mesons_heavy}. In appendix \ref{sec:states} we include a list of all the observed states whose trajectories we use. Below we discuss some of the predicted states in further detail and point out specific open issues.
					
	\subsection{Light mesons}
	The trajectories of the light mesons, made of \(u\), \(d\), and \(s\) quarks are well established. The trajectories of mesons containing only \(u\) and \(d\) quarks are nearly linear, while the \(s\) quark adds a deviation from linearity thanks to its mass.	In this sector our predictions are the surest, since the HISH model describes well both the masses and widths of light mesons, and they chiefly concern highly excited states.
	
	Among the light mesons, the pion is notable in that it does not belong on a Regge trajectory. This can be explained as the pions can be thought of as the pseudo-Goldstone bosons associated with chiral symmetry breaking and as a result have an uncharacteristically light mass. On the other hand, the excited states of the pion, both orbital and radial fall neatly enough on HMRTs with the usual meson slope.
	
	One point that remains unclear is the trajectory of the kaon. It is not obvious that the lightest kaons should lie on a trajectory, since they share the pseudo-Goldstone nature of the pions when considering three flavors.	Given the first excited state \(K_1(1270) 1^+\), the pseudoscalar kaon does appear to connect to it on a trajectory. The next state should be the \(K_2(1770)\), but there is another reported state with \(J^P = 2^-\) that is lower than that, \(K_2(1580)\). Since the lowest mass states for a given \(J^P\) should lie on the leading HMRT the existence of the \(K_2(1580)\) is puzzling. Similarly, the PDG lists only the states \(K_3(2320)\) for \(J^P= 3^+\) and \(K_4(2500)\) for \(4^-\) where there should also be lighter states for these quantum numbers, as listed in our tables, which are yet unobserved. The natural parity \(K^*\) mesons on the other hand fall very neatly on a trajectory, and there is no apparent reason for the \(K\) mesons to deviate too much from their own trajectory.	

We have omitted from this section the discussion of the spectrum of the \(f_0\) and \(f_2\) resonances. The quantum numbers of these states, \(I = 0\) with \(J^{PC} = 0^{++}\) or \(2^{++}\), are the same as those expected for the lowest glueball states. For this reason, their spectrum is discussed in conjunction with the glueball spectrum in section \ref{sec:glueballs}.
	
\subsection{Heavy-light mesons}
One of the strengths of the HISH model is that it provides a unified description of both light and heavy hadrons. The endpoint masses allow us to describe the massive modified Regge trajectories of hadrons containing \(b\) and \(c\) quarks, and the fits, whenever there are two or more mesons that can be placed on a trajectory, point to them having the same slope as the light mesons trajectory.
	
\begin{figure}[p!] \centering
	(a)\includegraphics[width=0.44\textwidth]{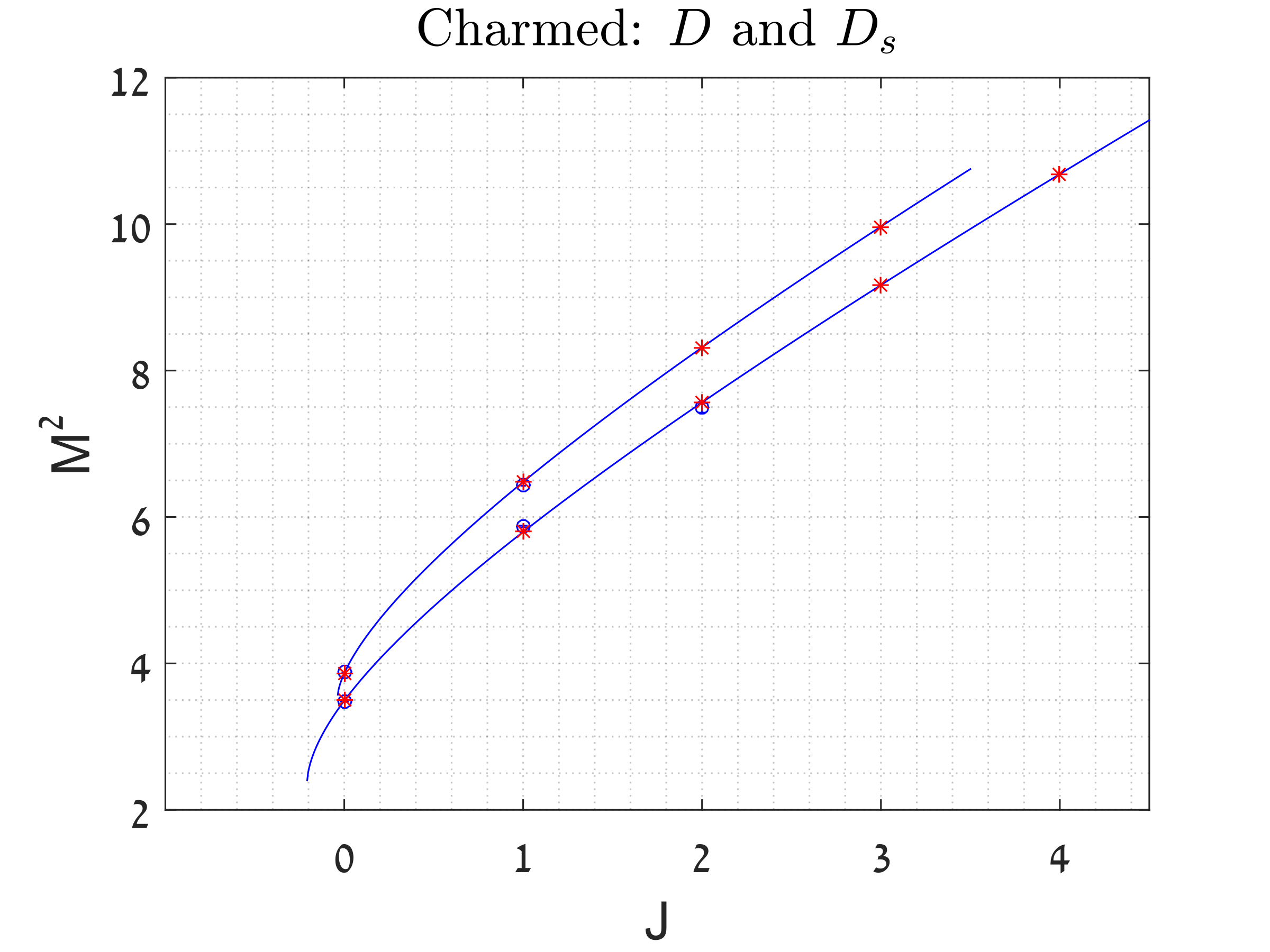} 
	(b)\includegraphics[width=0.44\textwidth]{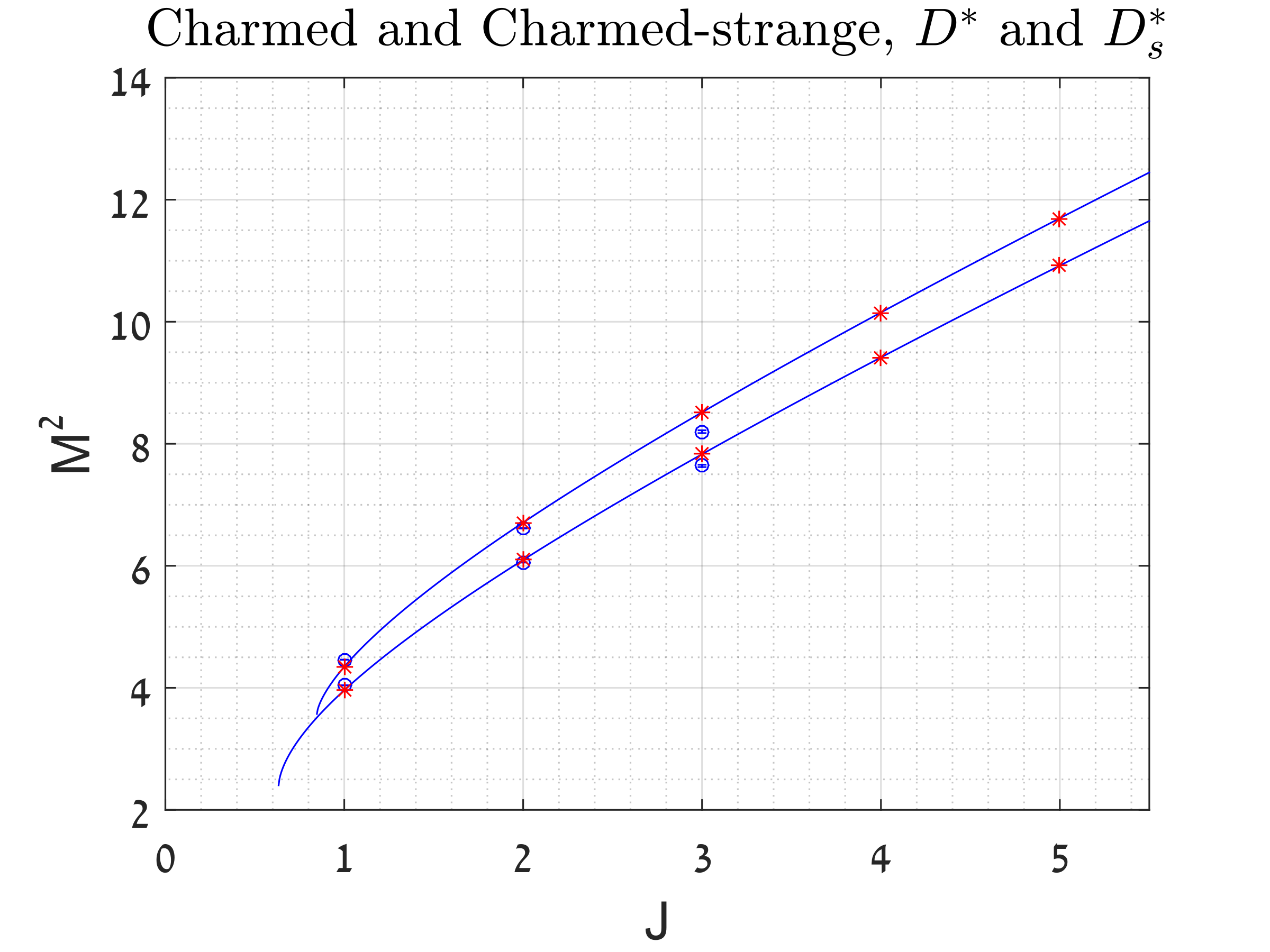}
	(c)\includegraphics[width=0.44\textwidth]{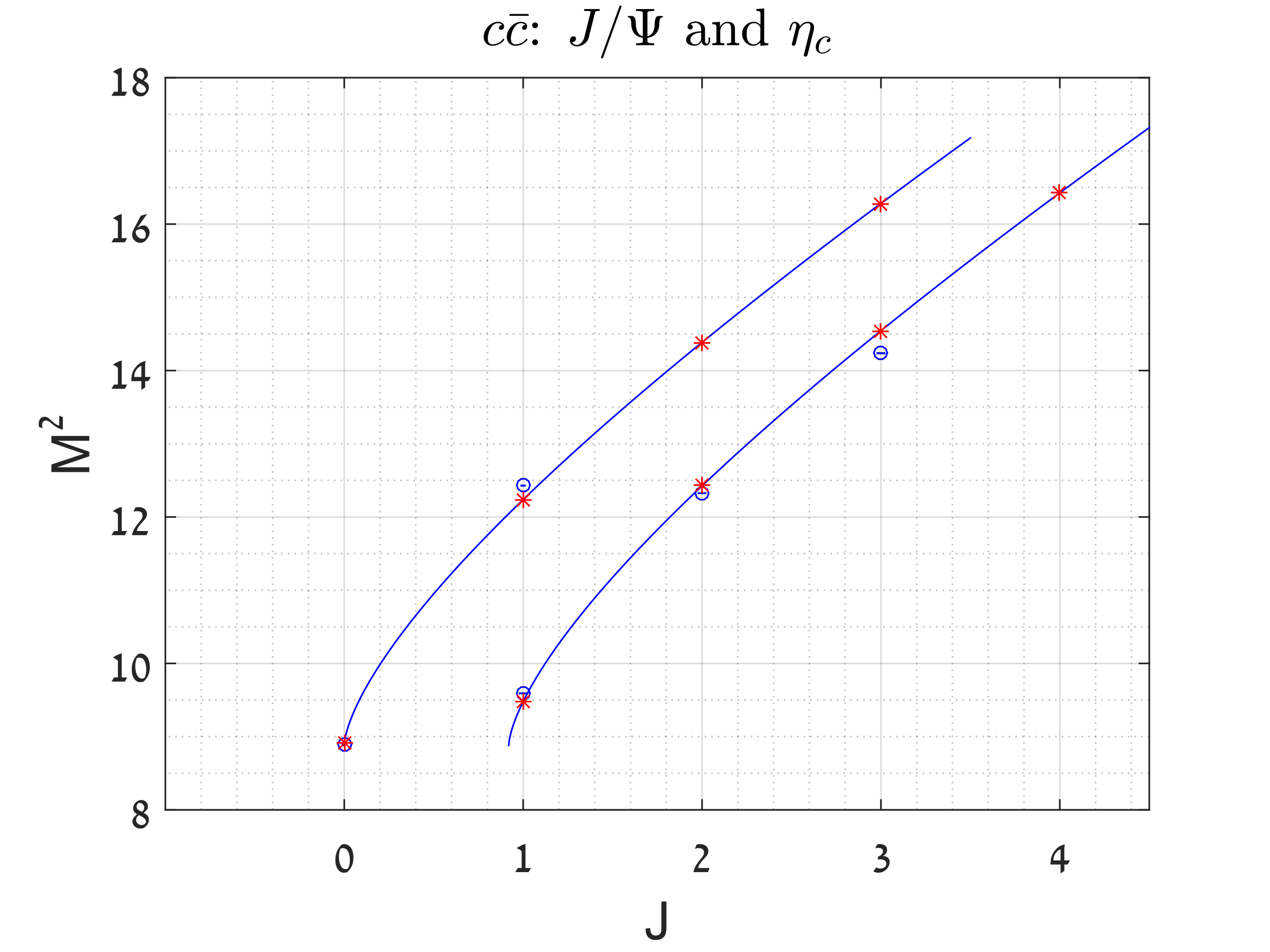} \\
	(d)\includegraphics[width=0.44\textwidth]{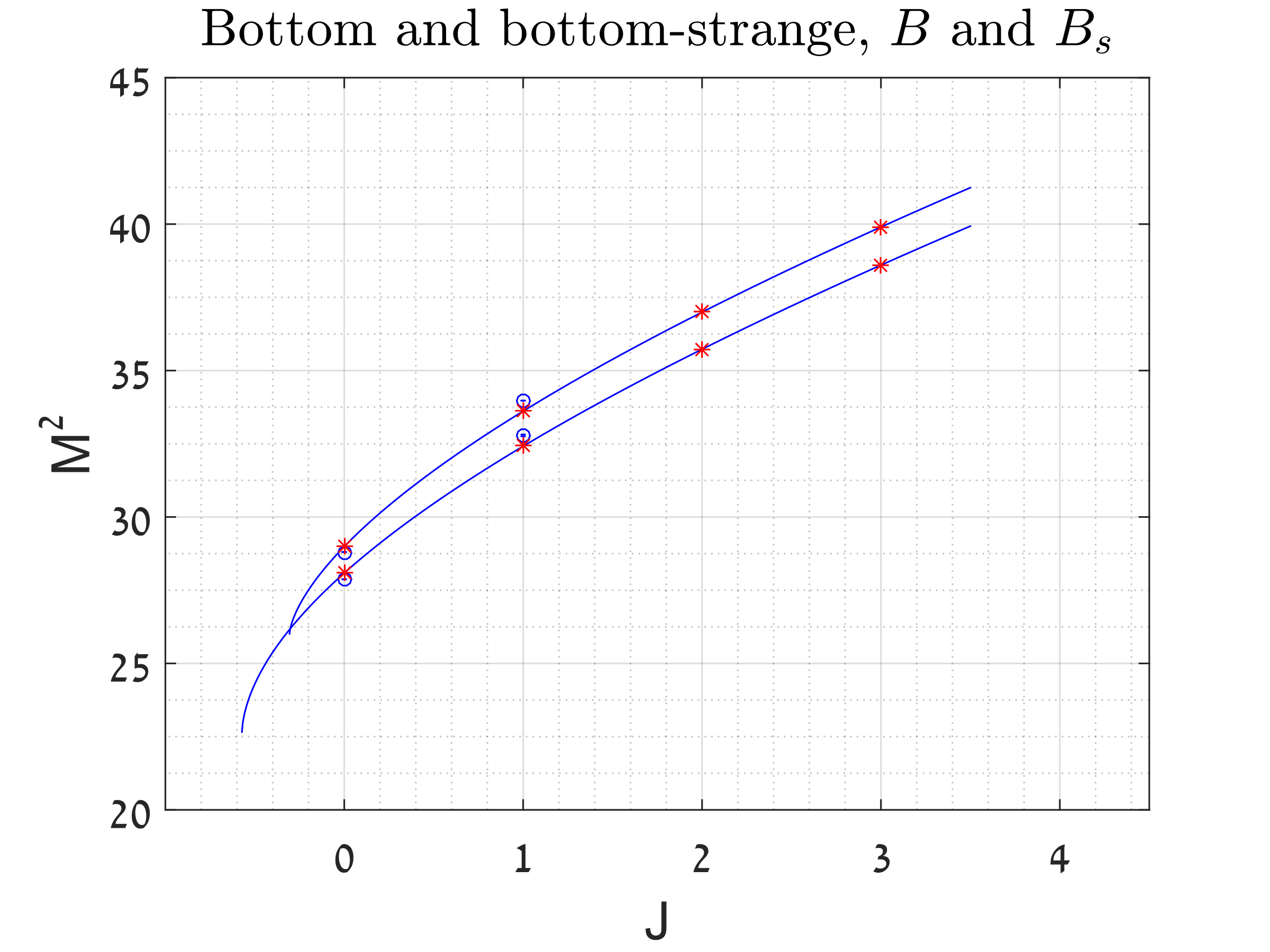}
	(e)\includegraphics[width=0.44\textwidth]{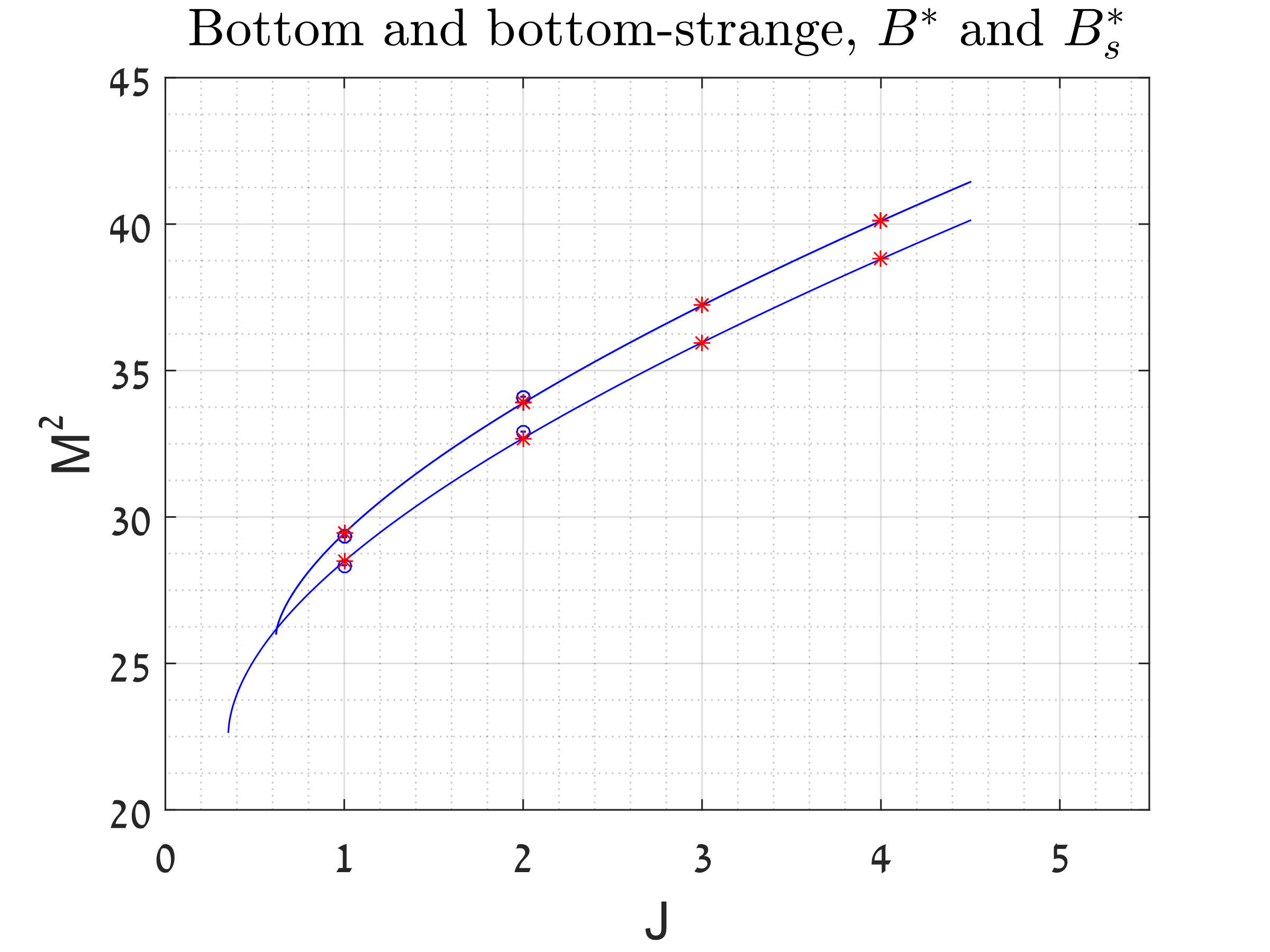}
	\caption{\label{fig:mesons_heavy}. The leading orbital modified Regge trajectories of heavy light mesons and charmonium, fitted with the common slope \(\alp = 0.88 \GEVm\). All the states fitted are listed in the appendix in table \ref{tab:all_mesons}. In blue are the data points, and the red asterisks along the trajectory are the theoretical values.}
\end{figure}

In previous work we have described the trajectories of the charmed mesons, including the decay widths of the \(D\) and \(D^*_s\) in \cite{Sonnenschein:2017ylo}. For the mesons containing a \(b\) quark and a light or \(s\) quark we can add now several examples of pairs of states that still fall on trajectories with the usual meson slope, as can be seen in figure \ref{fig:mesons_heavy}. For each of the \(B\), \(B^*\), \(B_s\), and \(B_s^*\) mesons we find such a pair and list the next two orbitally excited states on the leading HMRT.

Based on our trajectories  we can assign to some observed states their quantum numbers. The state \(D(2740)\) should be the \(2^-\) state. Similarly \(B_J(5970)\) is identified as the \(2^-\) bottom meson.
	
	\subsection{Charmonium and bottomonium} \label{sec:quarkonium}
	The spectrum of heavy quarkonia has been studied extensively. For current hadron phenomenology it is specially important to know the full spectrum since many of the most compelling exotic hadron candidates are found among the charmonium and bottomonium states.
	
	Here we start seeing some deviations from the universal Regge slope of the light and heavy-light mesons. The \((J,M^2)\) trajectories of the \(\ccb\) still have the normal meson slope of approximately \(0.9\) \GEV\, but in the \((n,M^2)\) plane we measured a slope of \(0.60 \GEVm\) \cite{Sonnenschein:2016ibx}. For the \(\bbb\) the slopes of orbital and radial trajectories were \(0.55 \GEVm\) and \(0.42	 \GEVm\) respectively \cite{Sonnenschein:2014jwa}. Explaining the lower slopes of these trajectories is another important puzzle. From the point of view of the HISH model, these deviations could mean that we start to feel the departure from the approximate picture of the holographic string as a string with massive endpoints in flat spacetime. It is possible that for the heavy quarkonia we have to solve the full holographic string model. But one would also need to explain why the disparity between the orbital and radial slopes grows so large, especially for the charmonium trajectories.
	
Also interesting is the \(B_c\) meson containing both charmed and bottom quarks. The two known states are the lightest of these, the \(B_c\) meson with \(J^P = 0^-\) at 6275 MeV as well as its first excited partner, also a pseudoscalar at 6842 MeV. The slope of a trajectory between them is 0.56 \GEV, which is between the \(\ccb\) and \(\bbb\) values. It would be interesting if more states are observed, and specifically the first \(J^P = 1^+\) state which would allow us to measure the slope of the orbital trajectory.

					\begin{figure}[p!] \centering
	(a)\includegraphics[width=0.44\textwidth]{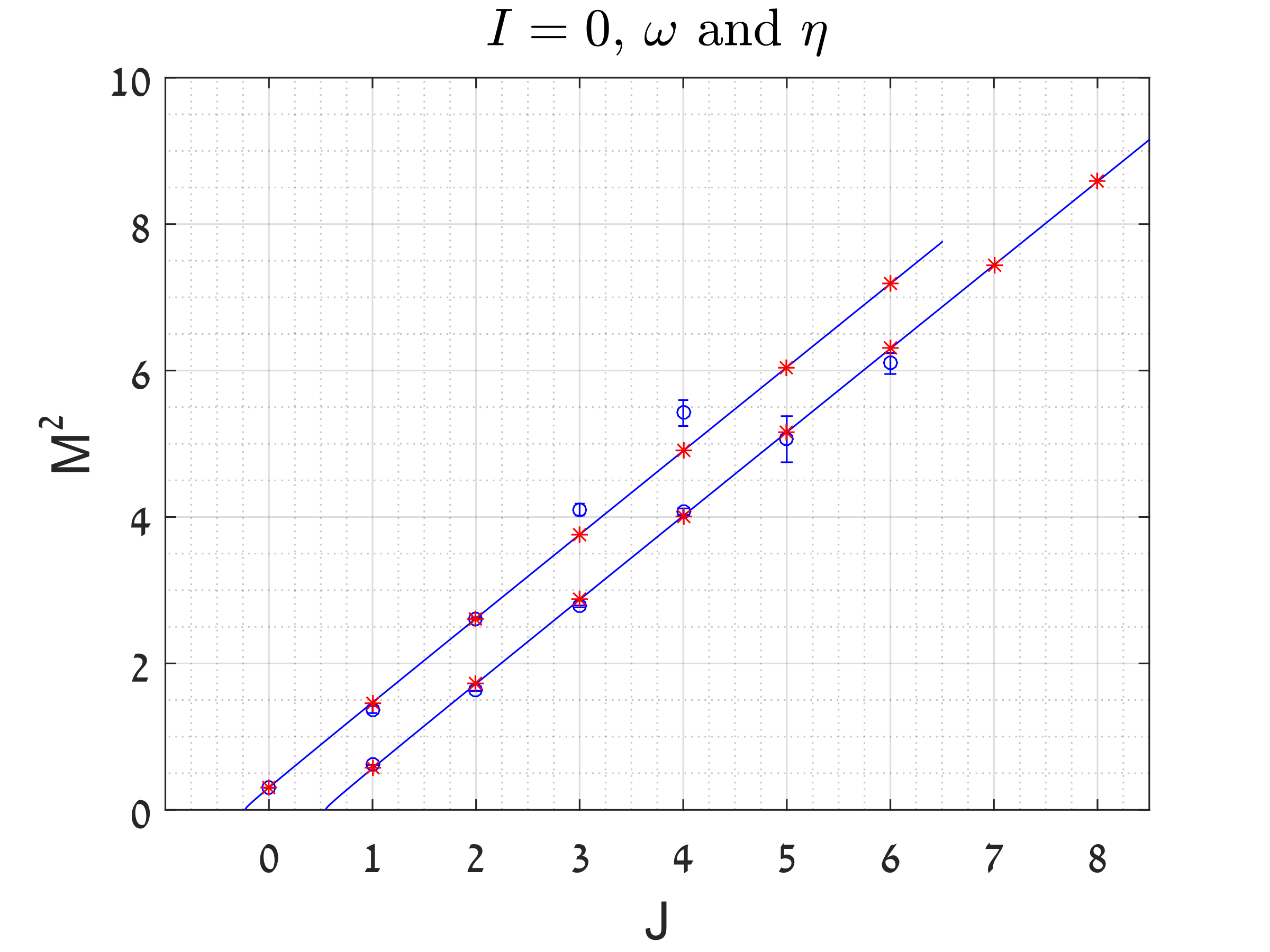}
	(b)\includegraphics[width=0.44\textwidth]{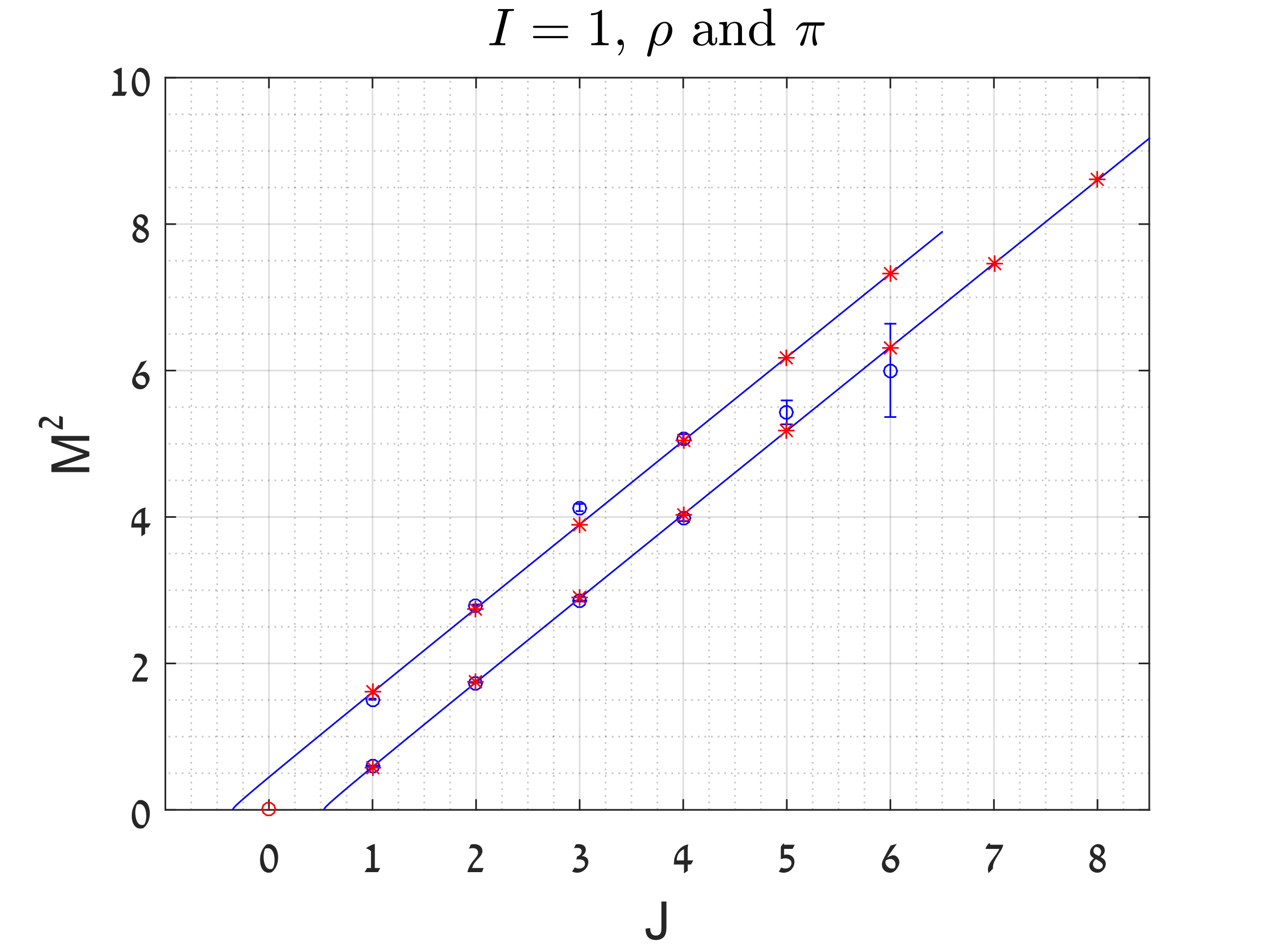} \\ 
	(c)\includegraphics[width=0.44\textwidth]{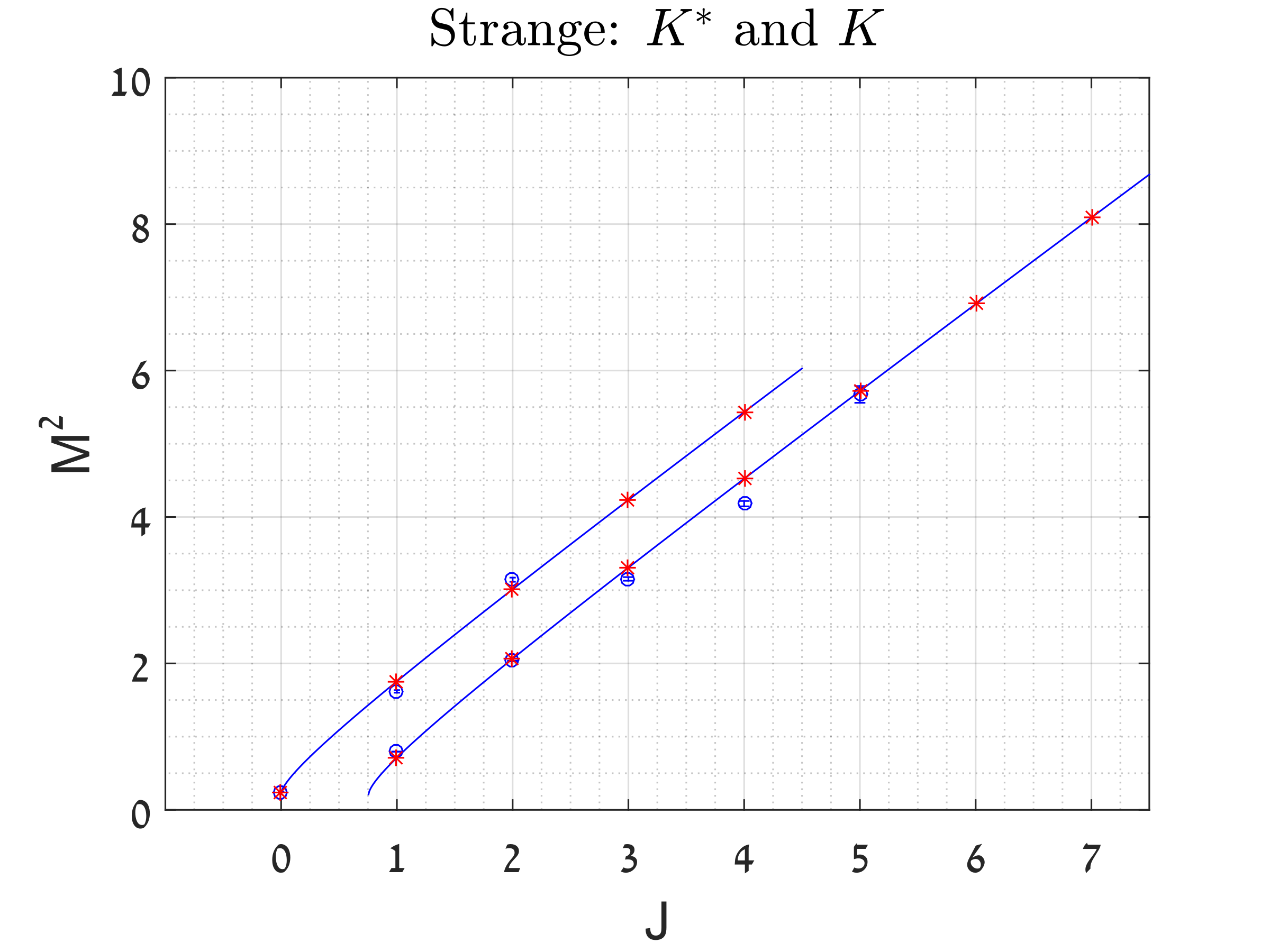} 
	(d)\includegraphics[width=0.44\textwidth]{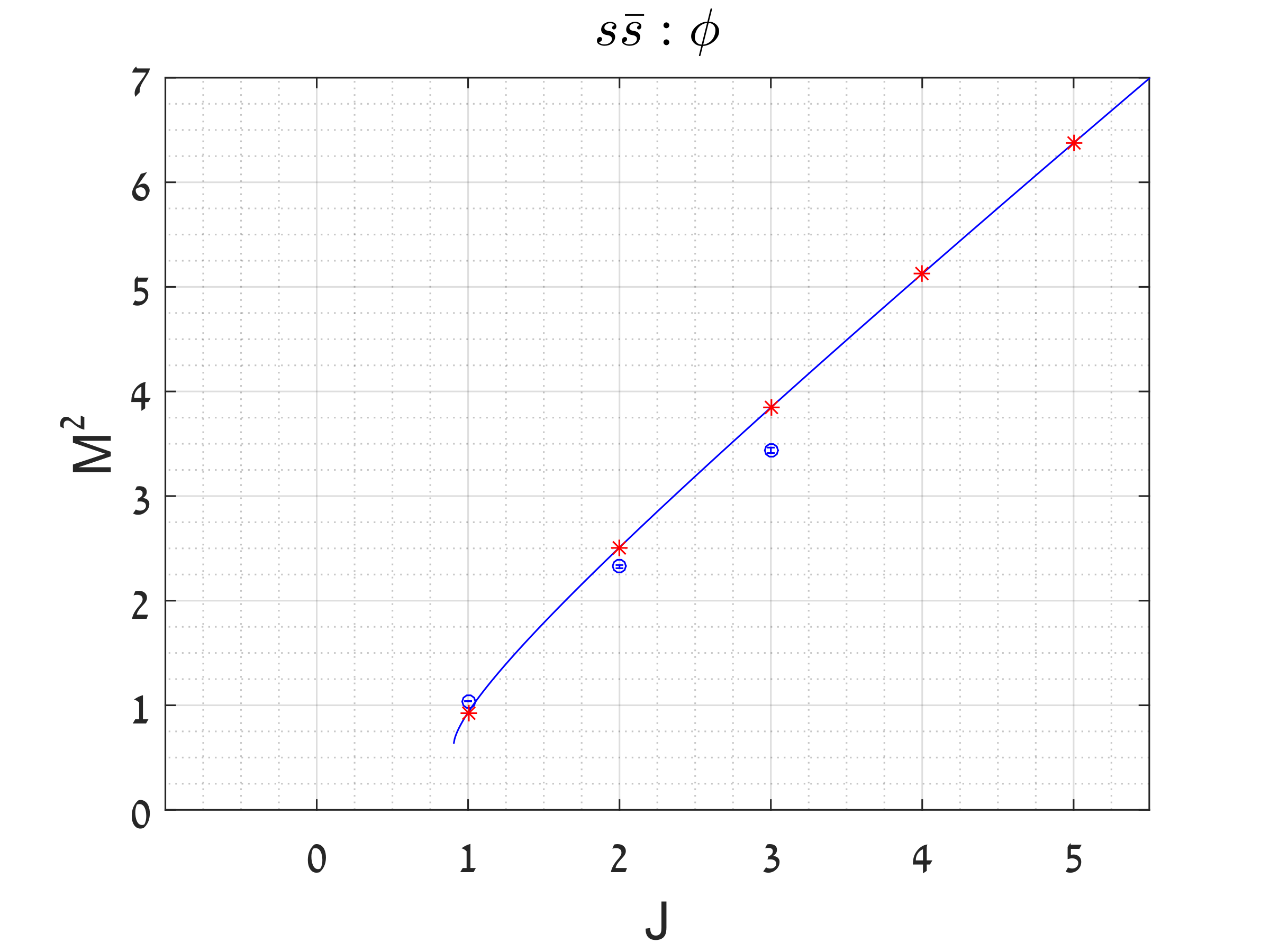} \\
	\caption{\label{fig:mesons_light}. The leading orbital modified Regge trajectories of light mesons, fitted with the common slope \(\alp = 0.88 \GEVm\). All the states fitted are listed in the appendix in table \ref{tab:all_mesons}. In blue are the data points, and the red asterisks along the trajectory are the theoretical values.}
\end{figure}

	\clearpage
					
					\section{Predictions for baryon states} \label{sec:baryons}
					\begin{table}[ht!] \centering
					\begin{tabular}{|c|c|ccc|ccc|} \hline
						
		Traj. & Quarks & \(J^P\) & Mass & Width & \(J^P\) & Mass & Width\\ \hline
						
		\(N\) & \(qqq\) & \jph{15}{-} & 2950 & 690 & \jph{17}{+} & 3050 & 580 \\
						
		\(\Delta\) & \(qqq\) & \jph{17}{-} & 3180 & 450 & \jph{19}{+} & 3160 & 490 \\
						
		\(\Lambda\) & \(qqs\) & \jph{11}{-} & 2610 & 120 &\jph{13}{+} & 2810 & 140 \\
				
		\(\Sigma\) & \(qqs\) & \jph{9}{+} & 2450 & 160 &\jph{11}{-} & 2660 & 180 \\
						
		\(\Sigma\) & \(qqs\) & \jph{9}{-} & 2310 & 200 & \jph{11}{+} & 2530 & 230 \\
						
		\(\Xi\) & \(qss\) & \jph{7}{-} & 2340 & - & \jph{9}{+} & 2570 & - \\
		
		\(\Omega\) & \(sss\) & \jph{5}{-} & 2070 & - & \jph{7}{+} & 2370 &  -\\
						
		\(\Lambda_c\) & \(qqc\) & \jph{7}{-} & 3140 & - & \jph{9}{+} & 3350 & - \\
		
		\(\Sigma_c\) & \(qqc\) & \jph3- & 2760 & - & \jph5+ & 3020 & -\\
		
		\(\Sigma_c\) & \(qqc\) & \jph5- & 2820 & - & \jph7+ & 3060 & -\\
		
		\(\Xi_c\) & \(qsc\) & \jph{5}{+} & 3070 & -& \jph{7}{-} & 3300 & -\\
		%\(\Xi_c\) & \(qsc\) & \jph{5}{+} & 3080 & 10--30 & \jph{7}{-} & 3315 & 10--30 \\
		
		\(\Omega_c\) & \((ss)c\) & \jph5+ & 3310 & - & \jph7- & 3540 & - \\
		
		\(\Omega_c\) & \(s(sc)\) & \jph5+ & 3350 & -& \jph7- & 3590 &  -\\
		
		\(\Omega_c\) & \((ss)c\) & \jph7+ & 3360 & -& \jph9- & 3580 &  -\\
		
		\(\Omega_c\) & \(s(sc)\) & \jph7+ & 3390 & - & \jph9- & 3620 &  -\\
		
		\(\Xi_{cc}\)  & \((qc)c\) & \jph{3}{-} & 3870 & Narrow? & \jph{5}{+} & 4090 & -\\
		
		\(\Xi_{cc}\)  & \(q(cc)\) & \jph{3}{-} & 4000 & - & \jph{5}{+} & 4270 & - \\
						
		\(\Lambda_b\) & \(qqb\) & \jph{5}{+} & 6140 & - & \jph{7}{-} & 6340  & -\\
		
		\(\Sigma_b\) & \(qqb\) & \jph{3}{-} & 6060 & - &  \jph{5}{+} & 6260  & -\\
		
		\(\Sigma^*_b\) & \(qqb\) & \jph{5}{-} & 6070 & - &  \jph{7}{+} & 6280  & -\\
		
		\(\Xi_b\) & \(qsb\) & \jph{3}{-} & 6060 & & \jph{5}{+} & 6280  & -\\
		
		\(\Omega_b\) & \((ss)b\) & \jph{3}{-} & 6340 & - & \jph{5}{+} & 6580  & -\\
		
		\(\Omega_b\) & \(s(sb)\) & \jph{3}{-} & 6310 & - & \jph{5}{+} & 6520  & -\\
		
		\hline \end{tabular}
		\caption{\label{tab:baryons_J} Predictions for the baryon next states in the \((J,M^2)\) plane. In some cases, such as the \(\Xi_{cc}\), we list the two different predictions one gets when assuming different configurations of the diquark. \(q\) signifies a light quark (\(u/d\)). We only list widths for the trajectories fitted in \cite{Sonnenschein:2017ylo}, for the rest there are too few data points.}
					\end{table}
					
					\begin{table}[h!] \centering
					\begin{tabular}{|c|c|c|cc|cc|} \hline
						
						Traj. & Quarks & \(J^P\) & \(n\) & Mass & \(n\) & Mass \\ \hline
						
						$N$ & \(qqq\) & \jph{1}{+} & 4 & 2330 & 5 & 2560 \\
						
						$N$ & \(qqq\) &\jph{3}{-} & 3 & 2380 & 4 & 2610 \\
						
						$N$ & \(qqq\) &\jph{5}{+} & 2 & 2260 & 3 & 2490 \\
						
						$N$ & \(qqq\) &\jph{1}{-} & 2 & 2150 & 3 & 2400 \\
						
						$N$ & \(qqq\) &\jph{3}{+} & 2 & 2290 & 3 & 2520 \\
						
						$N$ & \(qqq\) &\jph{5}{-} & 2 & 2270 & 3 & 2510 \\ 
						
						$\Delta$ & \(qqq\) &\jph{3}{+} & 3 & 2210 & 4 & 2450 \\ \hline
						
						$\Lambda_b$ & \(qqb\) & \jph1+ & 1 & 6070 & 2 & 6420 \\
						
						$\Lambda_b$ & \(qqb\) &\jph3- & 1 & 6290 & 2 & 6600 \\
						
						$\Sigma_b$ & \(qqb\) &\jph1+ & 1 & 6210 & 2 &  6530 \\
						
						$\Sigma_b^*$ & \(qqb\) &\jph3+ & 1 & 6230 & 2 & 6540 \\
						
						$\Xi_b$ & \(qsb\) &\jph1+ & 2 & 6560 &  3 & 6840 \\
						
						$\Omega_b$ & \((ss)b\) &\jph1+ & 1 & 6470 &  2 & 6790 \\
						
						$\Omega_b$ & \(s(sb)\) &\jph1+ & 1 & 6520 &  2 & 6870 \\
						
					\hline \end{tabular}
					\caption{\label{tab:baryons_n} Predictions for the baryon next states in the \((n,M^2)\) plane. The light baryon HMRTs have the common slope \(\alp=0.9\GEVm\). For the bottom baryons, we have one tentative measurement indicating \(\alp=0.55\GEVm\), which we use to make predictions for other excited bottom baryons.}
					\end{table}

In tables (\ref{tab:baryons_J}) and (\ref{tab:baryons_n}) are the predictions for higher \(J\) and \(n\) baryons respectively, based on the results of our fits. The orbital trajectories have baryons across the spectrum, from the light ones to the heaviest charmed and bottom baryons. The radial trajectories concern only the \(N\) and \(\Delta\) states, where the trajectories were best established and the slope measured.

The fits done in \cite{Sonnenschein:2017ylo} showed a larger deviation of the decay widths of baryons from our model's predictions when compared with the mesons, so the width predictions are less certain. In particular for some cases, like for the \(N\) and \(\Delta\) baryons, it looked like the width grows faster than what our model would predict based on linearity in the string length.

\subsection{Light and strange baryons}
The \(N\) and \(\Delta\) trajectories are well established with relatively many states on each trajectory \cite{Sonnenschein:2014bia}. One interesting effect that they exhibit, which is still not understood, is the splitting of the trajectories into two parallel lines for even and odd orbital angular momentum states. The strange baryons do not seem to exhibit this phenomenon. We list predictions for higher spin light and strange baryons in table \ref{tab:all_baryons}. There are also many \(N\) baryons which fall neatly on nearly linear radial trajectories, and predictions for higher states for them are in table \ref{tab:baryons_n}.

Moving on to the spectrum of the \(sss\), the excited \(\Omega\) baryons could have helped us determine the mass of the \(ss\) diquark directly, but there is not much data on the excited states. Our model predicts the first orbitally excited state of the \(\Omega\) to have a mass of around 2050 MeV, but there is no confirmed state there. The states that are known - \(\Omega(2250)\), \(\Omega(2380)\), and \(\Omega(2470)\) - could be higher excitations. The \(\Omega(2380)\) is at the right mass for the \(\jp{7}{+}\) state. One possibility is that the \(\Omega\) trajectories exhibit the same even-odd effect as the \(N\) and \(\Delta\). Then, one might claim that there is no missing state, and the \(\Omega(2250)\) as the \(\jp{5}{-}\) state belongs on a separate parallel trajectory of the odd states. On the other hand, this would make the even-odd effect larger for the \(\Omega\) than it is for the \(N\) and \(\Delta\). Spin-orbit interactions which would be related to this splitting are suppressed for higher quark masses, so this possibility is less likely.

\subsection{Charmed baryons}
The \(\Lambda_c\) and \(\Xi_c\) trajectories were examined in \cite{Sonnenschein:2014bia}. On the \(\Lambda_c\) trajectory we have three states, \(J^P = \jp1+\), \jp3- and \jp5+, from which we see that the light baryon slope of \(0.95 \GEVm\) applies also to the charmed baryons. This is confirmed also by the pair of charmed-strange \(\Xi_c\) states. For the \(\Xi_c\) we find that the charm quark is outside the diquark, as a fit with \(m_s\) on one endpoint and \(m_c\) on the other is preferred.

We use the known slope to predict higher states of the other types of charmed baryons. For the \(\Sigma_c\) only the \jp1+ and \jp3+ states are confirmed. The observed state \(\Sigma_c(2800)\) is at the right mass to be the first excitation of either one of those.

Recently, five narrow \(\Omega_c\) resonances with the quark content \(ssc\) were measured for the first time \cite{Aaij:2017nav,Karliner:2017kfm}. Their masses are consistent with the HISH prediction for the first orbital excitations based on the Regge trajectories starting with the \(\jp1+\) and \(\jp3+\) ground states. The five states are assigned \(J^P = \jp3-\) and \jp1- for the excitations of the \jp1+ ground state, and \jp5-, \jp3-, and \jp1- for the excitations of the \jp3+. There is a splitting due to spin-orbit and spin-spin interactions of \(\sim30\) MeV between states. We do not compute that in our model. What we can do, which is consistent with what we have done in other cases, is pair the \jp1+ with the state assumed to be  \jp3-, and likewise the \jp3+ with the \jp5- and continue the two resulting trajectories. The diquark in the \(\Omega_c\) is either \(ss\) or \(sc\). We have no other measurement of the mass of either diquark, so this adds further ambiguity, but from fitting the two chosen pairs of states, the \(s(sc)\) configuration is preferred. We list in the table the predictions of either option.

\subsection{Doubly charmed baryons}
For the doubly charmed baryons, only the ground states \(\Xi_{cc}^+\) and \(\Xi_{cc}^{++}\) are listed by the PDG. The status of the former is very uncertain, so we only use the \(\Xi_{cc}^{++}\) whose measured mass is \(3621.4\pm0.8\) MeV.

A priori the diquark in a doubly charmed baryon can be either \((uc)/(dc)\), or it can be \((cc)\). In the former configuration we will see masses of \(m_c = 1.5\) GeV on each side, while in the later we will have one light endpoint and one heavy one. The mass of the \((cc)\) diquark is naturally expected to be close to \(2m_c\), although that is not obvious in the holographic picture where the mass of the diquark is determined by the vertical segments of the strings and the tension of the baryonic vertex. The predicted mass for the next state depends strongly on the choice of configuration, and we list both options in table \ref{tab:baryons_J}. The configuration with \(m_c\) on either side gives lower masses of the excited states. For the orbital trajectory we take the usual slope of \(0.95 \GEVm\).

We may estimate the width by comparison with the single charm baryons. The first excited state of the \(\Xi_{cc}\) will be just heavy enough to decay to the ground state and one or two pions, so we can expect a width of a few MeV (for most direct comparison see the charmed-strange baryon \(\Xi_{c}(2815) \jp{3}{-}\) whose width is slightly less than 3 MeV). The second excited state will be above the threshold to decay to a charmed baryon plus charmed meson, \(\Lambda_c D\), which would be the preferred channel unless the diquark is \((cc)\). A decay width of around 20 MeV would be a reasonable estimate for the first above threshold state.

\subsection{Bottom baryons}
\begin{figure}[p!] \centering
	(a)\includegraphics[width=0.40\textwidth]{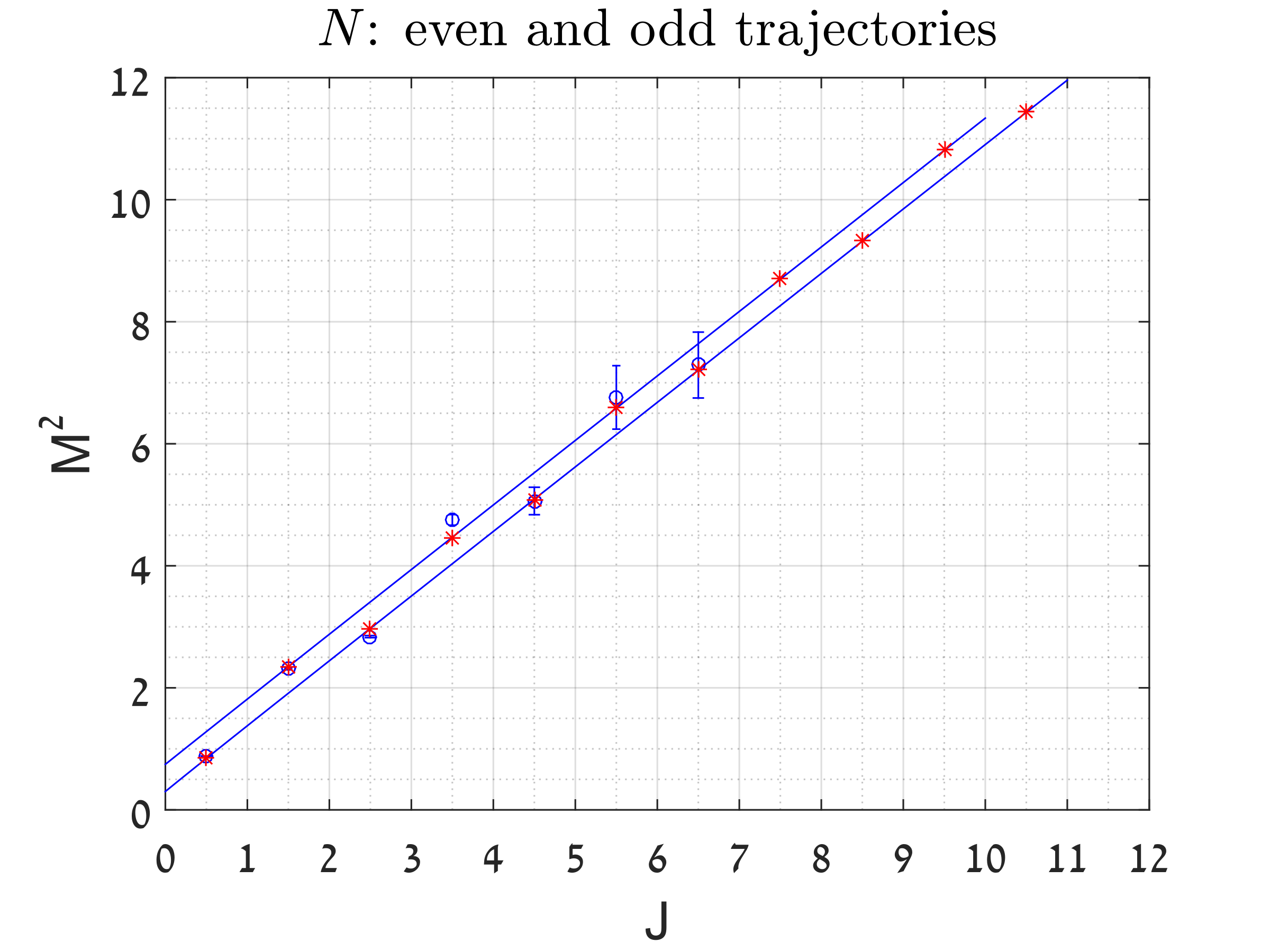}
	(b)\includegraphics[width=0.40\textwidth]{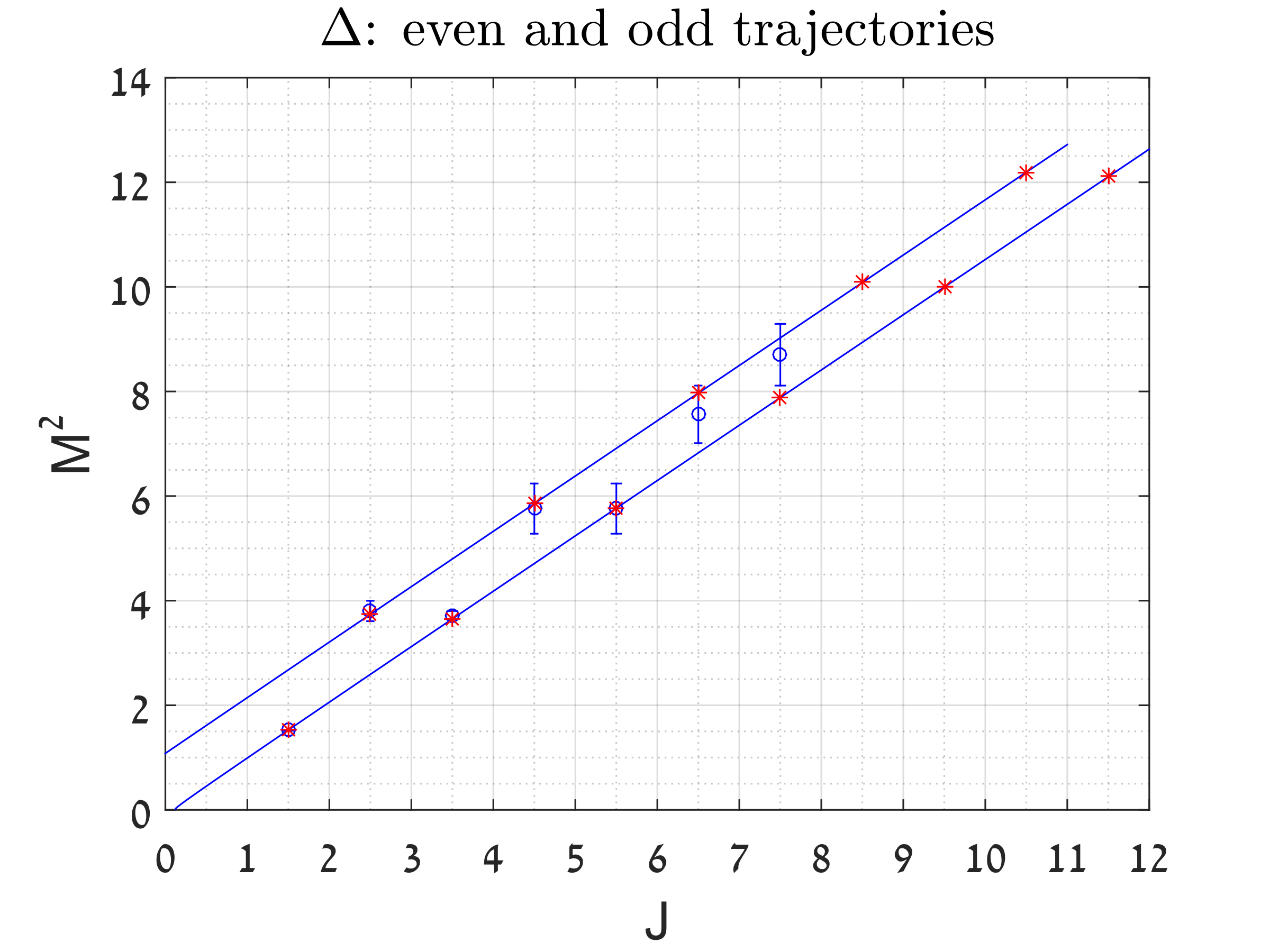} \\ 
	(c)\includegraphics[width=0.40\textwidth]{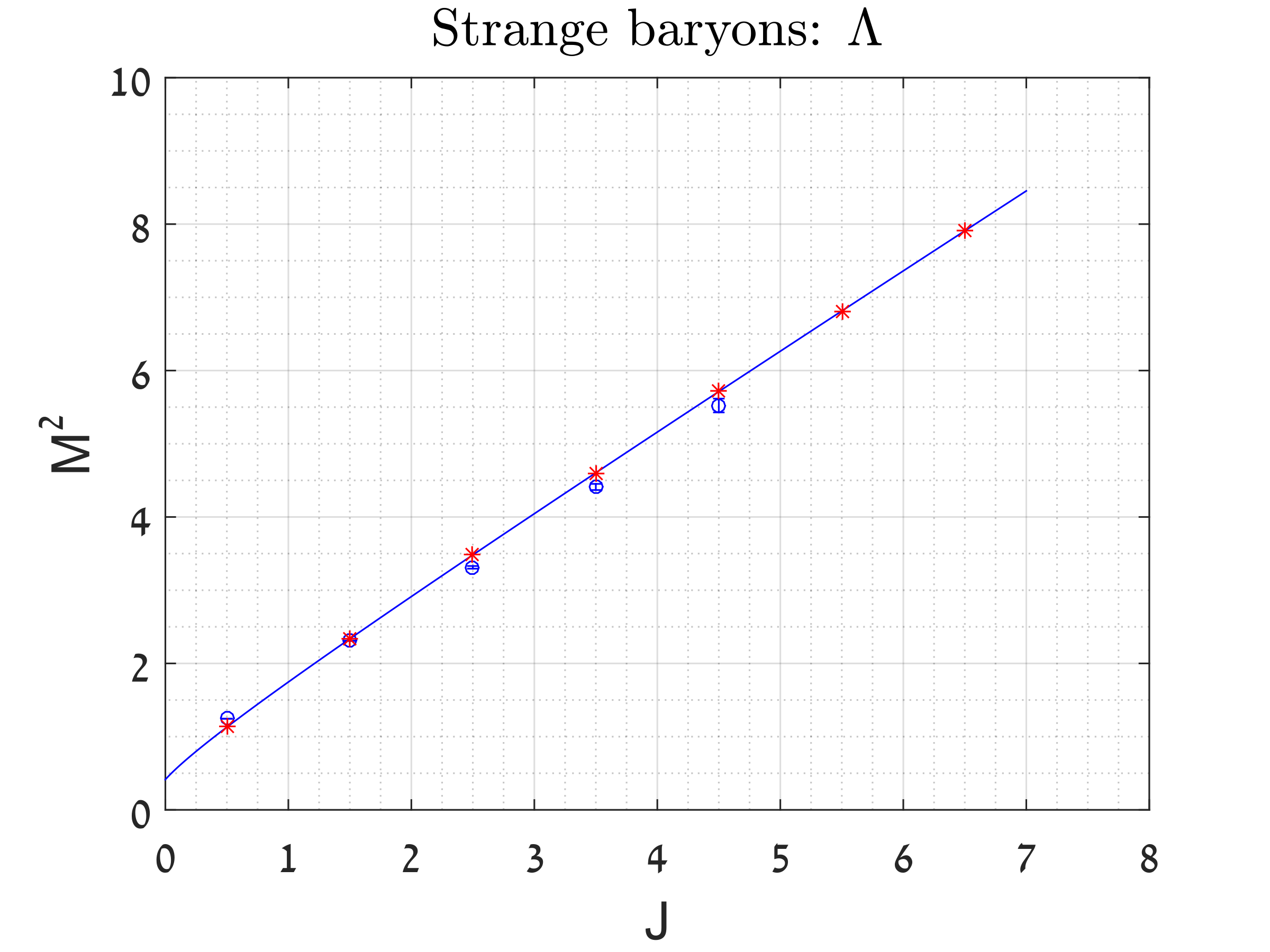} 
	(d)\includegraphics[width=0.40\textwidth]{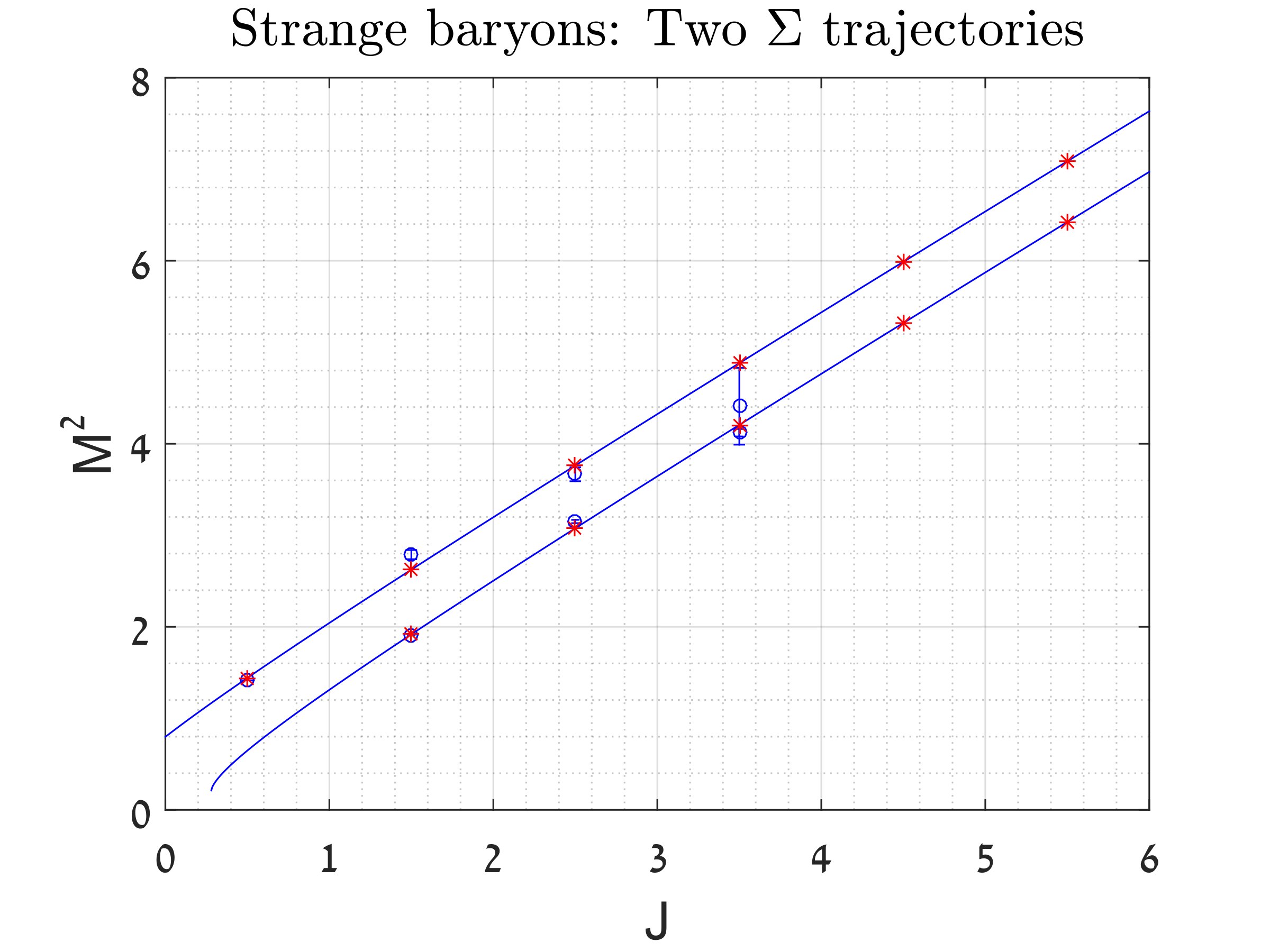} \\
	(e)\includegraphics[width=0.40\textwidth]{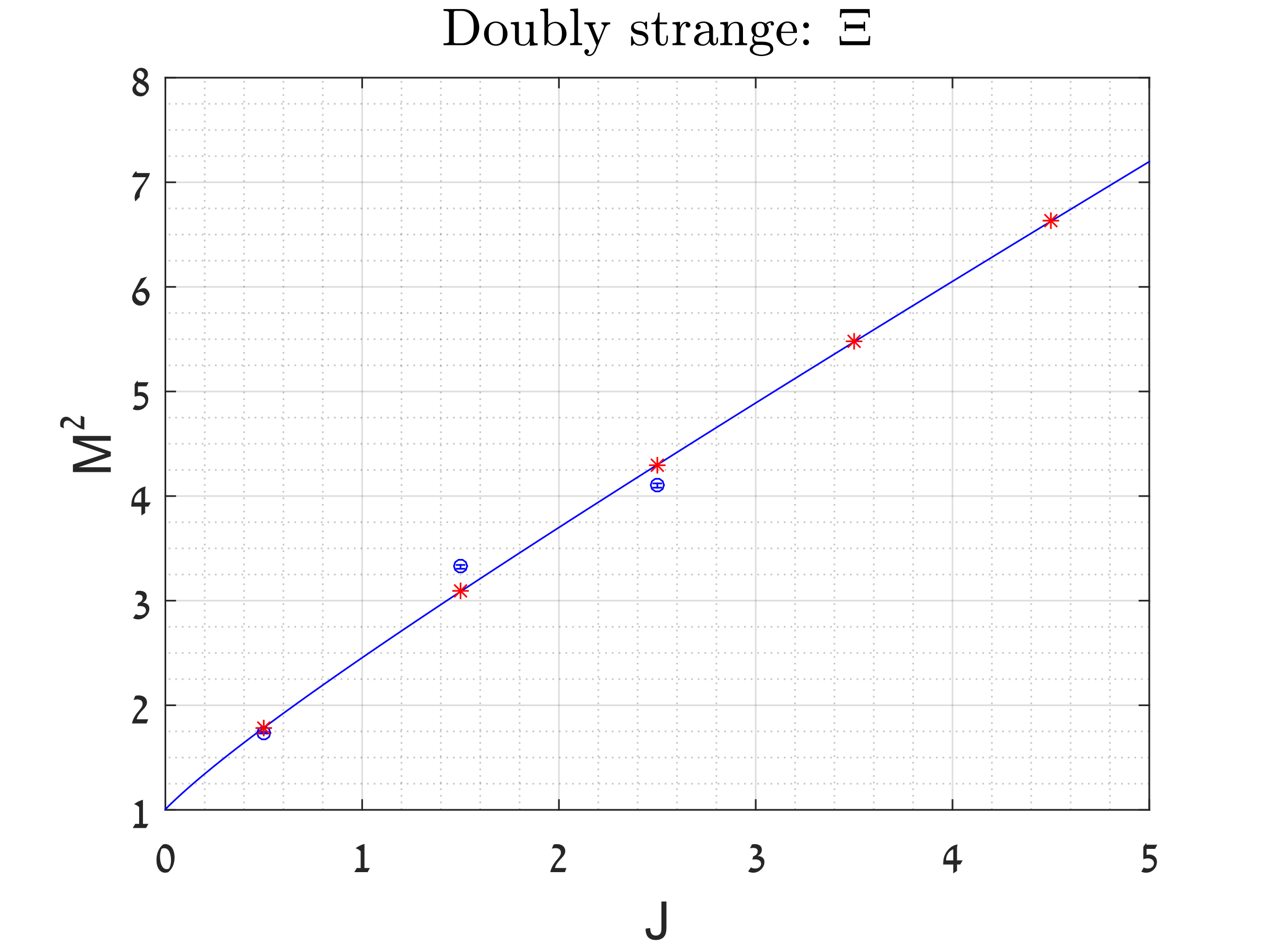} 
	(f)\includegraphics[width=0.40\textwidth]{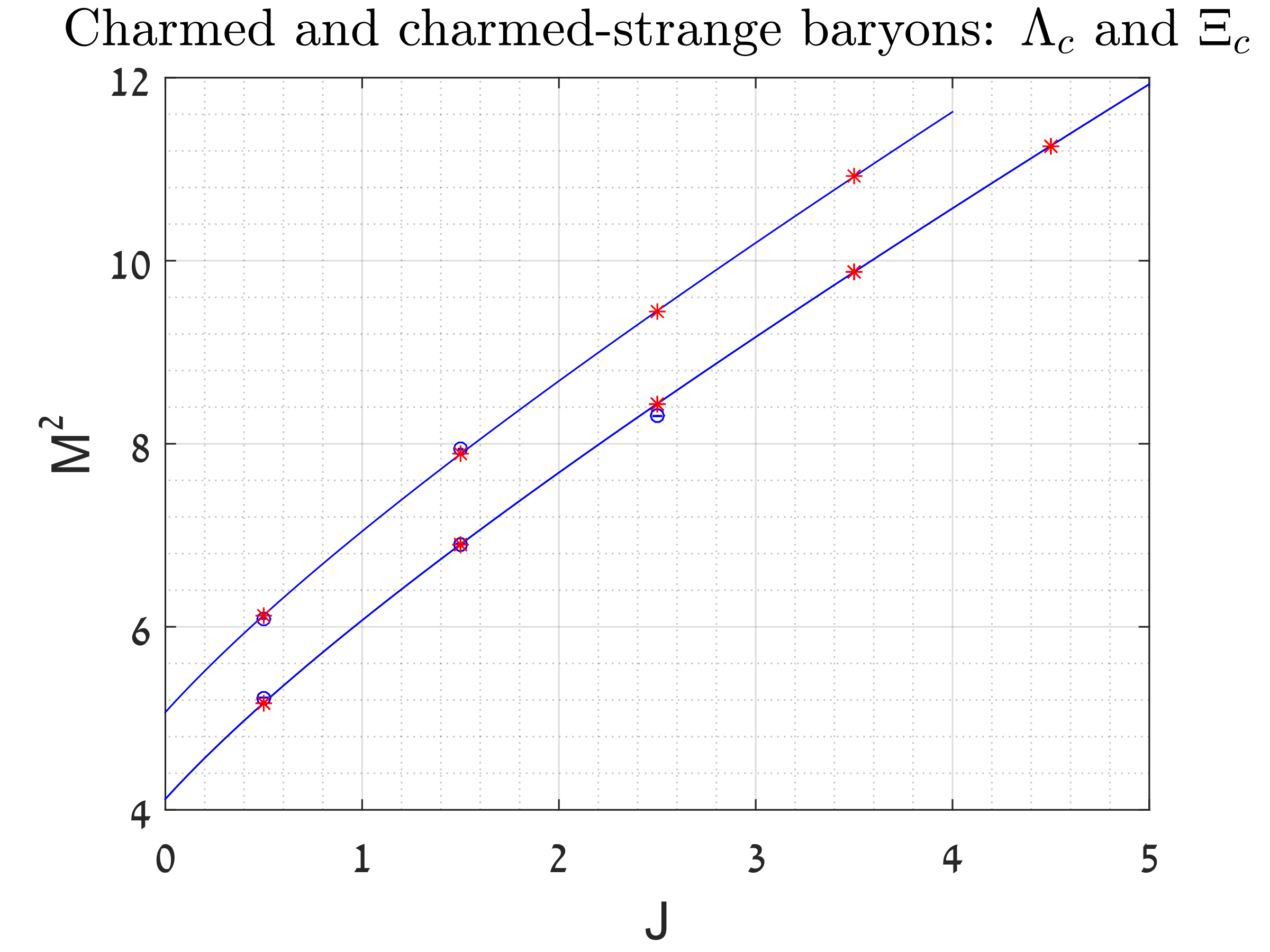} \\
	(g)\includegraphics[width=0.40\textwidth]{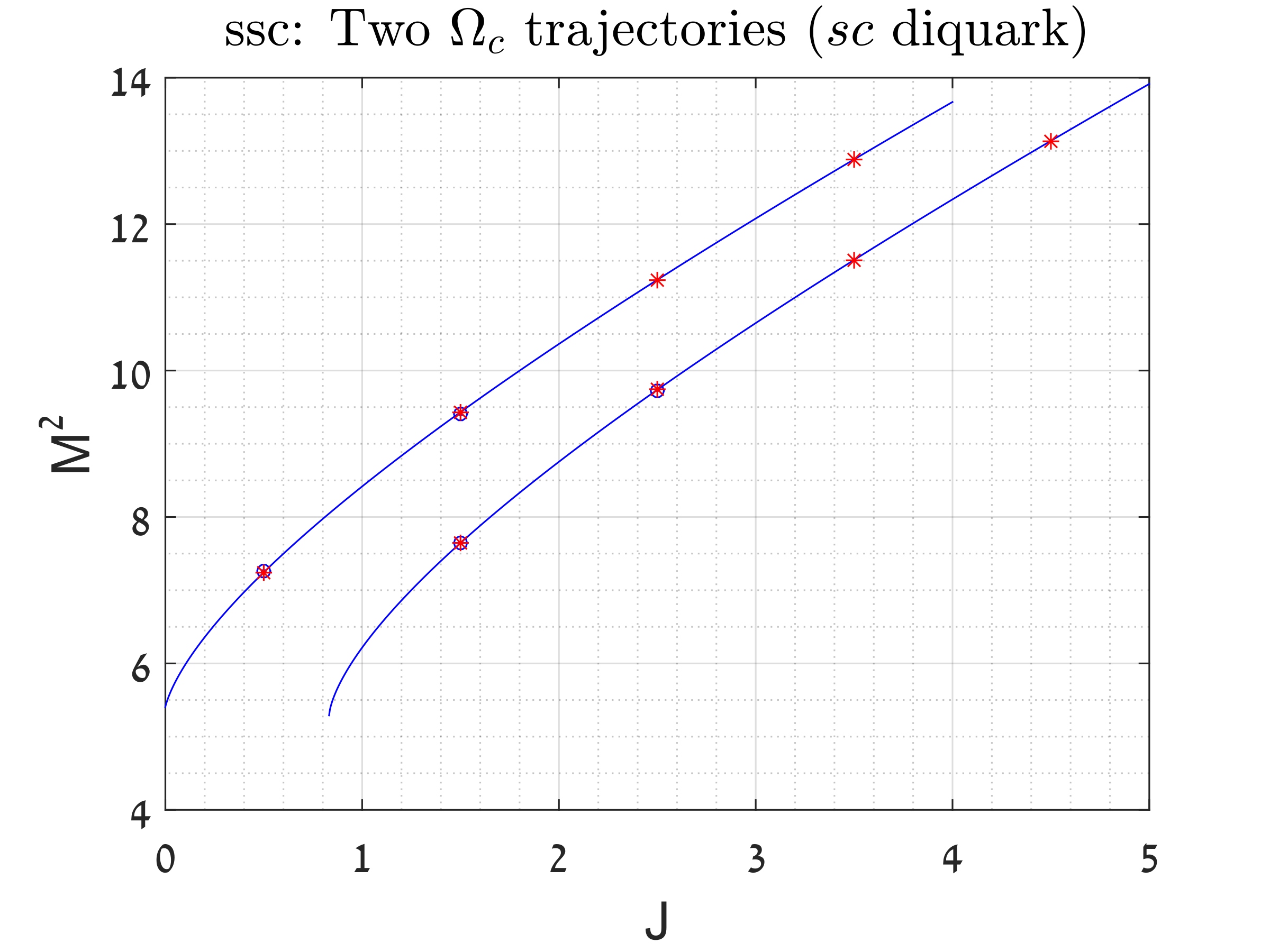}
	(h)\includegraphics[width=0.40\textwidth]{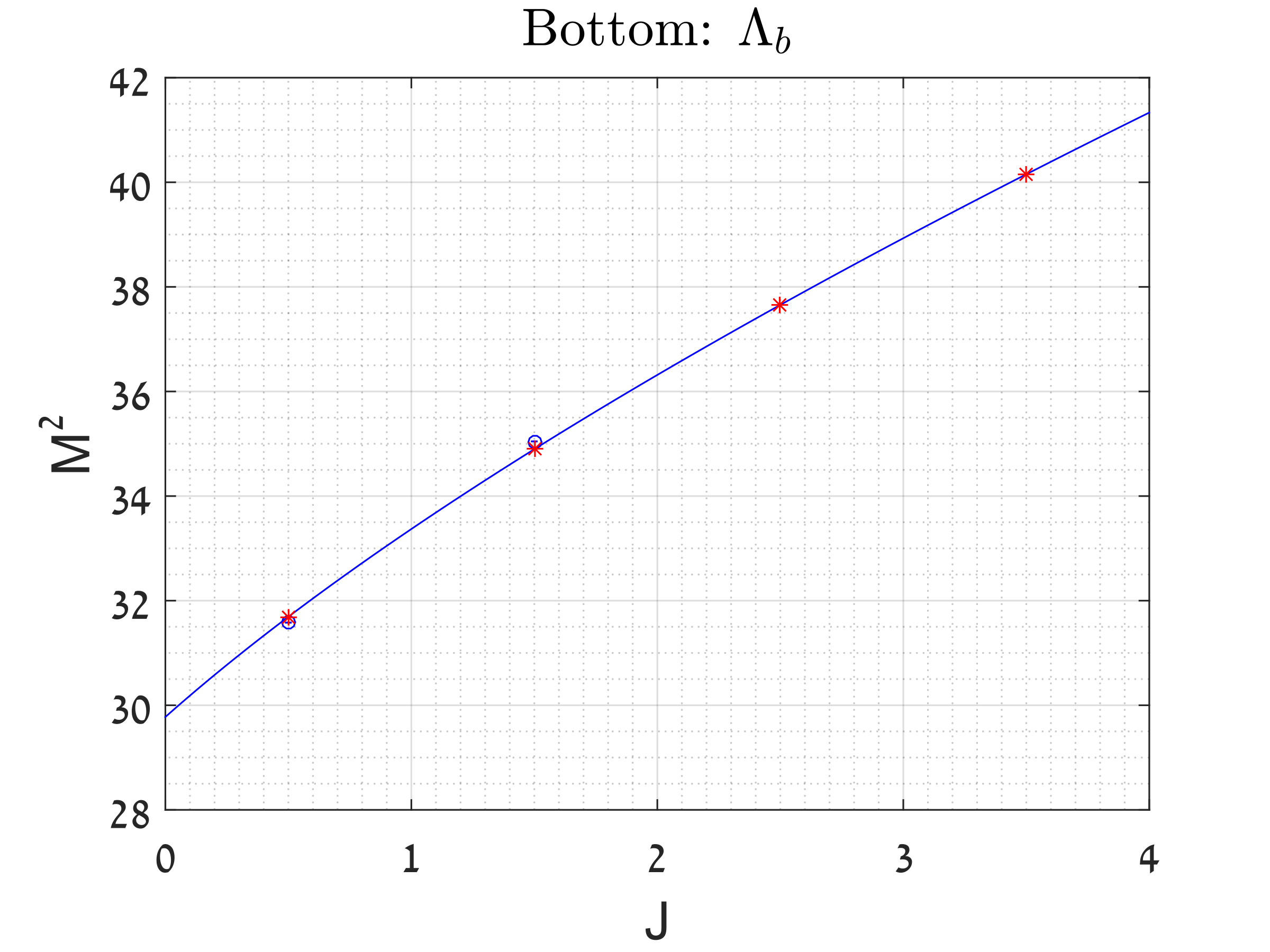}
	\caption{\label{fig:baryons_J}. The leading orbital trajectories of the baryons fitted with the common slope \(\alp = 0.95 \GEVm\). All the states fitted are listed in the appendix in table \ref{tab:all_baryons}.}
\end{figure}

The spectrum of excitations of the bottom baryons is largely unexplored. The search and identification of spectrum of bottom baryon excitations is ongoing \cite{Aliev:2018lcs,Chen:2018orb,Wang:2018fjm,Karliner:2018bms}. The only confirmed excitation is a \(\Lambda_b\) state, \(\Lambda_b^0(5930)\jp{3}{-}\), first observed in. It and the ground state \(\Lambda_b^0(5620) \jp{1}{+}\) are again connected by a line whose slope is \(\alp = 0.95 \GEVm\), as for the other baryons, as can be seen in figure \ref{fig:baryons_J}h. The other bottom baryons do not have yet their excited partners. We will assume they can also be accommodated on trajectories with the same slope, so we can predict where the excited states will be for the \(\Sigma_b\), \(\Xi_b\), and \(\Omega_b\). For the latter, first excited \(ssb\) state could be narrow, like the equivalent \(\Omega_c\). Here the predicted mass does not depend very strongly on whether we have the \((ss)b\) or \(s(sb)\) configuration, but the latter gives a lighter mass for the first excited state.

Recently, a \(\Sigma_b\) state was observed for the first time, \(\Sigma_b(6097)\) \cite{Aaij:2018tnn}. It has a mass slightly higher than the 6060 MeV predicted by us for the first orbital excitation of the \(\Sigma_b\), but can certainly be identified as the \(1P\) state. Another recent observation from LHCb is an excited \(\Xi_b^-\) resonance \cite{Aaij:2018yqz} at a mass of \(6226.9\pm2.4\) MeV. There it is speculated to be either the first orbital (\(1P\)) or first radial (\(2S\)) state. According to our model the first orbital excitation should be much lower than that, and in fact the measured mass is closer to our prediction for the next excitation (\(1D\)). If the \(\Xi_b(6227)\) is the first orbital excitation, then the mass of the radial trajectory connecting it and the ground state is \(0.55 \GEVm\), very similar to the slope we find for the radial trajectories of heavy quarkonia. If we assume that this is the characteristic slope of the bottom baryons' radial HMRTs, we can predict the \(2S\) states for other bottom baryons. Our predictions for these states are in table \ref{tab:baryons_n}.

% and the intercept \(a = -1.39\).

\clearpage
\section{Predictions for glueball states} \label{sec:glueballs}
In \cite{Sonnenschein:2015zaa} we have analyzed the spectra of the isoscalar resonances \(f_0\) (\(J^{PC}=0^{++}\)) and \(f_2\) (\(2^{++}\)). In appendix \ref{sec:states} we list all the known \(f_0\) and \(f_2\) states. It is not too unlikely that one of the known \(f_0\) resonances is  in fact the glueball scalar ground state. In our paper we have proposed to organize the different resonances into the Regge trajectories of mesons and those of glueballs. To identify the glueball, one should measure the excited states.

\subsection{The spectrum of glueballs}
\subsubsection{Scalar glueballs}
In \cite{Sonnenschein:2015zaa} four candidates for the glueball ground state were examined: \(f_0(980)\), \(f_0(1370)\), \(f_0(1500)\), and \(f_0(1710)\). We analyzed the spectrum of the \(f_0\) resonances by assigning them to radial trajectories. There were three types of trajectory, nearly linear trajectories for light mesons, mass corrected trajectories of \(\ssb\), and the linear trajectories with half the meson slope for the glueballs. For each glueball candidate, we assign the rest of the \(f_0\) to meson trajectories and predict where the next states should be.

Table \ref{tab:f0s} is a summary of the different possibilities. Each group of three rows is a different assignment, and we list the next two predicted states on each trajectory. Note that trajectories begin with \(n=0\) and that for the glueball, only even \(n\) are allowed. Recurring in the table is the state \(f_0(1800)\), whose nature and status are still unclear. For most of these assignments to be complete this state is needed in one of the meson trajectories.

Among the conclusions of \cite{Sonnenschein:2015zaa} were that an assignment with meson trajectories alone is not fully consistent with all data, and that the \(f_0(980)\) can be a light glueball, or it can be excluded from the trajectories as a \(K\bar K\) bound state, which many suppose it is. The assignments that were most consistent were those with either \(f_0(1370)\) or \(f_0(1500)\) as the glueball ground state.

We also include an estimate for the widths of the excited states. When calculating the width of a glueball, we assume the simplified relation \(\Gamma/M = Const.\),\footnote{In our previous work \cite{Sonnenschein:2015zaa} we stated that \(\Gamma \propto L^2 \propto M^2\) for the glueball, with the rationale that a closed string has to tear twice, each time with a probability proportional to \(L\), to decay. However, the full decay width of the closed string is still expected to be proportional to \(L\) \cite{Dai:1989cp}, so we correct this now.} and therefore the width of a state with mass \(M\) is calculated as
\be \Gamma = M\frac{\Gamma_0}{M_0} \ee
with \(M_0\) and \(\Gamma_0\) the experimentally measured mass and width of the candidate ground state.

	%\begin{table}[tp!] \centering
		%\begin{tabular}{|c|c|c|c|c|c|c|c|c|c|c|c|c|} \cline{2-13}
		%\multicolumn{1}{c|}{} & \multicolumn{4}{|c|}{Glueball} & \multicolumn{4}{|c|}{Light} & \multicolumn{4}{|c|}{\(\ssb\)} \\ \hline\hline
		%Assignment  & \multicolumn{4}{|c|}{980} & \multicolumn{4}{|c|}{1370, 1710, 2100} & \multicolumn{4}{|c|}{1500, 2020} \\ \hline
		%Next \(M\) 
		%\(\Gamma\)
		%
		%
		%\end{tabular}
			%\end{table}
			
		\begin{table}[tp!] \centering
				\begin{tabular}{|c|l|l|c|c|c|c|c|c|c|c|} \hline
					\multicolumn{3}{|c|}{Trajectories} & \multicolumn{6}{|c|}{Predicted states} \\ \hline
					
					\(\alp\) & Type     &  Assigned states & \(n\) & Mass\, & Width & \(n\) & Mass\, & Width \\ \hline\hline
					
							& Glueball & \textbf{980} & 2 & 2470 & 180 & 4 & 3350 & 240 \\ 
				
					0.78 & Light & 1370, 1710, 2100, 2330 & 4 & 2620 & 200 & 4 & 2850 & 250 \\ 
					
							& \(\ssb\) & 1500, 2020 & 2 & 2300 & 300 & 3 & 2590 & 300 \\ \hline\hline

							& Glueball & \textbf{1370} & 2 & 2510 & \(>700\) & 4 & 3290 & \(>900\) \\ 
					
					0.89 & Light & 1500, *1800, 2100, 2330 & 4 & 2580 & 200 & 5 & 2790 & 250 \\ 
					
							& \(\ssb\) & 1710, 2200 & 2 & 2390 & 200 & 3 & 2630 & 250 \\ \hline\hline

							& Glueball & \textbf{1500} & 2 & 2600 & 180 & 4 & 3350 & 240 \\ 
					
					0.89 & Light & 1370, *1800, 2020, 2330 & 4 & 2540 & 350 & 5 & 2760 &  400 \\ 
					
							& \(\ssb\) & 1710, 2100 & 2 & 2360 & 250 & 3 & 2610 & 250 \\ \hline\hline

							& Glueball & \textbf{1710} & 2 & 2800 & 220 & 4 & 3570 & 280 \\ 
					
					0.82 & Light & 1370, *1800, 2100, 2330 & 4 & 2610 & 300 & 5 & 2840 & 300 \\ 
					
							& \(\ssb\) & 1500, 2020, 2200 & 3 & 2270 & 350 & 4 & 2550 & 350\\ \hline
					
					\end{tabular} \caption{\label{tab:f0s} The different assignments of the \(f_0\) into radial trajectories and predicted higher states. The slope \(\alp\) (in units of \GEV) was fitted for each assignment separately, as was done in \cite{Sonnenschein:2015zaa}, but is common to all three types of trajectories. Widths are provided as estimates, based on proportionality of the width to the string length.}
			\end{table}

In the following tables, \ref{tab:980}-\ref{tab:1710} we list separately the predicted glueball states, including orbital excitations. We only list states positive under parity and charge conjugation, \(PC = ++\). The states are calculated by taking linear trajectories with the slope 0.40--0.45 GeV\(^{-2}\) in the \((J,M^2)\), or a lower value, 0.35--0.40 GeV\(^{-2}\), in the \((n,M^2)\) plane. The errors take into account both the experimental uncertainty in \(M_0\) and \(\Gamma_0\) and the uncertainty in the fitting parameter \(\alp\).
					
\begin{table}[h!] \centering
\begin{tabular}{|c|c|c|} \hline
n 	&	 Mass 	&	 Width \\ \hline
0 	&	990\plm20 	&	 70\plm30	\\ \hline
% 1	&	1910\plm65 	&	 135\plm60	\\ \hline
2	&	2515\plm85 	&	 180\plm75	\\ \hline
%3	&	2995\plm100 	&	 210\plm90	\\ \hline
4	&	3415\plm115 	&	 240\plm105	\\ \hline
%5	&	3785\plm125 	&	 270\plm115	\\ \hline
6	&	4120\plm140 	&	 290\plm125	\\ \hline
%7	&	4430\plm150 	&	 315\plm135	\\ \hline
%8	&	4725\plm160 	&	 335\plm145	\\ \hline
%9	&	5000\plm165 	&	 355\plm150	\\ \hline
%10	&	5260\plm175 	&	 370\plm160	\\ \hline
\end{tabular} \qquad
\begin{tabular}{|c|c|c|} \hline
$J$ 	&	 Mass 	&	 Width \\ \hline
0 	&	990\plm20 	&	 70\plm30	\\ \hline
%1	&	1825\plm55 	&	 130\plm55	\\ \hline
2	&	2385\plm70 	&	 170\plm75	\\ \hline
%3	&	2835\plm85 	&	 200\plm85	\\ \hline
4	&	3225\plm95 	&	 230\plm100	\\ \hline
%5	&	3570\plm105 	&	 250\plm110	\\ \hline
6	&	3885\plm115 	&	 275\plm120	\\ \hline
%7	&	4175\plm125 	&	 295\plm125	\\ \hline
%8	&	4450\plm130 	&	 315\plm135	\\ \hline
%9	&	4705\plm140 	&	 335\plm145	\\ \hline
%10	&	4950\plm145 	&	 350\plm150	\\ \hline
\end{tabular}
\caption{\label{tab:980} Predictions using linear trajectories with glueball slope using \(f_0(980)\) as the ground state.}
\end{table}

\begin{table}[h!] \centering
	\begin{tabular}{|c|c|c|} \hline
n 	&	 Mass 	&	 Width \\ \hline
0 	&	1350\plm150 	&	 350\plm150	\\ \hline
%1	&	2120\plm120 	&	 550\plm245	\\ \hline
2	&	2675\plm120 	&	 695\plm310	\\ \hline
%3	&	3135\plm125 	&	 815\plm360	\\ \hline
4	&	3535\plm130 	&	 915\plm405	\\ \hline
%5	&	3895\plm140 	&	 1010\plm450	\\ \hline
6	&	4220\plm150 	&	 1095\plm485	\\ \hline
%7	&	4525\plm160 	&	 1175\plm520	\\ \hline
%8	&	4810\plm165 	&	 1250\plm555	\\ \hline
%9	&	5080\plm175 	&	 1315\plm585	\\ \hline
%10	&	5340\plm180 	&	 1385\plm615	\\ \hline
\end{tabular} \qquad
\begin{tabular}{|c|c|c|} \hline
$J$ 	&	 Mass 	&	 Width \\ \hline
0 	&	1350\plm150 	&	 350\plm150	\\ \hline
%1	&	2045\plm120 	&	 530\plm235	\\ \hline
2	&	2555\plm110 	&	 660\plm295	\\ \hline
%3	&	2980\plm110 	&	 775\plm345	\\ \hline
4	&	3350\plm115 	&	 870\plm385	\\ \hline
%5	&	3685\plm120 	&	 955\plm425	\\ \hline
6	&	3995\plm130 	&	 1035\plm460	\\ \hline
%7	&	4275\plm135 	&	 1110\plm490	\\ \hline
%8	&	4545\plm140 	&	 1180\plm525	\\ \hline
%9	&	4795\plm150 	&	 1245\plm550	\\ \hline
%10	&	5035\plm155 	&	 1305\plm580	\\ \hline
\end{tabular}
\caption{\label{tab:1370} Predictions using linear trajectories with glueball slope using \(f_0(1370)\) as the ground state.}
\end{table}

\begin{table}[h!] \centering
	\begin{tabular}{|c|c|c|} \hline
n 	&	 Mass 	&	 Width \\ \hline
0 	&	1505\plm6 	&	 109\plm7	\\ \hline
%1	&	2220\plm80 	&	 160\plm10	\\ \hline
2	&	2755\plm95 	&	 200\plm15	\\ \hline
%3	&	3205\plm110 	&	 230\plm15	\\ \hline
4	&	3595\plm120 	&	 260\plm20	\\ \hline
%5	&	3950\plm135 	&	 285\plm20	\\ \hline
6	&	4275\plm145 	&	 310\plm20	\\ \hline
%7	&	4575\plm155 	&	 330\plm25	\\ \hline
%8	&	4860\plm165 	&	 350\plm25	\\ \hline
%9	&	5125\plm170 	&	 370\plm25	\\ \hline
%10	&	5380\plm180 	&	 390\plm30	\\ \hline
\end{tabular} \qquad
\begin{tabular}{|c|c|c|} \hline
$J$ 	&	 Mass 	&	 Width \\ \hline
0 	&	1505\plm6 	&	 109\plm7	\\ \hline
%1	&	2150\plm70 	&	 155\plm10	\\ \hline
2	&	2640\plm80 	&	 190\plm15	\\ \hline
%3	&	3055\plm90 	&	 220\plm15	\\ \hline
4	&	3415\plm100 	&	 245\plm20	\\ \hline
%5	&	3745\plm110 	&	 270\plm20	\\ \hline
6	&	4050\plm120 	&	 295\plm20	\\ \hline
%7	&	4330\plm130 	&	 315\plm20	\\ \hline
%8	&	4590\plm135 	&	 335\plm25	\\ \hline
%9	&	4840\plm145 	&	 350\plm25	\\ \hline
%10	&	5080\plm150 	&	 370\plm25	\\ \hline
\end{tabular}
\caption{\label{tab:1500} Predictions using linear trajectories with glueball slope using \(f_0(1500)\) as the ground state.}
\end{table}

\begin{table}[h!] \centering
	\begin{tabular}{|c|c|c|} \hline
n 	&	 Mass 	&	 Width \\ \hline
0 	&	1720\plm6 	&	 135\plm8	\\ \hline
%1	&	2370\plm90 	&	 185\plm15	\\ \hline
2	&	2880\plm100 	&	 225\plm15	\\ \hline
%3	&	3310\plm115 	&	 260\plm20	\\ \hline
4	&	3690\plm125 	&	 290\plm20	\\ \hline
%5	&	4035\plm135 	&	 315\plm20	\\ \hline
6	&	4355\plm145 	&	 340\plm25	\\ \hline
%7	&	4650\plm155 	&	 365\plm25	\\ \hline
%8	&	4930\plm165 	&	 385\plm25	\\ \hline
%9	&	5190\plm175 	&	 410\plm30	\\ \hline
%10	&	5445\plm180 	&	 425\plm30	\\ \hline
\end{tabular} \qquad
\begin{tabular}{|c|c|c|} \hline
\(J\) 	&	 Mass 	&	 Width \\ \hline
0 	&	1720\plm6 	&	 135\plm8	\\ \hline
%1	&	2305\plm80 	&	 180\plm10	\\ \hline
2	&	2770\plm85 	&	 215\plm15	\\ \hline
%3	&	3165\plm95 	&	 250\plm15	\\ \hline
4	&	3515\plm105 	&	 275\plm20	\\ \hline
%5	&	3835\plm115 	&	 300\plm20	\\ \hline
6	&	4130\plm125 	&	 325\plm20	\\ \hline
%7	&	4410\plm130 	&	 345\plm25	\\ \hline
%8	&	4665\plm140 	&	 365\plm25	\\ \hline
%9	&	4915\plm145 	&	 385\plm25	\\ \hline
%10	&	5145\plm150 	&	 405\plm25	\\ \hline
\end{tabular}
\caption{\label{tab:1710} Predictions using linear trajectories with glueball slope using \(f_0(1710)\) as the ground state.}
\end{table}

\subsubsection{Tensor glueballs}
The first excited state of the glueball is expected to have \(J^{PC} = 2^{++}\). Again, we have many \(f_2\) resonances with the quantum number expected of the glueball. Sorting the \(f_2\) states into trajectories, the situation is somewhat simpler than with the scalar, as here we have two states that belong on meson trajectories in the \((J,M^2)\) plane. The \(f_2(1270)\) belongs to the trajectory of the \(\omega\) meson, and the \(f^\prime_2(1525)\) is \(s\bar{s}\) and sits on the \(\phi\) trajectory. Their decay modes and other properties are also well known and there is no real doubt about their nature. We take these two states as the heads of trajectories, and sort the remaining states.

In \cite{Sonnenschein:2015zaa} we organized the \(f_2\) resonances into the trajectories shown in table \ref{tab:f2s}. There are the meson trajectories headed by \(f_2(1270)\) and \(f^\prime_2(1525)\), as well as another pair of states on a light quark trajectory. Among the remaining \(f_2\) states listed by the PDG are the \(f_2(2010)\), which was observed to decay to \(\phi\phi\) (despite the very small phase space). It could perhaps be classified as a \(\phi\phi\) bound state, in an analogous fashion to the \(f_0(980)\) being \(K\bar K\). The most interesting states after that remain the \(f_2(1430)\) and \(f_J(2220)\). While the latter has been considered a candidate for the glueball and has been the object of some research (see many papers citing \cite{Bai:1996wm}), the former is rarely addressed, despite its curious placement in the spectrum between the lightest \(2^{++}\) light and \(\ssb\) mesons. It seems a worthwhile experimental question to clarify its status - and its quantum numbers, as the most recent observation \cite{Vladimirsky:2001ek} can not confirm whether it is a \(0^{++}\) or \(2^{++}\) state, a fact which led to at least one suggestion \cite{Vijande:2004he} that the \(f_2(1430)\) could be itself the scalar glueball.

The predictions for the \(2^{++}\) glueball were given in tables \ref{tab:980}-\ref{tab:1710}, assuming the tensor is an excited partner of the scalar ground state. Then it is expected in the range 2385--2770, depending on which is the scalar ground state. In table \ref{tab:f2s}, we give the predicted states of \(2^{++}\) mesons in that range. There are multiple meson states expected in that range as well as glueballs. We also list the predicted excited states for the glueball candidates \(f_2(1430)\) and \(f_J(2220)\), assuming they are the head of a new trajectory with half the meson slope.

	\begin{table}[tp!] \centering
				\begin{tabular}{|l|l|c|c|c|c|c|c|c|c|c|} \hline
					\multicolumn{2}{|c|}{Trajectories} & \multicolumn{9}{|c|}{Predicted states} \\ \hline
					
					Type     &  Assigned states & \(n\) & Mass\, & Width & \(n\) & Mass\, & Width  & \(n\) & Mass\, & Width \\ \hline\hline
					
							Light & 1270, 1640, 1950 & 3 & 2260 & 400 & 4 & 2510 & 450 & 5 & 2730 & 500 \\ 
				
					 \(\ssb\) & 1525, 2010, 2300 & 3 & 2540 & 150 & 4 & 2780 & 200 & 5 & 2990 & 200  \\ 
					
							2nd light & 1810, 2150 & 2 & 2380 & 200 & 3 & 2610 & 250 & 4 & 2830 & 250 \\ \hline\hline
							
							Glueball? & \(f_2(1430)\) & 2 & 2610 & - & 4 & 3390 & - & 6 & 4030 & - \\ \hline\hline
							
							Glueball? & \(f_J(2220)\) & 2 & 3110 & - & 4 & 3790 & - & 6 & 4370 & - \\ \hline
					\end{tabular} \caption{\label{tab:f2s} The meson trajectories of \(f_2\) as discussed in \cite{Sonnenschein:2015zaa}, with their excited states. The fitted slope here is \(0.85 \GEVm\), and half this value is used for the glueball trajectories. Widths are provided as estimates, based on proportionality of the width to the string length. However, the glueball candidates have very small widths, 13 MeV for the \(f_2(1430)\) and 23 MeV for the \(f_J(2220)\). The predicted widths for the excited states based on our model are between 30 to 50 MeV, but it is unlikely that they remain so narrow.}
			\end{table}

\subsection{Glueball decays} \label{sec:glueball_decays}
In \cite{Sonnenschein:2017ylo}, we have used several measurements from meson decays to measure the strangeness suppression factor \(\lambda_s\), which is defined as the relative probability to create a pair of \(s\bar s\) compared with a pair of \(u\bar u\) or \(d \bar d\). In holography pair creation is realized by the string fluctuating in the radial direction to reach a flavor brane. The probability is suppressed exponentially in the distance from the wall to the flavor brane, which itself is proportional to the mass of the quark. Namely there is a suppression factor of the form \(\exp(-cm^2_q/T)\), for some model dependent constant \(c\) \cite{Peeters:2005fq}. This type of exponential suppression in \(m_q^2\) is familiar also from Schwinger pair production in the QCD flux tube \cite{Casher:1978wy}.

The cleanest measurements of \(\lambda_s\) in \cite{Sonnenschein:2017ylo} were from the radiative decays of the \(\ccb\) and \(\bbb\). They are also most relevant to the glueball decays since we can think of radiative decays as the string closing and annihilating the quarks, then tearing again to create lighter quarks. The measured result was \(\lambda_s = 0.36\pm0.12\).

For the glueball, the closed string must tear twice to decay, and so we can create zero, one, or two pairs of \(s\) and \(\bar s\) in the decay. The glueball could then decay to a pair of light quark mesons, a pair of a strange and an anti-strange meson, or a pair of \(\ssb\), with each creation of an \(\ssb\) pair adding a factor of \(\lambda_s\). The presence of these three types of two body decays, with the given hierarchy between them, is unique to closed strings, and hence could be useful in identifying glueballs.

To give a numeric example of the expected ratios, consider a glueball candidate of a mass of 2.2 GeV, where we assume a specific value of the mass in order to evaluate the phase space factors explicitly. A glueball of this mass could be an excited scalar glueball, or a tensor \cite{Sonnenschein:2015zaa}. It is above threshold for decays into two \(\ssb\) mesons. So we can predict the existence of several decay channels and the relative ratios between them. For instance, for decays into a pair of chargeless vector mesons, we predict the ratio between the decay widths
\be \Gamma(GB\to\omega\omega) : \Gamma(GB\to K^{*0}K^{*0}) : \Gamma(GB\to\phi\phi) = 1 : 0.83\times \lambda_s : 0.53\times \lambda_s^2\,. \ee
Taking \(\lambda_s = 0.36\), then \(\lambda_s^2=0.13\) and the numbers are
\be \Gamma(GB\to\omega\omega) : \Gamma(GB\to K^{*0}K^{*0}) : \Gamma(GB\to\phi\phi) = 1 : 0.30 : 0.07. \ee
If the glueball is a higher mass than 2.2 GeV, then the phase space is less restricted. Even in the 2.2 GeV case, the ratio of 7\% to create the \(\phi\phi\) means all three decay channels should be observable. Other predictions for holographic glueball decays are in \cite{Brunner:2015oqa,Brunner:2015yha,Brunner:2016ygk,Brunner:2018wbv,Yadav:2018zxw}.
%%%%%%%%%%%%%%%%%%%%%%%%%%%%%%%%%%%%%%
\section{Predictions for tetraquark states} \label{sec:tetra}
Our paper \cite{Sonnenschein:2016ibx} discussed at length the stringy tetraquark, the system of a diquark and an anti-diquark joined by a string. Like all hadrons, these states should lie on HMRTs. We discuss in particular the charmonium sector, since there we find a candidate for exactly such a state. We predict more of these states, which should be identifiable by their decays to a baryon anti-baryon pair. The focus on exotic hadron phenomenology is among the charmonium and bottomonium states \cite{Esposito:2014rxa,Chen:2016qju,Ali:2017jda}. In section \ref{sec:quarkonium} we discussed the spectrum of quarkonia.

\subsection{Tetraquarks in the charmonium sector}
Of particular interest to us is the state \(Y(4630)\). It was observed by Belle in 2008, in the process \(e^+e^-\to \gamma \Lambda^+_c\Lambda^-_c\), with a significance of \(8\sigma\) \cite{Pakhlova:2008vn}. The parameters measured there for the \(Y(4630)\) were
\be J^{PC} = 1^{--}\,,\,\, M_{Y(4630)} = 4634^{+9}_{-11}\,, \,\, \Gamma_{Y(4630)} = 92^{+41}_{-32} \,.\ee
Near the \(Y(4630)\), there is another reported resonance, the \(Y(4660)\). It was observed by both Belle \cite{Wang:2007ea} and BaBar \cite{Aubert:2007zz}, also has \(J^{PC} = 1^{--}\), and its measured mass and width are
\be M_{Y(4660)} = 4665\pm10\,, \qquad \Gamma_{Y(4660)} = 53\pm16\,. \ee
The channel in which it was observed is \(e^+e^-\to\gamma\pi^+\pi^-\Psi(2S)\). It may well be that the two resonances \(Y(4630)\) and \(Y(4660)\) are really one and the same. The PDG lists the \(Y(4630)\) and \(Y(4660)\) measurements under a single entry for the \(Y(4660)\). What we have then is a single state with both decay channels, \(\Lambda_c\overline{\Lambda}_c\) and \(\Psi(2S)\pi^+\pi^-\). If this is the case, then the \(\Lambda_c\bar{\Lambda}_c\) is still the dominant decay, so this does not interfere so much with the tetraquark picture \cite{Cotugno:2009ys,Guo:2010tk}.

If the \(Y(4630)\) is indeed a tetraquark, it should be part of a HMRT. Using the values of \(m_{cu}\approx m_c\) and \(\alp\) of the \(\ccb\) trajectories, we can use the known mass of the state \(Y(4630)\) and extrapolate from it to higher (and perhaps lower) states along the trajectory. We use the values
\be m_c = 1490 \MEV\,, \alp\!_J = 0.86 \GEVm\,,\qquad \alp\!_n = 0.59 \GEVm\,. \ee
where there are two different slopes, one for orbital trajectories in \(J\) and one for radial trajectories in \(n\).

By extrapolating the trajectory backwards, to lower masses, we can check if the \(Y(4630)\) is an excited state of an already known resonance. In the angular momentum plane, the preceding state would be a scalar  \(J^{PC} = 0^{++}\). There is no currently known scalar resonance at the mass predicted by the HMRT, of about 4480 MeV. In the \((n,M^2)\) plane we find the \(Y(4360)\), another vector resonance that is at the right mass for the \(Y(4630)\) to be its first excited state. Its mass and width are
\be M_{Y(4360)} = 4354\pm10\,, \qquad \Gamma_{Y(4360)} = 78\pm16\,. \ee
It was observed in decays to \(\Psi(2S)\pi^+\pi^-\) like the \(Y(4660)\). The problem with assigning the \(Y(4360)\) as the unexcited partner of the \(Y(4630)\) is in its width. The widths of the two states are roughly the same even though the lower state, being beneath the \(\Lambda_c\bar{\Lambda}_c\) threshold, should be significantly narrower than the above threshold state.

Whether or not there is a lower, unexcited version of \(Y(4630)\), we still predict higher excitations of the stringy tetraquark. For the higher states, the \(\Lambda_c\bar{\Lambda}_c\) decays should be dominant, and their masses should be such that they fall on a HMRT.   The predictions are in table \ref{tab:pred_y_c}.

\begin{table}[h!] \centering
	\begin{tabular}{|c|c|c|} \hline
\(n\)	&	 Mass 	&	 Width \\ \hline\hline
``-2'' & 4060\plm50 & Narrow \\ \hline
``-1'' & 4360\plm50 & Narrow \\ \hline\hline
0 & \(\bf{4634^{+9}_{-11}}\)	&	\(\bf{92^{+41}_{-32}}\)	\\ \hline\hline
1 &	4870\plm50 	&	 150--250	\\ \hline
2	&	5100\plm60 	&	 200--300	\\ \hline
3	&	5305\plm60 	&	 220--320	\\ \hline
4	&	5500\plm60 	&	 250--350	\\ \hline
\end{tabular} \qquad\qquad
\begin{tabular}{|c|c|c|} 
\multicolumn{3}{c}{} \\ \hline
\(J^{PC}\) &	  Mass 	&	 Width \\ \hline\hline
\(0^{++}\) & 4485\plm40					& Narrow \\ \hline\hline
\(1^{--}\) & \(\bf{4634^{+9}_{-11}}\)					& \(\bf{92^{+41}_{-32}}\)	\\ \hline\hline
\(2^{++}\) & 4800\plm40 	&	 150--250	\\ \hline
\(3^{--}\) & 4960\plm40 	&	 180--280	\\ \hline
\(4^{++}\) & 5100\plm45 	&	 200--300	\\ \hline
\(5^{--}\) & 5260\plm45 	&	 250--350	\\ \hline
\end{tabular} \caption{\label{tab:pred_y_c} Trajectories of the \(Y(4630)\). Based on the experimental mass and width of the \(Y(4630)\) we extrapolate to higher excited states on the trajectory. Uncertainties are based on both experimental errors and uncertainties in the fit parameters. The excited states are expected to decay into \(\Lambda_c\overline{\Lambda}_c\). Some possible lower states, with masses below the \(\Lambda_c\overline{\Lambda}_c\) threshold, are also included. \(Y(4360)\), observed to decay to \(\Psi(2S)\pi^+\pi^-\), is a candidate for the \(n = -1\) state. See text for details.}
\end{table}

\subsection{Tetraquarks in the bottomonium sector}
In the spectroscopy of heavy quarkonia, one usually expects analogies to exist between the \(\ccb\) and \(\bbb\) spectra. Tetraquarks and other exotics should be no exception. Therefore, we propose to search for a state directly analogous to the \(Y(4630)\), which we will call the \(Y_b\), that decays primarily to \(\Lambda_b\bar{\Lambda}_b\). We will assume that, like the \(Y(4630)\), the \(Y_b\) state is located a little above the relevant baryon-antibaryon threshold of \(\Lambda_b\overline{\Lambda}_b\).

The mass of the \(\Lambda_b^0\) is \(5619.51\pm0.23\) MeV. As an estimate - take a mass of
\be \begin{split} & M(Y_b) \approx 2M(\Lambda_b^0)+40\MEV \\
 & \Delta M(Y_b) \approx 40\MEV \end{split} \ee
The \(Y_b\) would also have its own Regge trajectory, so, based on the mass we choose for it, we can predict the rest of the states that lie on the trajectory. The trajectories are calculated using the slopes of the \(\Upsilon\) trajectories, which are \(\alp \approx 0.64\GEVm\) in the \((J,M^2)\) plane, and \(\alp\approx0.46\GEVm\) in the \((n,M^2)\) plane. The string endpoint mass of the \(b\) quark is \(m_b = 4730\) MeV \cite{Sonnenschein:2014jwa}. The resulting masses are listed in table \ref{tab:pred_y_b}.

The masses of the states that would precede the above threshold \(Y_b\) tetraquark are also listed in table \ref{tab:pred_y_b}. The observed resonance \(Y_b(10890)\) is a potential match to be the ``\(n = -2\)'' state. This resonance was observed to decay to \(\pi^+\pi^-\Upsilon(nS)\) and has a mass of \(10888.4\pm0.3\) \cite{Olsen:2014qna,Abe:2007tk,Ali:2009pi}. It seems analogous to the \(Y(4360)\) from the charmonium sector, which decayed to \(\pi^+\pi^-\Psi(2S)\) and was at the right mass to be an unexcited partner of the tetraquark candidate \(Y(4630)\). If the \(Y(10890)\) is on the trajectory, then we also predict another resonance following it, which being similarly below the baryon-antibaryon threshold should have similar properties, at a mass of \(\approx11080\) MeV.

\begin{table}[h!] \centering
	\begin{tabular}{|c|c|} \hline
\(n\)	&	 Mass 	\\ \hline\hline
``-2'' & 10870\plm40 \\ \hline
``-1'' & 11080\plm40 \\ \hline\hline
0 & 11280\plm40	\\ \hline
1 &	11470\plm40 \\ \hline
2	&	11650\plm40 \\ \hline
3	&	11820\plm40 \\ \hline
4	&	11980\plm40 \\ \hline
\end{tabular} \qquad\qquad
\begin{tabular}{|c|c|} \hline
\(J^{PC}\) &	  Mass 	\\ \hline\hline
\(0^{++}\) & 11140\plm40	\\ \hline
\(1^{--}\) & 11280\plm40	\\ \hline
\(2^{++}\) & 11420\plm40 	\\ \hline
\(3^{--}\) & 11550\plm40 	\\ \hline
\(4^{++}\) & 11670\plm40	\\ \hline
\(5^{--}\) & 11790\plm40 	\\ \hline
\end{tabular} 
\caption{\label{tab:pred_y_b} Predictions for the states of the \(Y_b\), a tetraquark containing \(\bbb\) and decaying to \(\Lambda_b\bar{\Lambda}_b\). The mass of the first state is taken near threshold, masses of higher states are on the HMRTs that follow from the ground state.}
\end{table}

\subsection{Lighter tetraquarks}
One of the puzzles in exotic hadron spectroscopy is the absence of exotics in the light quark sector. We propose to search for the hidden strange analogue of the \(Y(4630)\), a vector resonance decaying to \(\Lambda\bar{\Lambda}\), and the excited states on its trajectory. We predict again a trajectory of states beginning with a near-threshold one. The masses are in table \ref{tab:pred_y_s}. We take a ground state mass of
\be M(Y_s) \approx (2M(\Lambda)+40)\pm40\MEV \ee
with the usual slope for light quark (\(u\), \(d\), \(s\)) hadrons of 0.9 \GEV in the \((J,M^2)\) plane. In the \((n,M^2)\) the slope is lower, and we take 0.8 \GEV. The mass of the \(s\) quark is taken to be 400 MeV. Again we repeat the exercise of extrapolating the trajectory backwards, listing two preceding states. The vector resonance \(\rho(1570)\) is at the right mass to be a match.

\begin{table}[h!] \centering
	\begin{tabular}{|c|c|} \hline
\(n\)	&	 Mass 	\\ \hline\hline
``-2'' & 1570\plm40	\\ \hline
``-1'' & 1960\plm40	\\ \hline\hline
0 & 2270\plm40	\\ \hline
1 &	2540\plm40 		\\ \hline
2	&	2790\plm40 		\\ \hline
3	&	3010\plm40 		\\ \hline
4	&	3220\plm40 		\\ \hline
\end{tabular} \qquad\qquad
\begin{tabular}{|c|c|} \hline
\(J^{PC}\) &	  Mass 	 \\ \hline\hline
\(0^{++}\) & 1990\plm40			\\ \hline
\(1^{--}\) & 2270\plm40			\\ \hline
\(2^{++}\) & 2510\plm40 		\\ \hline
\(3^{--}\) & 2730\plm40 		\\ \hline
\(4^{++}\) & 2940\plm40			\\ \hline
\(5^{--}\) & 3130\plm40 		\\ \hline
\end{tabular} 

\caption{\label{tab:pred_y_s} Predictions for the states of the \(Y_s\), a tetraquark containing \(\ssb\) and decaying to \(\Lambda\bar{\Lambda}\).}
\end{table}

We note that while the natural way to search for molecules is by looking in the vicinity of different two hadron thresholds \cite{Karliner:2015ina}, a good way to identify tetraquarks is to look above the two baryon threshold. It is there that we can look for particles that decay into a baryon anti-baryon pair, the stringy tetraquark's natural and distinctive mode of decay, but it is also a region where data is still quite scarce. And while the predictions above are for the \(Y(4630)\), and its immediate analogues \(Y_b\) and \(Y_s\), it should be noted that there could be similar states with non-symmetric decay channels, like \(\Lambda_c\overline \Sigma_c\) for other hidden charm tetraquarks, \(\Lambda_c\overline\Lambda\) being charmed and strange, etc.. For resonances found in such channels we will also expect trajectories, which can be easily delineated using the HISH model.

%%%%%%%%%%%%%%%%%%%%%%
%%%%%%%%%%%%%%%%%%%%%%

\section{Summary} \label{sec:summary}
The Regge trajectories are a central feature of the hadron spectrum, and are still widely used today to characterize the spectrum of hadrons \cite{Anisovich:2000kxa,Kataev:2004ve,Gershtein:2006ng,Ebert:2009ub,Ebert:2009ua,Ebert:2011jc,Kher:2017wsq,Afonin:2018bej,Chen:2018bbr,Jia:2018vwl}, although these descriptions usually deal only with light hadrons. Not only do we have the linear or near-linear trajectories of the light hadrons such as the \(\rho\) meson or the nucleons, but the picture also extends to the HISH mass modified Regge trajectories of hadrons including heavy quarks. Charmed and bottom mesons can be sorted into trajectories like the light mesons, with the endpoint masses of the quarks describing the deviation from linearity of these trajectories. The Regge slope (or equivalently string tension) measured for these trajectories is almost always the same for heavy hadrons as it is for the light ones, which is strong evidence for the validity of the stringy picture for all the different types of hadrons. Furthermore, the stringy picture appears to remain valid even when describing low spin, unexcited states, contrary to the usual lore that is should hold  only for excited states long flux tubes form, which can then be given an effective string description.

Using the measured endpoint masses of the quarks - \(m_{u/d} \approx 60\) MeV, \(m_s = 400\) MeV, \(m_c = 1490\) MeV, and \(m_b = 4700\) MeV - and the (near) universal Regge slope of \(\alp = 0.88 \GEVm\) for mesons or \(0.95 \GEVm\) for baryons, it is easy to place various hadrons on trajectories, find the intercept for each trajectory, and then predict higher excited states by extrapolating the line. Given the slope and endpoint masses, we can use any single state to predict the spectrum of its excited partners. In tables \ref{tab:mesons_J} and \ref{tab:mesons_n} we collected all our predictions for excited meson states. Tables \ref{tab:baryons_J} and \ref{tab:baryons_n} collect the results for baryons.

Some exotic hadrons are also expected to behave as strings. The glueball can be described as a folded closed string and as such will have a trajectory with half the slope of the mesons. A tetraquark can be thought of as a string connecting a diquark to an anti-diquark, and will also have a trajectory like that of the baryons. In both these cases, it could be possible to identify the exotics by their trajectories. And this is done not by the properties of a single state, but by finding a whole spectrum of exotic states. For the glueballs the first excited states are expected in the range of 2.5-3.0 GeV, where not much is known. The excited glueballs are expected to have a characteristic decay pattern as described in section \ref{sec:glueball_decays}. To identify tetraquarks we have suggested to look for states above the threshold for baryon-antibaryon decays of the tetraquark, for instance above the \(\approx4570\) MeV threshold for \(\Lambda_c\overline\Lambda_c\) decays to identify charmonium-like tetraquark candidates. Our predictions determine where these states should be found. If one accepts diquarks as natural objects in QCD, as they appear in all baryons in our model, then the tetraquark states must surely also be part of the spectrum.

The observation and understanding of higher excited states is crucial if we want to gain more insight into the stringy nature of hadrons. In particular, measurements excitations of heavy and doubly heavy baryons will teach us more about diquarks and the diquark structure of the baryons. In this note we assumed that a naive approximation of the diquark mass (as a string endpoint, as always) is \(m_{(q_1 q_2)} \approx m_{q_1}+m_{q_2}\), even though in the holographic picture there is no reason that this simple sum rule should hold. It would certainly be interesting if one would discover significant deviations from this.

A natural question related to the predictions of the HISH model is what are the priorities in attempting to verify the predictions. It seems to us that the highest priority should be given to detecting  glueballs and disentangling them from flavorless mesons. The main point here is that to identify them it is not enough to detect one state but rather one has to find, preferably three or more states that reside on a glueball trajectory. Only then by identifying the correct slope (of around $0.45 \GEVm$) we will be able to declare a glueball state. The second priority we would assign to verifying or disproving the HMRT associated with tetraquarks where certain of the states residing on it decay predominately to a baryon and an antibaryon. As mentioned above the $Y(4630)$ is such a state and it is important to verify the existence of its HMRT partners. There are also potential narrow below thresholds state that may belong to the trajectory, which may be seen in the \(\Psi(nS)\pi\pi\) channels. In particular, the Belle II experiment is well suited to observed this type of resonances. The next priority is to detect the HMRT states that include charm and bottom flavors and in particular baryonic configurations that have not been observed so far with three charmed quarks and two and three bottom quarks. 
 
Among the challenges still facing the HISH model are
\begin{itemize}
\item
As shown in appendix \ref{app:a} the measured  intercept, defined by the dependence of the orbital angular momentum on $M^2$, is always negative. The first determination of the intercept in the HISH model was carried out in \cite{Sonnenschein:2018aqf}. In symmetric systems where the two endpoints masses are the same the outcome intercept was found out to be always positive. For asymmetric strings, for instance for one massless and one massive endpoint the intercept can be negative. However it is obvious that this cannot explain the negative intercept of mesons with identical quarks at their ends. The analysis of \cite{Sonnenschein:2018aqf} did not include endpoint electric charges and spins. The latter change the boundary conditions of the string and hence the eigenfrequencies of fluctuations, and thus also the intercept.
\item
The basic structure of the HISH model is independent on the particular holographic background that is implemented. However, there are certain properties of the model that do vary from model to model. One such a property is the location along the holographic radial direction of the baryonic vertex \cite{Seki:2008mu,Dymarsky:2010ci,Sonnenschein:2014bia}. This affects the determination of the mass of the diquark. As was mentioned above, our current understanding of the HISH model  does not  allow us to  determine  unambiguously the mass of  a diquark built from two identical quarks. The mass of the diquark can vary between the mass of the quark and twice this mass, and can be heavier if the baryonic vertex contributes more than the strings to the endpoint mass. In addition the HISH model cannot yet determine the structure of the diquark in the cases that there are several options for the structure. These two ambiguities of the HISH model deserve further investigation.
\item
Since for light-light and heavy-light hadrons the HISH model with a universal slope fits nicely the experimental data, it seemed  that  the exploration of the differences between the HISH model and the stringy holographic model is not so crucial. Not unexpectedly for heavy quarkonia there are significant  deviations from a universal slope. This obviously calls for a quantitative study of holographic hadrons and the limitations of the HISH model. A related question is to try and understand the differences between the orbital and radial slops and between the baryon slope and the one for mesons.
\item
Here in this note we have included only predictions about tetraquark exotic hadrons. However, as was discussed in \cite{Sonnenschein:2016ibx} there is in principle a much richer spectrum of exotic hadrons.  For instance one can add to each baryonic vertex a pair (or several pairs) of strings with opposing orientations, namely connecting it to a quark and an antiquark. This will create a pentaquark (or higher multi-quark) states. These, if exist, are genuine exotic hadrons and not molecules of several hadrons.  
\item
A major challenge for the HISH model is to determine scattering amplitudes of hadrons from their stringy pictures, and to compare them to the measured ones. This subject is currently under investigation.
\end{itemize}

\section*{Acknowledgments}
We would like to thank Marek Karliner, Shmuel Nussinov, and Abner Soffer for useful discussions and for their remarks on the manuscript. This work was supported in part by a center of excellence supported by the Israel Science Foundation (grant number 2289/18).

\bibliographystyle{JHEP}
\bibliography{Decays}

\providecommand{\href}[2]{#2}\begingroup\raggedright\begin{thebibliography}{10}

\bibitem{Sonnenschein:2016pim}
J.~Sonnenschein, {\it {Holography Inspired Stringy Hadrons}},  {\em Prog. Part.
  Nucl. Phys.} {\bf 92} (2017) 1--49,
  [\href{http://xxx.lanl.gov/abs/1602.00704}{{\tt arXiv:1602.00704}}].

\bibitem{Sonnenschein:2014jwa}
J.~Sonnenschein and D.~Weissman, {\it {Rotating strings confronting PDG
  mesons}},  {\em JHEP} {\bf 1408} (2014) 013,
  [\href{http://xxx.lanl.gov/abs/1402.5603}{{\tt arXiv:1402.5603}}].

\bibitem{Sonnenschein:2014bia}
J.~Sonnenschein and D.~Weissman, {\it {A rotating string model versus baryon
  spectra}},  {\em JHEP} {\bf 1502} (2015) 147,
  [\href{http://xxx.lanl.gov/abs/1408.0763}{{\tt arXiv:1408.0763}}].

\bibitem{Sonnenschein:2015zaa}
J.~Sonnenschein and D.~Weissman, {\it {Glueballs as rotating folded closed
  strings}},  {\em JHEP} {\bf 12} (2015) 011,
  [\href{http://xxx.lanl.gov/abs/1507.01604}{{\tt arXiv:1507.01604}}].

\bibitem{Sonnenschein:2016ibx}
J.~Sonnenschein and D.~Weissman, {\it {A tetraquark or not a tetraquark: A
  holography inspired stringy hadron (HISH) perspective}},
  \href{http://xxx.lanl.gov/abs/1606.02732}{{\tt arXiv:1606.02732}}.

\bibitem{Sonnenschein:2017ylo}
J.~Sonnenschein and D.~Weissman, {\it {The decay width of stringy hadrons}},
  {\em Nucl. Phys.} {\bf B927} (2018) 368--454,
  [\href{http://xxx.lanl.gov/abs/1705.10329}{{\tt arXiv:1705.10329}}].

\bibitem{Collins:book}
P.~Collins, {\em {An Introduction to Regge Theory and High Energy Physics}}.
\newblock Cambridge Univeristy Press, 1977.

\bibitem{Selem:2006nd}
A.~Selem and F.~Wilczek, {\it {Hadron systematics and emergent diquarks}},  in
  {\em {Proceedings, Ringberg Workshop on New Trends in HERA Physics 2005:
  Ringberg Castle, Tegernsee, Germany, October 2-7, 2005}}, pp.~337--356, 2006.
\newblock \href{http://xxx.lanl.gov/abs/hep-ph/0602128}{{\tt hep-ph/0602128}}.

\bibitem{Seki:2008mu}
S.~Seki and J.~Sonnenschein, {\it {Comments on Baryons in Holographic QCD}},
  {\em JHEP} {\bf 01} (2009) 053,
  [\href{http://xxx.lanl.gov/abs/0810.1633}{{\tt arXiv:0810.1633}}].

\bibitem{Dymarsky:2010ci}
A.~Dymarsky, D.~Melnikov, and J.~Sonnenschein, {\it {Attractive Holographic
  Baryons}},  {\em JHEP} {\bf 06} (2011) 145,
  [\href{http://xxx.lanl.gov/abs/1012.1616}{{\tt arXiv:1012.1616}}].

\bibitem{Bhanot:1980fx}
G.~Bhanot and C.~Rebbi, {\it {SU(2) String Tension, Glueball Mass and
  Interquark Potential by Monte Carlo Computations}},  {\em Nucl.Phys.} {\bf
  B180} (1981) 469.

\bibitem{BoschiFilho:2002ta}
H.~Boschi-Filho and N.~R.~F. Braga, {\it {QCD / string holographic mapping and
  glueball mass spectrum}},  {\em Eur. Phys. J.} {\bf C32} (2004) 529--533,
  [\href{http://xxx.lanl.gov/abs/hep-th/0209080}{{\tt hep-th/0209080}}].

\bibitem{Lucini:2004my}
B.~Lucini, M.~Teper, and U.~Wenger, {\it {Glueballs and k-strings in SU(N)
  gauge theories: Calculations with improved operators}},  {\em JHEP} {\bf
  0406} (2004) 012, [\href{http://xxx.lanl.gov/abs/hep-lat/0404008}{{\tt
  hep-lat/0404008}}].

\bibitem{Meyer:2004jc}
H.~B. Meyer and M.~J. Teper, {\it {Glueball Regge trajectories and the pomeron:
  A Lattice study}},  {\em Phys.Lett.} {\bf B605} (2005) 344--354,
  [\href{http://xxx.lanl.gov/abs/hep-ph/0409183}{{\tt hep-ph/0409183}}].

\bibitem{Athenodorou:2010cs}
A.~Athenodorou, B.~Bringoltz, and M.~Teper, {\it {Closed flux tubes and their
  string description in D=3+1 SU(N) gauge theories}},  {\em JHEP} {\bf 02}
  (2011) 030, [\href{http://xxx.lanl.gov/abs/1007.4720}{{\tt
  arXiv:1007.4720}}].

\bibitem{Rosner:1968si}
J.~L. Rosner, {\it {Possibility of baryon - anti-baryon enhancements with
  unusual quantum numbers}},  {\em Phys. Rev. Lett.} {\bf 21} (1968) 950--952.

\bibitem{Krokovny}
P.~Krokovny, {\it {Proceedings of the International Workshop on QCD Exotics
  ,7–12 June 2015, Shandong University, Jinan, China}}, .

\bibitem{Esposito:2014rxa}
A.~Esposito, A.~L. Guerrieri, F.~Piccinini, A.~Pilloni, and A.~D. Polosa, {\it
  {Four-Quark Hadrons: an Updated Review}},  {\em Int. J. Mod. Phys.} {\bf A30}
  (2015) 1530002, [\href{http://xxx.lanl.gov/abs/1411.5997}{{\tt
  arXiv:1411.5997}}].

\bibitem{Maiani:2014aja}
L.~Maiani, F.~Piccinini, A.~D. Polosa, and V.~Riquer, {\it {The Z(4430) and a
  New Paradigm for Spin Interactions in Tetraquarks}},  {\em Phys. Rev.} {\bf
  D89} (2014) 114010, [\href{http://xxx.lanl.gov/abs/1405.1551}{{\tt
  arXiv:1405.1551}}].

\bibitem{Aharony:2002up}
O.~Aharony, {\it {The NonAdS / nonCFT correspondence, or three different paths
  to QCD}},  in {\em {Progress in string, field and particle theory}},
  pp.~3--24, 2002.
\newblock \href{http://xxx.lanl.gov/abs/hep-th/0212193}{{\tt hep-th/0212193}}.

\bibitem{PandoZayas:2003yb}
L.~A. Pando~Zayas, J.~Sonnenschein, and D.~Vaman, {\it {Regge trajectories
  revisited in the gauge / string correspondence}},  {\em Nucl.Phys.} {\bf
  B682} (2004) 3--44, [\href{http://xxx.lanl.gov/abs/hep-th/0311190}{{\tt
  hep-th/0311190}}].

\bibitem{Kruczenski:2004me}
M.~Kruczenski, L.~A. Pando~Zayas, J.~Sonnenschein, and D.~Vaman, {\it {Regge
  trajectories for mesons in the holographic dual of large-N(c) QCD}},  {\em
  JHEP} {\bf 0506} (2005) 046,
  [\href{http://xxx.lanl.gov/abs/hep-th/0410035}{{\tt hep-th/0410035}}].

\bibitem{Baker:2002km}
M.~Baker and R.~Steinke, {\it {Semiclassical quantization of effective string
  theory and Regge trajectories}},  {\em Phys. Rev.} {\bf D65} (2002) 094042,
  [\href{http://xxx.lanl.gov/abs/hep-th/0201169}{{\tt hep-th/0201169}}].

\bibitem{Schreiber:2004ie}
E.~Schreiber, {\it {Excited mesons and quantization of string endpoints}},
  \href{http://xxx.lanl.gov/abs/hep-th/0403226}{{\tt hep-th/0403226}}.

\bibitem{Bigazzi:2006jt}
F.~Bigazzi and A.~L. Cotrone, {\it {New predictions on meson decays from string
  splitting}},  {\em JHEP} {\bf 11} (2006) 066,
  [\href{http://xxx.lanl.gov/abs/hep-th/0606059}{{\tt hep-th/0606059}}].

\bibitem{Iengo:2006gm}
R.~Iengo and J.~G. Russo, {\it {Handbook on string decay}},  {\em JHEP} {\bf
  02} (2006) 041, [\href{http://xxx.lanl.gov/abs/hep-th/0601072}{{\tt
  hep-th/0601072}}].

\bibitem{Bigazzi:2007qa}
F.~Bigazzi, A.~L. Cotrone, L.~Martucci, and W.~Troost, {\it {Splitting of
  macroscopic fundamental strings in flat space and holographic hadron
  decays}},  {\em Mod. Phys. Lett.} {\bf A22} (2007) 1057--1074,
  [\href{http://xxx.lanl.gov/abs/hep-th/0703284}{{\tt hep-th/0703284}}].

\bibitem{Armoni:2009zq}
A.~Armoni and A.~Patella, {\it {Degeneracy Between the Regge Slope of Mesons
  and Baryons from Supersymmetry}},  {\em JHEP} {\bf 07} (2009) 073,
  [\href{http://xxx.lanl.gov/abs/0901.4508}{{\tt arXiv:0901.4508}}].

\bibitem{Hellerman:2013kba}
S.~Hellerman and I.~Swanson, {\it {String Theory of the Regge Intercept}},
  {\em Phys.Rev.Lett.} {\bf 114} (2015), no.~11 111601,
  [\href{http://xxx.lanl.gov/abs/1312.0999}{{\tt arXiv:1312.0999}}].

\bibitem{Zahn:2013yma}
J.~Zahn, {\it {The excitation spectrum of rotating strings with masses at the
  ends}},  {\em JHEP} {\bf 12} (2013) 047,
  [\href{http://xxx.lanl.gov/abs/1310.0253}{{\tt arXiv:1310.0253}}].

\bibitem{Hellerman:2014cba}
S.~Hellerman, S.~Maeda, J.~Maltz, and I.~Swanson, {\it {Effective String Theory
  Simplified}},  {\em JHEP} {\bf 1409} (2014) 183,
  [\href{http://xxx.lanl.gov/abs/1405.6197}{{\tt arXiv:1405.6197}}].

\bibitem{Dubovsky:2015zey}
S.~Dubovsky and V.~Gorbenko, {\it {Towards a Theory of the QCD String}},  {\em
  JHEP} {\bf 02} (2016) 022, [\href{http://xxx.lanl.gov/abs/1511.01908}{{\tt
  arXiv:1511.01908}}].

\bibitem{Dubovsky:2016cog}
S.~Dubovsky and G.~Hernandez-Chifflet, {\it {Yang--Mills Glueballs as Closed
  Bosonic Strings}},  {\em JHEP} {\bf 02} (2017) 022,
  [\href{http://xxx.lanl.gov/abs/1611.09796}{{\tt arXiv:1611.09796}}].

\bibitem{Rossi:2016szw}
G.~Rossi and G.~Veneziano, {\it {The string-junction picture of multiquark
  states: an update}},  {\em JHEP} {\bf 06} (2016) 041,
  [\href{http://xxx.lanl.gov/abs/1603.05830}{{\tt arXiv:1603.05830}}].

\bibitem{Aharony:2013ipa}
O.~Aharony and Z.~Komargodski, {\it {The Effective Theory of Long Strings}},
  {\em JHEP} {\bf 05} (2013) 118,
  [\href{http://xxx.lanl.gov/abs/1302.6257}{{\tt arXiv:1302.6257}}].

\bibitem{Sonnenschein:2018aqf}
J.~Sonnenschein and D.~Weissman, {\it {Quantizing the rotating string with
  massive endpoints}},  {\em JHEP} {\bf 06} (2018) 148,
  [\href{http://xxx.lanl.gov/abs/1801.00798}{{\tt arXiv:1801.00798}}].

\bibitem{Friedmann:2009mx}
T.~Friedmann, {\it {No Radial Excitations in Low Energy QCD. I. Diquarks and
  Classification of Mesons}},  {\em Eur. Phys. J.} {\bf C73} (2013), no.~2
  2298, [\href{http://xxx.lanl.gov/abs/0910.2229}{{\tt arXiv:0910.2229}}].

\bibitem{Aaij:2017nav}
{\bf LHCb} Collaboration, R.~Aaij et~al., {\it {Observation of five new narrow
  $\Omega_c^0$ states decaying to $\Xi_c^+ K^-$}},  {\em Phys. Rev. Lett.} {\bf
  118} (2017), no.~18 182001, [\href{http://xxx.lanl.gov/abs/1703.04639}{{\tt
  arXiv:1703.04639}}].

\bibitem{Karliner:2017kfm}
M.~Karliner and J.~L. Rosner, {\it {Very narrow excited $\Omega_c$ baryons}},
  \href{http://xxx.lanl.gov/abs/1703.07774}{{\tt arXiv:1703.07774}}.

\bibitem{Aliev:2018lcs}
T.~M. Aliev, K.~Azizi, Y.~Sarac, and H.~Sundu, {\it {Structure of the
  $\Xi_b(6227)^-$ resonance}},  {\em Phys. Rev.} {\bf D98} (2018), no.~9
  094014, [\href{http://xxx.lanl.gov/abs/1808.08032}{{\tt arXiv:1808.08032}}].

\bibitem{Chen:2018orb}
B.~Chen, K.-W. Wei, X.~Liu, and A.~Zhang, {\it {Role of newly discovered
  $\Xi_b(6227)^-$ for constructing excited bottom baryon family}},  {\em Phys.
  Rev.} {\bf D98} (2018), no.~3 031502,
  [\href{http://xxx.lanl.gov/abs/1805.10826}{{\tt arXiv:1805.10826}}].

\bibitem{Wang:2018fjm}
K.-L. Wang, Q.-F. Lü, and X.-H. Zhong, {\it {Interpretation of the newly
  observed $\Sigma_b(6097)^{\pm}$ and $\Xi_b(6227)^-$ states as the $P$-wave
  bottom baryons}},  \href{http://xxx.lanl.gov/abs/1810.02205}{{\tt
  arXiv:1810.02205}}.

\bibitem{Karliner:2018bms}
M.~Karliner and J.~L. Rosner, {\it {Scaling of P-wave excitation energies in
  heavy-quark systems}},  {\em Phys. Rev.} {\bf D98} (2018), no.~7 074026,
  [\href{http://xxx.lanl.gov/abs/1808.07869}{{\tt arXiv:1808.07869}}].

\bibitem{Aaij:2018tnn}
{\bf LHCb} Collaboration, R.~Aaij et~al., {\it {Observation of two resonances
  in the $\Lambda_b^0 \pi^\pm$ systems and precise measurement of
  $\Sigma_b^\pm$ and $\Sigma_b^{*\pm}$ properties}},
  \href{http://xxx.lanl.gov/abs/1809.07752}{{\tt arXiv:1809.07752}}.

\bibitem{Aaij:2018yqz}
{\bf LHCb} Collaboration, R.~Aaij et~al., {\it {Observation of a new $\Xi_b^-$
  resonance}},  {\em Phys. Rev. Lett.} {\bf 121} (2018), no.~7 072002,
  [\href{http://xxx.lanl.gov/abs/1805.09418}{{\tt arXiv:1805.09418}}].

\bibitem{Dai:1989cp}
J.~Dai and J.~Polchinski, {\it {The Decay of Macroscopic Fundamental Strings}},
   {\em Phys. Lett.} {\bf B220} (1989) 387.

\bibitem{Bai:1996wm}
{\bf BES Collaboration} Collaboration, J.~Bai et~al., {\it {Studies of xi
  (2230) in J / psi radiative decays}},  {\em Phys.Rev.Lett.} {\bf 76} (1996)
  3502--3505.

\bibitem{Vladimirsky:2001ek}
V.~Vladimirsky, V.~Grigorev, O.~Erofeeva, Y.~Katinov, V.~Lisin, et~al., {\it
  {Resonance maximum in the system of two K(S) mesons at 1450-MeV}},  {\em
  Phys.Atom.Nucl.} {\bf 64} (2001) 1895--1897.

\bibitem{Vijande:2004he}
J.~Vijande, F.~Fernandez, and A.~Valcarce, {\it {Constituent quark model study
  of the meson spectra}},  {\em J.Phys.} {\bf G31} (2005) 481,
  [\href{http://xxx.lanl.gov/abs/hep-ph/0411299}{{\tt hep-ph/0411299}}].

\bibitem{Peeters:2005fq}
K.~Peeters, J.~Sonnenschein, and M.~Zamaklar, {\it {Holographic decays of
  large-spin mesons}},  {\em JHEP} {\bf 0602} (2006) 009,
  [\href{http://xxx.lanl.gov/abs/hep-th/0511044}{{\tt hep-th/0511044}}].

\bibitem{Casher:1978wy}
A.~Casher, H.~Neuberger, and S.~Nussinov, {\it {Chromoelectric Flux Tube Model
  of Particle Production}},  {\em Phys. Rev.} {\bf D20} (1979) 179--188.

\bibitem{Brunner:2015oqa}
F.~Br{\"u}nner, D.~Parganlija, and A.~Rebhan, {\it {Glueball Decay Rates in the
  Witten-Sakai-Sugimoto Model}},  {\em Phys. Rev.} {\bf D91} (2015), no.~10
  106002, [\href{http://xxx.lanl.gov/abs/1501.07906}{{\tt arXiv:1501.07906}}].

\bibitem{Brunner:2015yha}
F.~Brünner and A.~Rebhan, {\it {Nonchiral enhancement of scalar glueball decay
  in the Witten-Sakai-Sugimoto model}},  {\em Phys. Rev. Lett.} {\bf 115}
  (2015), no.~13 131601, [\href{http://xxx.lanl.gov/abs/1504.05815}{{\tt
  arXiv:1504.05815}}].

\bibitem{Brunner:2016ygk}
F.~Brünner and A.~Rebhan, {\it {Holographic QCD predictions for production and
  decay of pseudoscalar glueballs}},  {\em Phys. Lett.} {\bf B770} (2017)
  124--130, [\href{http://xxx.lanl.gov/abs/1610.10034}{{\tt
  arXiv:1610.10034}}].

\bibitem{Brunner:2018wbv}
F.~Brünner, J.~Leutgeb, and A.~Rebhan, {\it {A broad pseudovector glueball
  from holographic QCD}},  {\em Phys. Lett.} {\bf B788} (2019) 431--435,
  [\href{http://xxx.lanl.gov/abs/1807.10164}{{\tt arXiv:1807.10164}}].

\bibitem{Yadav:2018zxw}
V.~Yadav and A.~Misra, {\it {M-Theory Exotic Scalar Glueball Decays to Mesons
  at Finite Coupling}},  {\em JHEP} {\bf 09} (2018) 133,
  [\href{http://xxx.lanl.gov/abs/1808.01182}{{\tt arXiv:1808.01182}}].

\bibitem{Chen:2016qju}
H.-X. Chen, W.~Chen, X.~Liu, and S.-L. Zhu, {\it {The hidden-charm pentaquark
  and tetraquark states}},  {\em Phys. Rept.} {\bf 639} (2016) 1--121,
  [\href{http://xxx.lanl.gov/abs/1601.02092}{{\tt arXiv:1601.02092}}].

\bibitem{Ali:2017jda}
A.~Ali, J.~S. Lange, and S.~Stone, {\it {Exotics: Heavy Pentaquarks and
  Tetraquarks}},  {\em Prog. Part. Nucl. Phys.} {\bf 97} (2017) 123--198,
  [\href{http://xxx.lanl.gov/abs/1706.00610}{{\tt arXiv:1706.00610}}].

\bibitem{Pakhlova:2008vn}
{\bf Belle} Collaboration, G.~Pakhlova et~al., {\it {Observation of a
  near-threshold enhancement in the e+e- ---> Lambda+(c) Lambda-(c) cross
  section using initial-state radiation}},  {\em Phys. Rev. Lett.} {\bf 101}
  (2008) 172001, [\href{http://xxx.lanl.gov/abs/0807.4458}{{\tt
  arXiv:0807.4458}}].

\bibitem{Wang:2007ea}
{\bf Belle} Collaboration, X.~L. Wang et~al., {\it {Observation of Two Resonant
  Structures in e+e- to pi+ pi- psi(2S) via Initial State Radiation at Belle}},
   {\em Phys. Rev. Lett.} {\bf 99} (2007) 142002,
  [\href{http://xxx.lanl.gov/abs/0707.3699}{{\tt arXiv:0707.3699}}].

\bibitem{Aubert:2007zz}
{\bf BaBar} Collaboration, B.~Aubert et~al., {\it {Evidence of a broad
  structure at an invariant mass of 4.32- $GeV/c^{2}$ in the reaction $e^{+}
  e^{-} \to \pi^{+} \pi^{-} \psi_{2S}$ measured at BaBar}},  {\em Phys. Rev.
  Lett.} {\bf 98} (2007) 212001,
  [\href{http://xxx.lanl.gov/abs/hep-ex/0610057}{{\tt hep-ex/0610057}}].

\bibitem{Cotugno:2009ys}
G.~Cotugno, R.~Faccini, A.~D. Polosa, and C.~Sabelli, {\it {Charmed
  Baryonium}},  {\em Phys. Rev. Lett.} {\bf 104} (2010) 132005,
  [\href{http://xxx.lanl.gov/abs/0911.2178}{{\tt arXiv:0911.2178}}].

\bibitem{Guo:2010tk}
F.-K. Guo, J.~Haidenbauer, C.~Hanhart, and U.-G. Meissner, {\it {Reconciling
  the X(4630) with the Y(4660)}},  {\em Phys. Rev.} {\bf D82} (2010) 094008,
  [\href{http://xxx.lanl.gov/abs/1005.2055}{{\tt arXiv:1005.2055}}].

\bibitem{Olsen:2014qna}
S.~L. Olsen, {\it {A New Hadron Spectroscopy}},  {\em Front. Phys.(Beijing)}
  {\bf 10} (2015), no.~2 121--154,
  [\href{http://xxx.lanl.gov/abs/1411.7738}{{\tt arXiv:1411.7738}}].

\bibitem{Abe:2007tk}
{\bf Belle} Collaboration, K.~F. Chen et~al., {\it {Observation of anomalous
  Upsilon(1S) pi+ pi- and Upsilon(2S) pi+ pi- production near the Upsilon(5S)
  resonance}},  {\em Phys. Rev. Lett.} {\bf 100} (2008) 112001,
  [\href{http://xxx.lanl.gov/abs/0710.2577}{{\tt arXiv:0710.2577}}].

\bibitem{Ali:2009pi}
A.~Ali, C.~Hambrock, I.~Ahmed, and M.~J. Aslam, {\it {A case for hidden
  $b\bar{b}$ tetraquarks based on $e^+e^- \to b\bar{b}$ cross section between
  $\sqrt{s}=10.54$ and 11.20 GeV}},  {\em Phys. Lett.} {\bf B684} (2010)
  28--39, [\href{http://xxx.lanl.gov/abs/0911.2787}{{\tt arXiv:0911.2787}}].

\bibitem{Karliner:2015ina}
M.~Karliner and J.~L. Rosner, {\it {New Exotic Meson and Baryon Resonances from
  Doubly-Heavy Hadronic Molecules}},  {\em Phys. Rev. Lett.} {\bf 115} (2015),
  no.~12 122001, [\href{http://xxx.lanl.gov/abs/1506.06386}{{\tt
  arXiv:1506.06386}}].

\bibitem{Anisovich:2000kxa}
A.~Anisovich, V.~Anisovich, and A.~Sarantsev, {\it {Systematics of q anti-q
  states in the (n, M**2) and (J, M**2) planes}},  {\em Phys.Rev.} {\bf D62}
  (2000) 051502, [\href{http://xxx.lanl.gov/abs/hep-ph/0003113}{{\tt
  hep-ph/0003113}}].

\bibitem{Kataev:2004ve}
A.~L. Kataev, {\it {QCD sum rules and radial excitations of light pseudoscalar
  and scalar mesons}},  {\em Phys. Atom. Nucl.} {\bf 68} (2005) 567--572,
  [\href{http://xxx.lanl.gov/abs/hep-ph/0406305}{{\tt hep-ph/0406305}}]. [Yad.
  Fiz.68,597(2005)].

\bibitem{Gershtein:2006ng}
S.~S. Gershtein, A.~K. Likhoded, and A.~V. Luchinsky, {\it {Systematics of
  heavy quarkonia from Regge trajectories on (n,M**2) and (M**2,J) planes}},
  {\em Phys. Rev.} {\bf D74} (2006) 016002,
  [\href{http://xxx.lanl.gov/abs/hep-ph/0602048}{{\tt hep-ph/0602048}}].

\bibitem{Ebert:2009ub}
D.~Ebert, R.~N. Faustov, and V.~O. Galkin, {\it {Mass spectra and Regge
  trajectories of light mesons in the relativistic quark model}},  {\em Phys.
  Rev.} {\bf D79} (2009) 114029, [\href{http://xxx.lanl.gov/abs/0903.5183}{{\tt
  arXiv:0903.5183}}].

\bibitem{Ebert:2009ua}
D.~Ebert, R.~N. Faustov, and V.~O. Galkin, {\it {Heavy-light meson spectroscopy
  and Regge trajectories in the relativistic quark model}},  {\em Eur. Phys.
  J.} {\bf C66} (2010) 197--206, [\href{http://xxx.lanl.gov/abs/0910.5612}{{\tt
  arXiv:0910.5612}}].

\bibitem{Ebert:2011jc}
D.~Ebert, R.~N. Faustov, and V.~O. Galkin, {\it {Spectroscopy and Regge
  trajectories of heavy quarkonia and $B_c$ mesons}},  {\em Eur. Phys. J.} {\bf
  C71} (2011) 1825, [\href{http://xxx.lanl.gov/abs/1111.0454}{{\tt
  arXiv:1111.0454}}].

\bibitem{Kher:2017wsq}
V.~Kher, N.~Devlani, and A.~K. Rai, {\it {Excited state mass spectra, Decay
  properties and Regge trajectories of charm and charm-strange mesons}},  {\em
  Chin. Phys.} {\bf C41} (2017), no.~7 073101,
  [\href{http://xxx.lanl.gov/abs/1704.00439}{{\tt arXiv:1704.00439}}].

\bibitem{Afonin:2018bej}
S.~S. Afonin and T.~D. Solomko, {\it {Large-$N_c$ masses of light mesons from
  QCD sum rules for nonlinear radial Regge trajectories}},  {\em Int. J. Mod.
  Phys.} {\bf A33} (2018), no.~12 1850069,
  [\href{http://xxx.lanl.gov/abs/1805.02553}{{\tt arXiv:1805.02553}}].

\bibitem{Chen:2018bbr}
J.-K. Chen, {\it {Concavity of the meson Regge trajectories}},  {\em Phys.
  Lett.} {\bf B786} (2018) 477--484,
  [\href{http://xxx.lanl.gov/abs/1807.11003}{{\tt arXiv:1807.11003}}].

\bibitem{Jia:2018vwl}
D.~Jia and W.-C. Dong, {\it {Regge-like spectra of excited singly heavy
  mesons}},  \href{http://xxx.lanl.gov/abs/1811.04214}{{\tt arXiv:1811.04214}}.

\bibitem{PDG:2018}
{\bf Particle Data Group} Collaboration, M.~Tanabashi et~al., {\it {Review of
  Particle Physics}},  {\em Phys. Rev.} {\bf D98} (2018), no.~3 030001.

\end{thebibliography}\endgroup

\clearpage

\appendix

\section{The states used in the fits} \label{sec:states}
In this appendix we provide tables with all states used in our fits, and which form the basis for the predictions presented in this note. All states are as listed in the Review of Particle Physics published by the Particle Data Group (PDG) \cite{PDG:2018}.

In tables \ref{tab:all_mesons} and \ref{tab:all_mesons_n} we list all the meson states that we placed on HMRTs, for orbital and radial trajectories respectively. The selection and identification of states was discussed in detail in \cite{Sonnenschein:2014jwa}. Separately, in tables \ref{tab:allf0} and \ref{tab:allf2} are listed all the flavorless \(f_0\) (\(J^{PC}=0^{++}\)) and \(f_2\) (\(2^{++}\)) states, among which the glueball could be found.

Tables \ref{tab:all_baryons} and \ref{tab:all_baryons_n} list all the baryon states used in the fits. The selection of states was discussed in \cite{Sonnenschein:2014bia}.

\begin{table}[tpb] \centering
	\begin{tabular}[t]{|c|c|l|c|} \hline
		Traj. & \(J^{PC}\) & State & \\ \hline\hline
		
		\(\pi/b\) & \(1^{+-}\) & \(b_1(1235)\) & $\bullet$ \\

		          & \(2^{-+}\) & \(\pi_2(1670)\) & $\bullet$\\

		          & \(3^{+-}\) & \(b_3(2030)\) & f. \\

							& \(4^{-+}\) & \(\pi_4(2250)\) & f. \\ \hline

		\(\rho/a\)& \(1^{--}\) & \(\rho\) & $\bullet$\\

							& \(2^{++}\) & \(a_2(1320)\) & $\bullet$ \\

							& \(3^{--}\) & \(\rho_3(1690)\) & $\bullet$ \\

							& \(4^{++}\) & \(a_4(2040)\) & $\bullet$ \\

							& \(5^{--}\) & \(\rho_5(2350)\) & \\

							& \(6^{++}\) & \(a_6(2450)\) & \\ \hline

		\(\eta/h\)& \(0^{-+}\) & \(\eta\) & $\bullet$ \\

							& \(1^{+-}\) & \(h_1(1170)\) & $\bullet$ \\

							& \(2^{-+}\) & \(\eta_2(1645)\) & $\bullet$ \\

							& \(3^{+-}\) & \(h_3(2025)\) & f. \\

							& \(4^{-+}\) & \(\eta_4(2330)\) & f. \\ \hline

		\(\omega/f\)&\(1^{--}\) & \(\omega\) & $\bullet$ \\

							& \(2^{++}\) & \(f_2(1270)\) & $\bullet$ \\

							& \(3^{--}\) & \(\omega_3(1670)\) & $\bullet$ \\
							
							& \(4^{++}\) & \(f_4(2050)\) & $\bullet$ \\

							& \(5^{--}\) & \(\omega_5(2250)\) & f. \\

							& \(6^{++}\) & \(f_6(2510)\) & \\ \hline
							
		\(K\) & \(0^-\)    & \(K\) & $\bullet$ \\
           
					& \(1^+\)    & \(K_1(1270)\) & $\bullet$ \\

            & \(2^-\)   & \(K_2(1770)\)  & $\bullet$\\ \hline

		\(K^*\) & \(1^-\)    & \(K^*\) & $\bullet$ \\
           
					& \(2^+\)    & \(K^*_2(1430)\) & $\bullet$ \\

            & \(3^-\)   & \(K^*_3(1780)\)  & $\bullet$\\

           & \(4^+\)    & \(K^*_4(2045)\) & $\bullet$ \\ 

         &	 \(5^-\)    & \(K^*_5(2380)\) & \\ \hline

	   \(\phi/f'\) & \(1^{--}\) & \(\phi(1020)\) & $\bullet$ \\

	            	& \(2^{++}\) & \(f_2^\prime(1525)\) & $\bullet$ \\

           	& \(3^{--}\) & \(\phi_3(1850)\)  & $\bullet$ \\ \hline
	\end{tabular} \qquad
	\begin{tabular}[t]{|c|c|l|c|} \hline
		Traj. & \(J^{PC}\) & State & \\ \hline\hline
		
		\(D\)   & \(0^-\)    & \(D\) & $\bullet$ \\

	         	& \(1^+\)    & \(D_1(2420)\) & $\bullet$ \\ 

           	& \(2^-\)    & \(D(2740)^0\) & \\ \hline

		\(D^*\) & \(1^-\)    & \(D^*\) & $\bullet$ \\

	         	& \(2^+\)    & \(D^*_2(2460)\) &  $\bullet$ \\
						
						& \(3^-\)    & \(D^*_3(2750)\) & \\ \hline

		\(D_s\) & \(0^-\)    & \(D_s\) & $\bullet$ \\

	         	& \(1^+\)    & \(D_{s1}(2536)\) & $\bullet$ \\ \hline

		\(D^*_s\) & \(1^-\)    & \(D^*_s\) &  $\bullet$ \\

	            	& \(2^+\)    & \(D^*_{s2}(2573)\) &  $\bullet$ \\

	            	& \(3^-\)    & \(D^*_{s3}(2860)\) & \\ \hline
								
		\(\Psi\) & \(1^{--}\) & \(J/\Psi(1S)\) & $\bullet$ \\ 
		
							& \(2^{++}\) & \(\chi_{c2}(1P)\) & $\bullet$ \\

							& \(2^{--}{}^{[a]}\) &  \(\Psi_2(3823)\) & $\bullet$\\ \hline
							
		\(\eta_c\) & \(0^{-+}\) & \(\eta_c(1S)\) & $\bullet$ \\ 
		
							& \(1^{+-}\) &  \(h_c(1P)\) & $\bullet$ \\ \hline

		\(B\)   & \(0^-\)    & \(B\) &  $\bullet$ \\

	         	& \(1^+\)    & \(B_1(5721)\) & $\bullet$ \\ \hline

		\(B^*\) & \(1^-\)    & \(B^*\) & $\bullet$ \\

	         	& \(2^+\)    & \(B^*_2(5747)\) & $\bullet$ \\ \hline

		\(B_s\) & \(0^-\)    & \(B_s\) & $\bullet$ \\

	         	& \(1^+\)    & \(B_{s1}(5830)^0\) & $\bullet$ \\ \hline

		\(B^*_s\) & \(1^-\)    & \(B^*_s\) & $\bullet$ \\

	            	& \(2^+\)    & \(B^*_{s2}(5840)^0\) & $\bullet$ \\ \hline
								
		\(\Upsilon\) & \(1^{--}\) & \(\Upsilon(1S)\) & $\bullet$ \\ 
		
							& \(2^{++}\) & \(\chi_{b2}(1P)\) & $\bullet$ \\

							& \(2^{--}{}^{[a]}\) &  \(\Upsilon_2(1D)\) & $\bullet$\\ \hline
								
		\(\eta_b\) & \(0^{-+}\) & \(\eta_b(1S)\) & $\bullet$ \\ 
		
							& \(1^{+-}\) &  \(h_b(1P)\) & $\bullet$ \\ \hline

	\end{tabular}
	\caption{\label{tab:all_mesons} All the meson states on the orbital leading Regge trajectories used in our fits. With the sole exception of the \(b\bar b\), all belong on trajectories with the same slope, \(\alp = 0.88 \GEVm\). The fourth column indicates the status given to each state by the PDG. States marked with a bullet are well established, while those marked with an ``f.'' are listed as further states, which need confirmation. [a] The \(\Psi(3823)\) and \(\Upsilon(1D)\) with \(J^{PC} = 2^{--}\) are used as placeholders for the \(3^{--}\) state, from which they are expected to differ by a small amount due to spin-orbit interactions.}
\end{table}

\begin{table}[tpb] \centering
	\begin{tabular}[t]{|c|c|l|c|} \hline
		Traj. & \(I(J^{PC})\) & State & \\ \hline\hline
		
		\(\pi\) & \(1(0^{-+})\) & \(\pi(1300)\)& \(\bullet\) \\ 

		          & & \(\pi(1800)\) & \(\bullet\) \\

		          & & \(\pi(2070)\) & f. \\

							& & \(\pi(2360)\)	& f. \\ \hline

		\(\pi_2\) & \(1(2^{-+})\) & \(\pi_2(1670)\) & \(\bullet\) \\

							& & \(\pi_2(2005)\) & f. \\

							& &\(\pi_2(2285)\) & f.  \\ \hline

		\(a_1\) & \(1(1^{++})\) & \(a_1(1260)\) & \(\bullet\)  \\ 

							& & \(a_1(1640)\) & \\

							& & \(a_1(2095)\) & f. \\

							& & \(a_1(2270)\) &	f.  \\ \hline

		\(h_1\)		&  \(0(1^{+-})\) & \(h_1(1170)\) &	\(\bullet\) \\

							& & \(h_1(1595)\) &	\\
							
							& & \(h_1(1965)\) &	f.  \\
							
							& & \(h_1(2215)\) & f. \\ \hline
							
		\(\omega\) & \(0(1^{--})\) & \(\omega(782)\) & \(\bullet\) \\

							& & \(\omega(1420)\) & \(\bullet\) \\

							& & \(\omega(1650)\) &	\(\bullet\) \\

							& & \(\omega(1960)\) & f. \\	
						
							& & \(\omega(2290)\) & f.  \\ \hline
							
				\(\omega_3\) &\(0(3^{--})\) &	\(\omega_3(1670)\) & \(\bullet\) \\

							& & \(\omega_3(1950)\) & f. \\

		          & & \(\omega_3(2255)\) & f. \\ \hline

	\end{tabular} \qquad
	\begin{tabular}[t]{|c|c|l|c|} \hline
		Traj. & \(I(J^{PC})\) & State & \\ \hline\hline

		\(\phi\)	&	\(0(1^{--})\)	& \(\phi(1020)\) & \(\bullet\)  \\ 

							&	&	\(\phi(1680)\) & \(\bullet\) \\

							& & \(\phi(2170)\) & \(\bullet\)\\ \hline

		\(\eta_{c}\) & \(0(0^{-+})\) & \(\eta_{c}(1S)\) & \(\bullet\) \\
		
							& & \(\eta_{c}(2S)\) & \(\bullet\) \\ \hline
		
		\(\Psi\) & \(0(1^{--})\) & \(J/\Psi(1S)\) & \(\bullet\)  \\ 

							&	& \(\Psi(2S)\) & \(\bullet\) \\

							& & \(\Psi(4040)\) & \(\bullet\) \\

							&	& \(\Psi(4415)\) & \(\bullet\) \\ \hline
		
		\(\chi_{c2}\) & \(0(2^{++})\) & \(\chi_{c2}(1P)\) & \(\bullet\) \\ 
		
							& & \(\chi_{c2}(2P)\) & \(\bullet\) \\ \hline
		
		\(\Upsilon\) & \(0(1^{--})\) & \(\Upsilon(1S)\) & \(\bullet\) \\

							& & \(\Upsilon(2S)\) & \(\bullet\) \\

							& & \(\Upsilon(3S)\) & \(\bullet\) \\

							& & \(\Upsilon(4S)\) & \(\bullet\) \\

							& & \(\Upsilon(10860)\) & \(\bullet\) \\ 

							& & \(\Upsilon(11020)\)& \(\bullet\)  \\ \hline

		\(\chi_{b1}\) &	\(0(1^{++})\) & \(\chi_{b1}(1P)\) & \(\bullet\)\\

							& & \(\chi_{b1}(2P)\) & \(\bullet\) \\

							& & \(\chi_{b1}(3P)\) & \(\bullet\) \\ \hline
	\end{tabular}
	\caption{\label{tab:all_mesons_n} All the meson states on radial trajectories used in our fits. The fourth column indicates the status given to each state by the PDG. States marked with a bullet are well established, while those marked with an ``f.'' are listed as further states, which need confirmation.}
\end{table}

\begin{table}[t!] \centering
		\begin{tabular}{|l|l|l|l|l|} \hline
		\textbf{State} & \textbf{Mass} [MeV] & \textbf{Width} [MeV] & \textbf{Width/mass} & \textbf{Decay modes} \\ \hline\hline
		\(f_0(500)/\sigma\) & 400--550 & 400--700 & 1.16\plm0.36 & \(\pi\pi\) dominant\\ \hline
		\(f_0(980)\) & \(990\pm20\) & 40--100 & 0.07\plm0.03 & \(\pi\pi\) dominant, \(K\overline{K}\) seen \\ \hline
		\(f_0(1370)\) & 1200--1500 & 200--500 & 0.26\plm0.11 & \(\pi\pi\), \(4\pi\), \(\eta\eta\), \(K\overline{K}\) \\ \hline
		\(f_0(1500)\) & \(1505\pm6\) & \(109\pm7\) & 0.072\plm0.005 & \(\pi\pi\) \([35\%]\), \(4\pi\) \([50\%]\), \\
									&  & & & \(\eta\eta\)/\(\eta\eta\prime\) \([7\%]\), \(K\overline{K}\) \([9\%]\) \\ \hline		
		\(f_0(1710)\) & \(1720\pm6\) & \(135\pm8\) & 0.078\plm0.005 & \(K\overline{K}\), \(\eta\eta\), \(\pi\pi\) \\ \hline
		\(f_0(2020)\) & \(1992\pm16\) & \(442\pm60\) & 0.22\plm0.03 & \(\rho\pi\pi\), \(\pi\pi\), \(\rho\rho\), \(\omega\omega\), \(\eta\eta\) \\ \hline
		\(f_0(2100)\) & \(2103\pm8\) & \(209\pm19\) & 0.10\plm0.01 & \\ \hline
		\(f_0(2200)\) & \(2189\pm13\) & \(238\pm50\) & 0.11\plm0.02 & \\ \hline
		\(f_0(2330)\) & \(2325\pm35\) & \(180\pm70\) & 0.08\plm0.03 &  \\ \hline
		*\(f_0\)(1200--1600) & 1200--1600 & 200--1000 & 0.43\plm0.29 & \\ \hline
		*\(f_0\)(1800) & \(1795\pm25\) & \(95\pm80\) & 0.05\plm0.04 & \\ \hline
		*\(f_0\)(2060) & \(\sim2050\) & \(\sim120\) & \(\sim0.04\)--\(0.10\) & \\ \hline
		\end{tabular} \caption{\label{tab:allf0} All the \(f_0\) states as listed by the PDG. The last few states, marked here by asterisk, are classified as ``further states''.}
	\end{table}

		\begin{table}[t!] \centering
		\begin{tabular}{|l|l|l|l|l|} \hline
		\textbf{State} & \textbf{Mass} [MeV] & \textbf{Width} [MeV] & Width/mass & \textbf{Decay modes} \\ \hline\hline
		\(f_2(1270)\) & 1275.1\plm1.2 & 185.1\plm2.9 & 0.15\plm0.00 & \(\pi\pi\) \([85\%]\), \(4\pi\) \([10\%]\), \(KK\), \(\eta\eta\), \(\gamma\gamma\), ... \\ \hline
		\(f_2(1430)\) & 1453\plm4 & 13\plm5 & 0.009\plm0.006 & \(KK\), \(\pi\pi\) \\ \hline
		\(f^\prime_2(1525)\) & 1525\plm5 & 73\plm6 & 0.048\plm0.004 & \(KK\) \([89\%]\), \(\eta\eta\) \([10\%]\), \(\gamma\gamma\) [seen], ... \\ \hline
		\(f_2(1565)\) & 1562\plm13 & 134\plm8 & 0.09\plm0.01 & \(\pi\pi\), \(\rho\rho\), \(4\pi\), \(\eta\eta\), ... \\ \hline
		\(f_2(1640)\) & 1639\plm6 & 99\plm60 & 0.06\plm0.04 & \(\omega\omega\), \(4\pi\), \(KK\) \\ \hline
		\(f_2(1810)\) & 1815\plm12 & 197\plm22 & 0.11\plm0.01 & \(\pi\pi\), \(\eta\eta\), \(4\pi\), \(KK\), \(\gamma\gamma\) [seen] \\ \hline
		\(f_2(1910)\) & 1903\plm9 & 196\plm31 & 0.10\plm0.02 & \(\pi\pi\), \(KK\), \(\eta\eta\), \(\omega\omega\), ... \\ \hline
		\(f_2(1950)\) & 1944\plm12 & 472\plm18 & 0.24\plm0.01 & \(K^*K^*\), \(\pi\pi\), \(4\pi\), \(\eta\eta\), \(KK\), \(\gamma\gamma\), \(pp\) \\ \hline
		\(f_2(2010)\) & 2011\plm76 & 202\plm67 & 0.10\plm0.03 & \(KK\), \(\phi\phi\) \\ \hline
		\(f_2(2150)\) & 2157\plm12 & 152\plm30 & 0.07\plm0.01 & \(\pi\pi\), \(\eta\eta\), \(KK\), \(f_2(1270)\eta\), \(a_2\pi\), \(pp\) \\ \hline
		\(f_J(2220)\) & 2231.1\plm3.5 & 23\plm8 & 0.010\plm0.004 & \(\pi\pi\), \(KK\), \(pp\), \(\eta\eta^\prime\) \\ \hline
		\(f_2(2300)\) & 2297\plm28 & 149\plm41 & 0.07\plm0.02 & \(\phi\phi\), \(KK\), \(\gamma\gamma\) [seen] \\ \hline
		\(f_2(2340)\) & 2339\plm55 & 319\plm81 & 0.14\plm0.04 & \(\phi\phi\), \(\eta\eta\) \\ \hline
		*\(f_2(1750)\) & 1755\plm10 & 67\plm12 & 0.04\plm0.01 & \(KK\), \(\gamma\gamma\), \(\pi\pi\), \(\eta\eta\) \\ \hline
		*\(f_2(2000)\) & 2001\plm10 & 312\plm32 & 0.16\plm0.02 & \\ \hline
		*\(f_2(2140)\) & 2141\plm12 & 49\plm28 & 0.02\plm0.01 & \\ \hline
		*\(f_2(2240)\) & 2240\plm15 & 241\plm30 & 0.11\plm0.01 & \\ \hline
		*\(f_2(2295)\) & 2293\plm13 & 216\plm37 & 0.10\plm0.02 & \\ \hline
		\end{tabular} \caption{\label{tab:allf2} All the \(f_2\) states as listed by the PDG. The last few states, marked here by asterisk, are classified as ``further states''.}
	\end{table}

\begin{table}[t!] \centering
	\begin{tabular}[t]{|c|c|l|l|} \hline
	Traj. & \(J^{PC}\) & State & Status\\ \hline\hline
		\(N\) & \jph{1}{+}  & \(n/p\)     & **** \\
					& \jph{3}{-}  & \(N(1520)\) & **** 		\\
					& \jph{5}{+}  & \(N(1680)\) & ****		\\
					& \jph{7}{-}  & \(N(2190)\) & ****		\\
					& \jph{9}{+}  & \(N(2220)\) & ****		\\ 
					& \jph{11}{-} & \(N(2600)\) & ***  \\
					& \jph{13}{+} & \(N(2700)\) & ** 	\\ \hline
					
		\(\Delta\) & \jph{3}{+}  & \(\Delta(1232)\) & **** \\ 
					& \jph{5}{-}  & \(\Delta(1930)\) & ***  \\
					& \jph{7}{+}  & \(\Delta(1950)\) & ****	\\
					& \jph{9}{-}  & \(\Delta(2400)\) & **		\\
					& \jph{11}{+} & \(\Delta(2420)\) & ****	\\
					& \jph{13}{-} & \(\Delta(2750)\) & **\\
					& \jph{15}{+} & \(\Delta(2950)\) & **	\\ \hline
					
		\(\Lambda\) & \jph{1}{+}  & \(\Lambda\) & **** 	\\
					& \jph{3}{-}  & \(\Lambda(1520)\) & **** 	\\
					& \jph{5}{+}  & \(\Lambda(1820)\) & ****	\\
					& \jph{7}{-}  & \(\Lambda(2100)\) & ****	\\
					& \jph{9}{+}  & \(\Lambda(2350)\) & ***		\\ \hline
					
		\(\Sigma\) & 	\jph{1}{+} & \(\Sigma\) & **** \\
					& \jph{3}{-} & \(\Sigma(1670)\) & **** \\ 
					& \jph{5}{+} & \(\Sigma(1915)\) & **** \\ \hline
					
		\(\Sigma\) & \jph{3}{+} & \(\Sigma(1385)\) & **** \\
					& \jph{5}{-} & \(\Sigma(1775)\) & **** \\	
					& \jph{7}{+} & \(\Sigma(2030)\) & **** \\ \hline
					
	\end{tabular} \qquad
	\begin{tabular}[t]{|c|c|l|l|} \hline
	Traj. & \(J^{PC}\) & State & Status\\ \hline\hline
		
		\(\Xi\) & \jph{1}{+} & \(\Xi^0/\Xi^-\)  & **** \\
					& \jph{3}{-} & \(\Xi(1820)\) & ***\\
					& \jph{5}{+} & \(\Xi(2030)\) & *** \\ \hline
					
		\(\Omega\) & \jph3+ & \(\Omega^-\) & **** \\ \hline
					
		\(\Lambda_c\) & \jph{1}{+} & \(\Lambda_c^+\) & **** \\
					& \jph{3}{-} & \(\Lambda_c(2625)^+\) & *** \\
					& \jph{5}{+} & \(\Lambda_c(2880)^+\) & *** \\ \hline
					
		\(\Sigma_c\) & 	\jph{1}{+} & \(\Sigma_c(2455)\) & **** \\ \hline
					
		\(\Sigma_c\) & \jph{3}{+} & \(\Sigma_c(2520)\) & *** \\ \hline
					
		\(\Xi_c\) & \jph{1}{+} & \(\Xi_c^+/\Xi_c^0\) & *** \\
					& \jph{3}{-} & \(\Xi_c(2815)^+\) & *** \\ \hline
					
		\(\Omega_c\) & \jph{1}{+} & \(\Omega_c^0\) & *** \\
					& \jph{3}{-} & \(\Omega_c(3065)^0\) & *** \\ \hline
					
		\(\Omega_c\) & \jph{3}{+} & \(\Omega_c(2770)^0\) & *** \\
					& \jph{5}{-} & \(\Omega_c(3120)^0\) & *** \\ \hline
					
		\(\Xi_{cc}\) & \jph{1}{+} & \(\Xi_{cc}^{++}\) & *** \\ \hline
					
		\(\Lambda_b\) & \jph{1}{+} & \(\Lambda_b^0\) & *** \\
					& \jph{3}{-} & \(\Lambda_b(5920)^0\) & *** \\ \hline
		
		\(\Sigma_b\) & \jph{1}{+} & \(\Sigma_b\) & *** \\ \hline
		
		\(\Sigma_b^*\) & \jph{3}{+} & \(\Sigma_b^*\) & *** \\ \hline
		
		\(\Xi_b\) & \jph{1}{+} & \(\Xi_b^0,\Xi_b^-\) & *** \\ \hline
		
		\(\Omega_b\) & \jph{1}{+} & \(\Omega_b^-\) & *** \\ \hline
		
	\end{tabular}
	\caption{\label{tab:all_baryons} All the baryon states on the orbital leading Regge trajectories used in our fits, or chosen as heads of trajectories where no excited states are identified. All fitted trajectories have the same slope, \(\alp = 0.95 \GEVm\). The fourth column indicates the status given to each state by the PDG, ranging from * for unconfirmed states to **** for those whose properties are certain.}
\end{table}

\begin{table}[t!] \centering
	\begin{tabular}{|c|c|c|l|l|c|c|c|l|l|} \hline
		Traj. & \(I(J^P)\) & \(n\) & State & Status & Traj. & \(I(J^P)\) & \(n\) & State & Status \\ \hline
		\(N\) & \(\frac{1}{2}(\jph{1}{+})\) & \(0\) & \(n/p\) & **** & \(N\) & \(\frac{1}{2}(\jph{1}{-})\) & 0 & \(N(1535)\) & **** \\
		 & & \(1\) & \(N(1440)\) & ****	& & & 1 & \(N(1895)\) & ** \\ \cline{6-10}
		 & & \(2\) & \(N(1880)\) & **  & \(N\) & \(\frac{1}{2}(\jph{3}{+})\) & 0 & \(N(1720)\) & **** \\
		 & & \(3\) & \(N(2100)\) & *   & & & 1 & \(N(2040)\) & * \\ \cline{1-5} \cline{6-10}
		\(N\) & \(\frac{1}{2}(\jph{3}{-})\) & \(0\) & \(N(1520)\) & **** & \(N\) & \(\frac{1}{2}(\jph{5}{-})\) & 0 & \(N(1675)\) & **** \\
		 & & \(1\) & \(N(1875)\) & *** & & & 1 & \(N(2060)\) & ** \\ \cline{6-10}
		 & & \(2\) & \(N(2150)\) & **  & \(\Delta\) & \(\frac{3}{2}(\jph{1}{+})\) & 0 & \(\Delta(1232)\) & ****\\ \cline{1-5}
		\(N\) & \(\frac{1}{2}(\jph{5}{+})\) & \(0\) & \(N(1680)\) & **** & & & 1 & \(\Delta(1600)\) & *** \\
		 & & \(1\) & \(N(2000)\) & ** & & & 2 & \(\Delta(1920)\) & *** \\ \hline
	\end{tabular}
	\caption{\label{tab:all_baryons_n} The baryon states used in the \((n,M^2)\) trajectory fits and their assignments. The fourth column indicates the status given to each state by the PDG, ranging from * for unconfirmed states to **** for those whose properties are certain.}
\end{table}

\section{The intercept} \label{app:a}
One of the most important challenges for the HISH model is to have a full theoretical computation of the intercept of the modified Regge trajectory. In \cite{Sonnenschein:2018aqf} we have determined the leading order quantum corrections to the trajectory of a string with massive endpoints for long rotating strings. However, this is a correction to the result \(a=1\) of the usual boson string theory. To describe real world hadrons we need the intercept to be negative (recall that we define the intercept w.r.t. the trajectory of the orbital angular momentum as a function of the mass, and therefore all trajectories start from \(J_{orbital}=0\) and the intercept is always negative).

We also wish to understand the dependence of the phenomenological intercept as a function of the endpoint masses. Below, in table \ref{tab:a_meson} we list the fitted values for the intercepts of the various orbital HMRTs of mesons described in this work, arranged by the quark contents of the meson. The values listed do not give a very clear picture of the mass dependence. It should also be noted that the intercept is very sensitive to changes in the quark masses and slope, and much more so percentage-wise than any of the predictions for the observables, the hadron masses or widths, given throughout this note. 

\begin{table}[h!] \centering
	\begin{tabular}{|l|c|c|c|c|} \cline{2-5}
		\multicolumn{1}{c|}{} & \(u/d\) & \(s\) & \(c\) & \(b\) \\ \hline
		
		\(u/d\) &	-0.46		&	-0.29		&	-0.38		& -0.65		\\ \hline
		
		\(s\) &		-0.29		&	-0.14		&	-0.10		& -0.08		\\ \hline
		
		\(c\) &		-0.38		&	-0.10 	&	-0.08		& 	-			\\ \hline
		
		\(b\) &		-0.65		&	-0.40		&	-				&  -0.27 	\\ \hline
	\end{tabular} \qquad
	\begin{tabular}{|l|c|c|c|c|} \cline{2-5}
		\multicolumn{1}{c|}{} & \(u/d\) & \(s\) & \(c\) & \(b\) \\ \hline
		
		\(u/d\) & -0.26		&	-0.01		&	-0.25		& -0.54		\\ \hline
		
		\(s\) &		-0.01		&	-				&	-0.04		& -0.31		\\ \hline
		
		\(c\) &		-0.25		&	 -0.04	&	0.00		& -		\\ \hline
		
		\(b\) &	-0.54			&	-0.31		&	-		&  0.00 	\\ \hline
	\end{tabular}	\caption{\label{tab:a_meson} The intercepts of the various meson trajectories, arranged by quark content. On the left are the vector meson trajectories, on the right the pseudoscalar trajectories.}
\end{table}

%\clearpage

%\section{Checklist}
%These are all done:
%\begin{itemize}
%	\item  Higher states on the meson Regge trajectories of \cite{Sonnenschein:2014jwa}.
%	\item  Higher states on the baryon Regge trajectories of \cite{Sonnenschein:2014bia}.
%	\item  More mesons that were absent from the first paper: \(K\), \(D^*\), \(D_s\), \(B\), \(B^*\), \(B_s\), \(B_s^*\).
%	\item  More baryons: charmed, doubly charmed, bottom.
%	\item  Higher states on the glueball Regge trajectories - the spectra of \(f_0\) and \(f_2\) resonances in general.
%	\item  Higher tetraquark states: the \(Y(4630)\) trajectory.
%	\item  Conjectured analogs of the \(Y(4630)\): \(Y_b\), \(Y_s\) and their respective trajectories.
%	\item  Decays: we had a few predictions involving \(\lambda_s\) in glueball decays.
%\end{itemize}

%Question we did not address (yet):
%\begin{itemize}
	%\item The radial trajectories we presented, just based on selecting usable data, are all for flavorless mesons, or for light hadrons. Is there a reason for that?
%\item Estimation of widths for states that weren't included in \cite{Sonnenschein:2017ylo}.
	%\item predictions without head of the trajectory for instance states with $(ccc)$ and $(bb q)$?
%\end{itemize}

 %%%%%%%%%%%%%%%%%%%%%%%%%%%%%%%%%
\end{document}